\documentclass[12pt,a4paper]{article}

\usepackage{natbib}

\usepackage{amsmath, amssymb, amsthm, bm}
\usepackage{mathtools, booktabs, setspace}
\usepackage[colorlinks=true, linkcolor=blue, citecolor=blue]{hyperref}
\usepackage[left=2.5cm,right=2.5cm,top=2.5cm,bottom=3.2cm]{geometry}
\usepackage[most]{tcolorbox}
\usepackage{multirow}

\newcommand{\dd}{\,\mathrm{d}}
\newcommand{\diag}{\text{diag}}
\newcommand{\Span}{\text{span}}
\newcommand{\sign}{\text{sign}}
\DeclareMathOperator*{\tr}{tr}
\DeclareMathOperator*{\plim}{plim}
\DeclareMathOperator*{\argmin}{argmin}
\DeclareMathOperator*{\argmax}{argmax}

\newtheorem{theorem}{Theorem}
\newtheorem{proposition}{Proposition}
\newtheorem{assumption}{Assumption}
\newtheorem{lemma}{Lemma}
\theoremstyle{remark}

\begin{document}

  \title{\bf Approximate Factor Models for Functional Time Series}
  \author{Sven Otto\thanks{
  Corresponding author: Sven Otto, Institute of Econometrics and Statistics, University of Cologne, Albertus-Magnus-Platz, 50923 Cologne, Germany. Email: sven.otto@uni-koeln.de
    }\hspace{.2cm}\\
    Institute of Econometrics and Statistics, University of Cologne\\
    and \\
    Nazarii Salish \\
    Department of Economics, Universidad Carlos III de Madrid}
  \maketitle

  \doublespacing

\begin{abstract}
We propose a novel approximate factor model tailored for analyzing time-de\-pen\-dent curve data.
Our model decomposes such data into two distinct components: a low-dimensional predictable factor component and an unpredictable error term.
These components are identified through the autocovariance structure of the underlying functional time series.
The model parameters are consistently estimated using the eigencomponents of a cumulative autocovariance operator and an information criterion is proposed to determine the appropriate number of factors.
Applications to mortality and yield curve modeling illustrate key advantages of our approach over the widely used functional principal component analysis, as it offers parsimonious structural representations of the underlying dynamics along with gains in out-of-sample forecast performance.
\end{abstract}

\noindent
{\it Keywords:} Curve data, functional data analysis, identification, information criterion, \\ forecasting, yield curve modeling \vspace{2ex}

\noindent
{\it JEL Classification:}  C32, C38, C55, E43
\vfill

\newpage
\section{Introduction}  \label{sec:intro}
Over the past decades, approximate and dynamic factor models have emerged as powerful tools for analyzing high-dimensional datasets, offering a parsimonious framework that captures the essential underlying structure while filtering out irrelevant parts.
These models, first conceptualized by \cite{chamberlain1983} and expanded upon by  \cite{forni2000}, \cite{stock2002, stock2002b}, and \cite{bai2003}, have become fundamental tools in areas such as economic forecasting, monetary policy analysis, psychology, environmental sciences and social sciences.
For comprehensive reviews, see \cite{breitung2013} and \cite{stock2016}.

Building on the success of these models in handling high-dimensional data, the generalization of factor modeling techniques to functional or infinite-dimensional data structures has gained significant attention.
This transition to functional data analysis (FDA) is a natural extension, as this field focuses on datasets consisting of continuous curves or functions, as discussed extensively in \cite{ramsay2005}, \cite{horvath2012}, \cite{hsing2015}, and \cite{kokoszka2017}.
FDA is particularly valuable in economic applications, where data often take functional forms, such as term structures of bond yields, credit default swaps, income profiles, and inflation expectations, for which conventional multivariate methods may prove overly restrictive or inadequate.
In this paper, we contribute to the literature on factor models for functional data by proposing a novel approximate factor model that explicitly models the dynamics of time-dependent functional data (or functional time series).

The study of factor models in FDA has seen significant progress in recent years.
Early contributions focused on generalizing classical factor models to the functional setting.
For instance, \cite{hays2012} and \cite{liebl2013} introduced functional factor models with a discrete error component, while \cite{hormann2022, hormann2023} and \cite{ofner2024} demonstrated how discretely observed functional data follow an approximate factor model structure.
These works primarily utilized the functional counterpart of principal component analysis (PCA), building on directions of maximal variability for factor identification.
The same PCA-based approach has been extended to functional time series.
For example, \cite{hyndman2009} and \cite{aue2015} applied PCA techniques to model and forecast functional time series, while further extensions have addressed challenges related to nonstationarity and long-range dependence (see, e.g., \citealt{salish2019} and \citealt{Li2020}).
More recent contributions have refined these methods, identifying factors asymptotically through large panel structures of high-dimensional functional time series datasets (\citealt{hallin2023, tavakoli2023, leng2024, guo2025, li2025}).

The identification and estimation of factors in functional data analysis typically rely on selecting directions of highest variability via functional PCA.
While effective for dimension reduction, this approach may fail to capture serial dependence and predictability, both of which are integral to functional time series analysis.
As argued by \cite{forni2000}, \cite{panaretos2013}, and \cite{hormann2015}, a sole focus on variability can lead to an incomplete representation of the data's dynamic structure, potentially overlooking important serial dependencies.

We address this limitation by proposing an approximate factor model for functional time series that is explicitly designed to capture their low-dimensional dynamics.
As in classical factor models, we decompose a functional time series into two distinct components: a low-dimensional \emph{factor component} and a remainder term, the \emph{error component}.
Our key innovation is to identify the factors not from directions of maximal variability, as in PCA, but from the autocovariance structure of the process so that they capture serial dependence and predictability.
Consequently, the \emph{factor component} and the \emph{error component} can be interpreted as \emph{predictable} and \emph{unpredictable} parts of the process, respectively.
Because we only impose that the error component carries no information about the dynamics of the original functional time series,
we can relax conventional restrictions on the error term of a factor model: (i) the eigenvalues of the factor and error components may be of comparable magnitude; (ii) the error function may exhibit non-negligible correlations across domain points, even asymptotically, hence our use of the term “approximate” in the sense of \cite{chamberlain1983}; and (iii) weak cross-correlations between the factors and the error component are allowed, accommodating various forms of nonstationarity and heteroskedasticity.

The identification of the number of factors and of the dynamic subspace builds on the cumulative autocovariance operator introduced by \cite{bathia2010} and used in related work \citep{chen2022,diks2023,lihao2025}.
We adopt the same device but pursue a different and broader objective.
Existing approaches (e.g., \citealp{bathia2010,lihao2025}) focus primarily on identifying the dimensionality of the dynamic subspace and do not aim to provide a structural representation (or identification) of the dynamic component.
As a result, the low-dimensional dynamics are typically modeled ex post, in a separate step, without delivering an identified population predictor.
Our contribution can be viewed as a natural next step in this line of research.
Beyond recovering the dynamic subspace and its dimension, we provide a structural characterization of the model by identifying the loading functions, the factors, and their autoregressive dynamics.
This structural specification in turn yields an identified population predictor for the functional time series and allows us to evaluate its population prediction error.

Because of these identification results (for the factor dimensionality, the loading functions, and the factors’ dynamic process),  we are able to develop consistent estimators and model selection procedures.
We derive consistent estimation for the model’s primitives, including the autocovariance structure, and address the challenge that the factor dimension is typically unknown and must be inferred from the data.
To tackle this challenge, we propose an information criterion, drawing on ideas from the factor-modeling literature (e.g., \citealt{bai2002,hallin2007}).
Our criterion is based on the prediction error curve and assumes that the factor component follows a vector autoregressive (VAR) process with an unknown lag order.
We incorporate a carefully designed penalty term to mitigate over-selection, and we show that, under mild conditions, the criterion consistently selects both the number of factors and the lag order.
All procedures are implemented in an accompanying \textsf{R} package.\footnote{\url{https://github.com/ottosven/dffm}}
Note a related information criterion is proposed by \cite{aue2015} for selecting the number of functional principal components relevant for prediction, but its construction does not directly apply to our approximate factor setting (see Section~\ref{sec:estimK} for a detailed comparison).

Finally, to illustrate the practical utility of our model, we apply it to mortality and yield curve modeling and forecasting.
In both applications, we compare our approach with functional PCA, which has arguably become the main applied workhorse model in functional time series analysis since the seminal work of \cite{aue2015}.
Additionally, for yield curves, we benchmark our model against the widely used dynamic Nelson-Siegel model, as reviewed in \cite{diebold2013}.
Our main findings highlight the importance of a careful structural representation of process dynamics -- that is, reducing the dimensionality of the original process in a meaningful way by extracting factors that drive the dynamics rather than merely capturing variability.
For instance, in the case of mortality rates, while the first three factors in our framework closely resemble those identified by PCA, the remaining factors differ.
This distinction proves important, as selecting factors based on their role in driving dynamics rather than explaining variability leads to improved forecasting performance.
In the case of yield-curve modeling, our results indicate that the number of factors required to capture yield curve dynamics varies over time, with more factors needed during periods of economic uncertainty than in stable periods.
These findings suggest that yield curves have a richer and more adaptive dynamic structure than typically considered in the literature.

The paper is structured as follows:
Section \ref{sec:model} develops the approximate functional factor model, providing detailed discussions on all necessary assumptions and the identification of its components.
In Section \ref{sec:estimation}, we introduce the estimator for the functional component, discuss the information criterion for jointly estimating the number of factors and their dynamics, and show their consistency. We also provide guidance on practical implementation.
Section \ref{sec:simulations} presents a Monte Carlo simulation to assess the model's performance and the proposed estimation methods in finite samples.
In Section \ref{sec:applications}, we apply the method to mortality rate and yield curves. Finally, Section \ref{sec:conc} concludes the paper.

To ease the readability of the paper, we collect the key notations below.
Consider $H=L^2([a,b])$, the space of square-integrable functions $f:[a,b] \rightarrow \mathbb{R}$ satisfying $\int_a^b f(r)^2 dr < \infty$ for $a < b$. $H$ forms a Hilbert space equipped with the inner product $\langle f, g \rangle = \int_a^b f(r)g(r) dr$ and the norm $\|f\| = \sqrt{\langle f, f \rangle}$, for $f,g \in H$.
Any square-integrable kernel function $\rho: [a,b] \times [a,b] \rightarrow \mathbb{R}$ defines an integral operator $\mathcal{R}: H \rightarrow H$, $f(\cdot) \mapsto \int_a^b \rho(\cdot,s)f(s) ds$ with squared Hilbert-Schmidt norm $\Vert\mathcal R\Vert_{\mathcal{S}}^2 = \int_a^b \int_a^b \rho(r,s)^2 \dd s \dd r < \infty$.
The image space of $\mathcal R$ is $Im(\mathcal R) = \{ g \in H: g(r) = \int_a^b \rho(r,s) f(s) \dd s \ \text{for some} \ f \in H\}$, and the rank of $\mathcal R$ is the dimension of its image space.
An eigenvalue-eigenfunction pair $(\xi, v)$ of $\mathcal R$ satisfies $\int_a^b \rho(r,s)v(s) ds = \xi v(r)$, for all $r \in [a,b]$. If $\rho(r,s)$ is symmetric and positive semi-definite, all eigenvalues are real, and eigenfunctions associated with distinct eigenvalues are orthogonal.
The adjoint operator $\mathcal R^*$ of $\mathcal R$ is the integral operator with kernel function $\rho^*(r,s) = \rho(s,r)$.
See \cite{hsing2015}, Sections 3 and 4, for a detailed exposition of the relevant operator theory.
To clarify vector and matrix norms, we use $\|\cdot\|_2$ for the Euclidean vector norm and $\|\cdot\|_M$ for the compatible Frobenius matrix norm.

\section{The Approximate Functional Factor Model}  \label{sec:model}

We consider a time series of curves $Y_1(r), \ldots, Y_T(r)$ defined on the domain $r \in [a,b]$.
Our main goal is to develop a framework that is capable of capturing the low-dimensional dynamic behavior of the given functional time series. To achieve this, we employ a general factor model framework:
\begin{align}
	Y_{t}(r) &= \mu(r)+\sum_{l=1}^K F_{l,t} \psi_{l}(r) + \epsilon_{t}(r) = \mu(r)+(\Psi(r))' F_t + \epsilon_t(r), \label{eq:factormodel}
\end{align}
where $t=1, \ldots, T$ and $r \in [a,b]$.
Here, $F_t = (F_{1,t}, \ldots, F_{K,t})'$ represents the $K\times 1$ vector of factors driving the dynamic part of $Y_t(r)$, while the error term $\epsilon_t(r)$ carries no relevant serial dependence signal, and $\mu(r)$ is the standard intercept function. The vector $\Psi(r) = (\psi_1(r), \ldots, \psi_K(r))'$ collects $K$ loading functions determining how each factor contributes to the curve series.
All components of this model, including the number of factors, $K$, the vector of loading functions, $\Psi(r)$, and the factors, $F_t$, are unobserved. To properly identify their role and estimate them, an additional set of conditions is necessary, which we discuss in detail below.

We begin by separating the factor component from the error term, which is a fundamental step in our modeling approach. Central to this step is the concept of serial dependence, which allows us to isolate the low-dimensional dynamic factor component from the noise. This concept is formalized through a set of restrictions outlined in the following assumption:

\begin{assumption} \label{as:components}
Model \eqref{eq:factormodel} holds true with
\begin{itemize}
	\item[(a)] $E[\epsilon_t(r) \mid  Y_{t-1}, Y_{t-2}, \ldots] = 0$ for all $t$ and $r \in [a,b]$;
	\item[(b)] $E[F_t]=0$ for all $t$, and $M := \sum_{\tau=1}^{q_0} \int_a^b (M_\tau(s))(M_{\tau}(s))' \dd s$ is positive definite for some integer $q_0\geq 1$, where ${M_\tau(s)} := \lim_{T \to \infty} T^{-1} \sum_{t=\tau+1}^T E[F_t Y_{t-\tau}(s)]$;
	\item[(c)] $\psi_1, \ldots, \psi_K$ are linearly independent and continuous functions.
\end{itemize}
\end{assumption}
By Assumption \ref{as:components}(a), the error term $\epsilon_t$ has no correlation with lagged $Y_t$, implying that it cannot be predicted from the past observations of $Y_t$.
In contrast, the factors globally correlate with at least one lag of the original process, $Y_t$, as specified in Assumption \ref{as:components}(b).
To be more specific, the positive definiteness of the matrix $M$ implies that $\lim_{T\to \infty} \sum_{\tau=1}^{q_0} \int_a^b ( T^{-1} \sum_{t=1}^T E[F_{l,t}  Y_{t-\tau}(r)] )^2 \dd r > 0$, ensuring that the global covariance between $F_{l,t}$ and $Y_{t-\tau}$ is non-zero for any $l=1, \ldots, K$ for at least one lag $\tau=1,\ldots,q_0$, and there is no cancellation of this covariance by the integration.
To ensure broad applicability of our framework, we adopt the concept of global covariance, as defined by \cite{white2001}, (i.e., $\lim_{T \to \infty} T^{-1} \sum_{t=\tau+1}^TE[F_{t}  Y_{t-\tau}(r)]$ is used instead of $E[F_{t}  Y_{t-\tau}(r)]$).
This concept accommodates potential heteroskedasticity and local nonstationarities within the original process $Y_t$, which are common in economic applications.
The role of $q_0$ is similar to that of portmanteau tests for autocorrelation, ensuring each factor correlates with at least one of the first $q_0$ lags of $Y_t$.
\cite{lam2012} and \cite{zhang2019} argue that a small $q_0$ is sufficient in practice, and in non-seasonal setups, $q_0 = 1$ can be chosen.
It is important to note that $q_0$ serves only to identify the factor model parameters and does not specify the type or features of temporal dependencies imposed on the process $Y_t$.
Finally, from Assumption \ref{as:components}(a) and (b), the intercept $\mu(r)$ is identified as the mean function $E[Y_t(r)]$.
This renders the first two terms in \eqref{eq:factormodel}, expressed as
\begin{equation} \label{eq:factorcomponent}
	\chi_t(r) := \mu(r) + (\Psi(r))'F_t,
\end{equation}
predictable from the original process $Y_t(r)$, justifying the terminology introduced in the introduction section for the corresponding parts as \emph{predictable} and \emph{unpredictable}.

The set of loading functions forms a basis for the factor space $H_F := \Span(\psi_1, \ldots, \psi_K)$, where $Y_t$ exhibits autocovariances.
The linear independence of the loading functions,  as dictated by Assumption \ref{as:components}(c), implies that $H_F$ is $K$-dimensional.
Essentially, this indicates the presence of $K$ distinct directions along which $Y_t$ displays temporal dependence.
Consequently, identifying $H_F$ plays a pivotal role in understanding the autocovariance structure of $Y_t$, leading to the subsequent step in our model: identifying the loading functions and determining their number.

The structure of the global autocovariance of $Y_t$ depends solely on the structure of the autocovariance of the factors and the directions represented by the loadings.
According to our model \eqref{eq:factormodel}, the global $\tau$-th order autocovariance function is expressed as:
\begin{equation} \label{eq:autocovariance}
c_\tau(r,s) := \lim_{T \to \infty} \frac{1}{T} \sum_{t=\tau+1}^T Cov[Y_t(r), Y_{t-\tau}(s)] = (\Psi(r))'(M_\tau(s)).
\end{equation}

Here, $c_\tau(r,s)$ is the kernel function of the global integral autocovariance operator $C_\tau$.
Heuristically, each autocovariance $C_\tau$ for $\tau=1,...,q_0$ captures some, but not necessarily all, of the directions of interest across which the original series exhibits serial dependence.
This fact follows from \eqref{eq:autocovariance}, as these directions are linear combinations of $\psi_1, \ldots, \psi_K$.
Specifically, the directions captured by $C_\tau$ are represented by its image, $Im(C_\tau)$, the space spanned by the right-singular functions of $C_\tau$.
Following standard results on singular value decomposition, these functions are the orthonormal eigenfunctions of the positive semi-definite operator $C_\tau C_\tau^*$, where $C_\tau^*$ denotes the adjoint operator of $C_\tau$.
Since $Im(C_\tau)$ may only capture a subset of the relevant directions (i.e., $Im(C_\tau) \subset H_F$), a combined analysis of such operators for $\tau=1,...,q_0$ becomes necessary.

We adopt the approach used in \cite{bathia2010}, introducing a cumulative operator $D = \sum_{\tau=1}^{q_0} C_\tau C_\tau^*$, with its kernel function given by:
\begin{equation*}
	d(r,s) := \sum_{\tau = 1}^{q_0} \int_a^b c_\tau(r,q) c_\tau(s,q) \dd q = (\Psi(r))'M(\Psi(s)).
\end{equation*}
The cumulative autocovariance operator $D$ is symmetric with $Im(D) = Im(D^*) = H_F$ and $rank(D) = K$, as implied by Assumption \ref{as:components}(b) and (c).
By construction, its collection of ordered orthogonal eigenfunctions, denoted here as $d_1,...,d_K$, span the entire space $H_F$, thereby capturing all directions in which $Y_t$ exhibits serial dependence.
However, the loading functions $\psi_1, \ldots, \psi_K$ also span $H_F$, indicating that they are linear combinations of $d_1,...,d_K$.
Hence, while $H_F$ can be properly identified through $d_1,...,d_K$, the loadings are identified up to some rotation of these eigenfunctions.

We employ the cumulative autocovariance operator \(D\) in the spirit of \citet{bathia2010}, but with a different objective.
In their signal–noise framework \(Y_t = X_t + \epsilon_t\), the main aim is to identify the dynamic subspace \(\mathcal{M}\) of \(X_t\) and its dimensionality, which they achieve via the range of \(D\).
In our setting, \(D\) serves the same role, so \(H_F = \mathcal{M}\).
We then build on this step and proceed further: we move from subspace recovery to a structural characterization of the dynamic component of \(Y_t\), identifying a full factor structure with explicit loading functions and factor dynamics.
This, in turn, yields an identified one-step-ahead population predictor and its population prediction error, providing the theoretical foundation for a consistent information criterion.

To achieve the exact identification, we resort here to the solution routinely used in conventional factor analysis, where loadings are assumed to be orthonormal and factors have a diagonal covariance matrix (see, e.g., \citealt{stock2002}, and \citealt{bai2013}).
Unlike classical factor analysis, we impose a diagonal structure on the positive definite matrix $M$, as our model aims to identify factors based on the cumulative autocovariance rather than the highest variability.
This leads us to the next set of restrictions outlined in the following assumption:
\begin{assumption} \label{as:rotation}
\
\begin{itemize}
\item[(a)] The loading functions $\psi_1, \ldots, \psi_K$ satisfy $\|\psi_l\|=1$ and $\langle \psi_l, \psi_m \rangle= 0$ for $l \neq m$.
\item[(b)] The matrix $M$ is diagonal with $M = \diag(\lambda_1, \ldots, \lambda_K)$ and $\lambda_1 > \ldots > \lambda_K > 0$.
\end{itemize}
\end{assumption}
These conditions fix the rotation of the loading functions in the factor space $H_F$, ensuring their exact identification.
To see this, by Assumption \ref{as:rotation}(b), the kernel function of the operator $D$ satisfies $d(r,s) = (\Psi(r))' M (\Psi(s)) = \sum_{l=1}^K \lambda_l \psi_l(r) \psi_l(s)$, implying that $\lambda_1, \ldots, \lambda_K$ are the descendingly ordered nonzero eigenvalues of $D$.
Then the loadings $\psi_1, \ldots, \psi_K$ are identified as the eigenfunctions $d_1,...,d_K$ of operator $D$ up to a sign change and their number as the rank of this operator.
It is noteworthy that $\psi_1, \ldots, \psi_K$ as eigenfunctions satisfy certain optimality properties given in the proposition below:

\begin{proposition}\label{prop:optimality}
Let $\mathcal{Q}(f)
  :=
  \lim_{T\to\infty}
  \sum_{\tau=1}^{q_0}
  \int_{a}^{b}
    (
      T^{-1}
      \sum_{t=\tau+1}^{T}
      E[
        \langle Y_t-\mu , f\rangle Y_{t-\tau}(r)]
    )^{2}
  \dd r$.
Then, under Assumptions \ref{as:components}–\ref{as:rotation}, the maximizing directions are obtained recursively as
$$
	\psi_1
  \;=\;
  \underset{\|f\|=1}{\arg\max}\;
  \mathcal{Q}(f), \qquad \psi_l
  \;=\;
  \argmax\limits_{\substack{f \in \Span(\psi_1,...,\psi_{l-1})^\perp \\ \|f\|=1}}
  \mathcal{Q}(f), \quad l=2, \ldots,K.
$$
Moreover, $\mathcal{Q}(g) = 0$ for every $g \in H_F^{\perp}$.

\end{proposition}

\noindent
According to Proposition \ref{prop:optimality}, the projection of $Y_t - \mu$ onto $H_F$, as defined by
\begin{equation}	\label{eq:projection}
Y_t^*(r):= \mu(r) + \sum_{l=1}^K\langle Y_t-\mu, \psi_l \rangle  \psi_l(r),
\end{equation}
captures all components of the functional time series that correlate with its past $q_0$ lags.
The projection coefficient $\langle Y_t - \mu, \psi_1\rangle$ is optimal in the sense that there are no other projection coefficients that have a higher dependency with these lags.
The second projection coefficient $\langle Y_t - \mu, \psi_2 \rangle$ is optimal among all projections orthogonal to $\psi_1$, and this pattern continues with subsequent coefficients.

As the loadings $\psi_1,\ldots,\psi_K$ are deterministic, the dynamic nature of $Y_t$ can be equivalently represented by the $K\times 1$ vector $F_t^* := (F_{1,t}^*, \ldots, F_{K,t}^*)'$, where $F_{l,t}^* := \langle Y_t - \mu, \psi_l \rangle$ are the scores of projection \eqref{eq:projection} for $l=1,\ldots,K$.
Further from model \eqref{eq:factormodel}, we deduce:
\begin{equation}\label{eq:IdentFactors}
  F_{l,t}^* = \langle Y_t-\mu, \psi_l \rangle = F_{l,t} + \langle \epsilon_t, \psi_l \rangle.
\end{equation}
This implies that the factors $F_{l,t}$ are partially identified as $F_{l,t}^*$, up to an unpredictable noise error term $\langle \epsilon_t, \psi_l \rangle$.
In other words, under Assumptions \ref{as:components} and \ref{as:rotation}, the factor model \eqref{eq:factormodel} cannot be distinguished from its orthogonalized form $Y_t(r) = \mu(r) + (\Psi(r))'F_t^* + \epsilon_t^*(r)$,
where $ \epsilon_t^*(r) := \epsilon_t(r) - \sum_{l=1}^K \langle \epsilon_t, \psi_l \rangle \psi_l(r)$.

Achieving complete factor identification may require imposing further restrictions on the model.
For example, one could assume that $\epsilon_t$ solely takes values in $H_F^\perp$ (the orthogonal complement of $H_F$), implying $F_t^* = F_t$ for all $t$.
Alternatively, a weaker version of this restriction, where the variance of $\epsilon_t$ in $H_F$ is asymptotically negligible, could be considered, as often done in conventional factor literature.
Additionally, adopting smoothness versus roughness type of restrictions could help to separate factors from the part of the error term in $H_F$, as proposed in \cite{descary2019}.
From a different perspective, one may postulate a parametric dynamic law directly for the latent factor process $F_t$ while allowing $\langle \epsilon_t,\psi_l\rangle\neq 0$; in such settings, autocovariance-based moment (GMM-type) estimators that exploit lagged autocovariances can be used to estimate the dynamic parameters, as in \citet{chen2022} and \citet{chang2024}.
We refrain from imposing additional restrictions in this study to maintain the generality of our framework and proceed with the identified score process $F_{l,t}^*$ in the subsequent analysis.
Furthermore, all results presented in the subsequent sections regarding the estimation of the model's primitives, the subspace $H_F$ and the suggested information criterion for the number of factors, $K$, do not require complete factor identification.

While Assumptions \ref{as:components} and \ref{as:rotation} establish the necessary restrictions to identify the role of each unobserved component in our model, additional constraints are required to specify time dependencies allowed for the process $Y_t$ as well as factors. These restrictions are required to establish the asymptotic properties of the estimators proposed in the subsequent sections.
\begin{assumption} \label{as:VAR} \
\begin{itemize}
	\item[(a)] The modified factors $F_t^*$ follow a $K$-variate $VAR(p)$ model, described by $A(L) F_t^* = \eta_t$, where $A(L) = I_K - \sum_{i=1}^p A_i L^i$ is the lag polynomial with $L$ as the backshift operator.
It is assumed that $\det(A(z))$ has all roots outside the unit circle and $A_p \neq 0$.
The innovations vector $\eta_{t}$ forms a martingale difference sequence with zero conditional mean and a positive definite global conditional covariance matrix $\Sigma_\eta$, i.e., $E[\eta_t | \eta_{t-1}, \eta_{t-2}, \ldots] = 0$ and $\lim_{T \to \infty} T^{-1} \sum_{t=1}^T E[\eta_t\eta_t' | \eta_{t-1}, \eta_{t-2}, \ldots] = \Sigma_\eta$.
Moreover,  $\lim_{T \to \infty} \sup_{i_1, i_2, i_3,i_4 \in \mathbb{N}} T^{-1} | \sum_{t,s=1}^T Cov[ \eta_{l_1,t-i_1} \eta_{l_2,t-i_2}, \eta_{l_3, s-i_3} \eta_{l_4, s-i_4}]| < \infty$ for all $l_1, l_2, l_3, l_4 \in\{ 1, \ldots, K\}$, where $\eta_{l,t}$ denotes the $l$-th element of $\eta_{t}$,
and $\sup_{t \in \mathbb Z} E[\eta_{l,t}^4] < \infty$ for any $l=1, \ldots, K$.
	\item[(b)] The error term satisfies $\lim_{T \to \infty} \sup_{s \in [a,b], h \geq 0} E[\|T^{-1/2} \sum_{t=h+1}^T F_{t}^* \epsilon^*_{t-h}(s) \|_2] < \infty$ and $\sup_{r \in [a,b]} \sup_{t \in \mathbb Z} E[(\epsilon_t^*(r))^4] < \infty$.
\end{itemize}
\end{assumption}

Vector autoregressions, as described in Assumption~\ref{as:VAR}(a), offer a general and convenient framework for modeling the time dependence of the modified factors and allow us to estimate the loadings and scores at parametric rates.
Assumption~\ref{as:VAR}(a) also clarifies the type of nonstationarity that can be accommodated by the global covariance concept introduced in Assumption~\ref{as:components}.
Although the modified factors $F_t^*$ follow a stable $\operatorname{VAR}(p)$ process with time-invariant autoregressive coefficients, the processes $F_t^*$, $\eta_t$, and $Y_t$ need not be covariance stationary in the usual sense of having date-invariant second moments. Taken together, Assumptions~\ref{as:components}(a)--(b) and~\ref{as:VAR}(a) ensure that the time-averaged limits defining the global counterparts of the relevant factor second moments and covariance functions of $Y_t$ are well defined.
These global quantities, rather than their date-specific counterparts, are central for identification and estimation.
Alternative weak-dependence conditions may instead be formulated using strong mixing or $L^4$-$m$-approximability; see \citet{hoermann2010}. Assumption~\ref{as:VAR}(b) further permits weak dependence between the factors and lagged errors, similar to that considered by \citet{bai2003}, provided that it is sufficiently small asymptotically.

\section{Estimation} \label{sec:estimation}
The previous section's identification results demonstrate that all unobserved components in the model can be expressed using the global autocovariances of the functional time series $Y_t$.
Most of these components can be estimated using method of moments estimators by substituting population moments with sample equivalents.
Section \ref{sec:estimPrim} elaborates on the consistency of these estimators.
However, estimating the number of factors presents a more challenging task, which is addressed in Section \ref{sec:estimK}, where we introduce an information criterion for consistently determining the number of factors.
Finally, Section \ref{sec:practical} provides practical implementation guidelines, including an estimation and prediction algorithm.

\subsection{Estimation of parameter functions} \label{sec:estimPrim}
Consider the sample mean function $\widehat \mu(r) = T^{-1} \sum_{t=1}^T Y_t(r)$,
the $\tau$-th order sample autocovariance function
$\widehat c_\tau(r,s) := T^{-1} \sum_{t=\tau+1}^T  (Y_t(r)-\widehat\mu(r))(Y_{t-\tau}(s)-\widehat\mu(s))$,
and the sample counterpart of the cumulative autocovariance function $d(r,s)$ given as
\begin{equation}
\widehat d(r,s) = \sum_{\tau=1}^{q_0} \int_a^b \widehat c_\tau(r,q) \widehat c_\tau(s,q) \dd q. \label{eq:d-estimator}
\end{equation}
The integral operators with kernel functions $\widehat c_\tau(r,s)$ and $\widehat d(r,s)$ are denoted as $\widehat C_\tau$ and $\widehat D$, respectively.
Let $\widehat \lambda_1 \geq \ldots \geq \widehat \lambda_T \geq 0$ be the eigenvalues of $\widehat D$, and let $\widehat \psi_1, \ldots, \widehat \psi_T$ be corresponding orthonormal eigenfunctions.
In practice, the eigenequation and the integral in \eqref{eq:d-estimator} are computed by numerical integration.
A ready-to-use implementation is provided in our accompanying R package.

\begin{theorem} \label{thm:consistency}
Under Assumptions \ref{as:components}--\ref{as:VAR},
\begin{itemize}
	\item[(a)] $\|\widehat \mu - \mu \| = O_P(T^{-1/2})$
	\item[(b)] $\|\widehat C_\tau - C_\tau \|_{\mathcal S} = O_P(T^{-1/2})$ for all $\tau = 1, \ldots, q_0$
	\item[(c)] $\|\widehat D - D \|_{\mathcal S} = O_P(T^{-1/2})$
	\item[(d)] $|\widehat \lambda_l - \lambda_l | = O_P(T^{-1/2})$ for all $l = 1, \ldots, K$ and $\widehat \lambda_l = O_P(T^{-1/2})$ for $l > K$.
	\item[(e)]
	$\|\widehat \psi_l - s_l \psi_l \| = O_P(T^{-1/2})$ for all $l = 1, \ldots, K$, where $s_l = \sign(\langle\widehat \psi_l, \psi_l \rangle)$
\end{itemize}
\end{theorem}

Theorem \ref{thm:consistency} implies that the parameter functions in model \eqref{eq:factormodel} are consistently estimated with parametric rates of convergence.
To estimate the factors themselves, given the estimated intercept and loading functions, we use the sample equivalents of the factors in \eqref{eq:IdentFactors}, defined as
$$\widehat F_{l,t} := \langle Y_t - \widehat \mu, \widehat \psi_l \rangle, \quad l=1, \ldots, K,\  t= 1, \ldots, T.$$
Using this projection coefficient as our factor estimator is further justified by the least squares principle, as it optimizes the model fit by minimizing
$$
	\Big\| Y_t - \widehat \mu - \sum_{k=1}^K f_{k,t} \widehat \psi_k \Big\|^2 = \sum_{k=1}^K \big( f_{k,t}^2 - 2 f_{k,t} \langle Y_t - \widehat \mu, \widehat \psi_k \rangle \big) + \big\| Y_t - \widehat \mu \|^2,
$$
where the minimum is attained when $f_{l,t} = \widehat F_{l,t}$.

Theorem \ref{thm:consistency}(e) highlights the relevance of the sign $s_l = \sign(\langle\widehat \psi_l, \psi_l \rangle)$ for our theoretical analysis, since the signs of the eigenfunctions are unidentified.
Conditional on the chosen signs for the eigenfunctions of $\widehat D$, the sign-adjusted quantities $s_l \psi_l$ and $s_l F_{l,t}^* = \langle Y_t - \mu, s_l \psi_l \rangle$ serve as the population equivalents of $\widehat \psi_l$ and $\widehat F_{l,t}$.
However, flipping the signs of the loadings also flips the signs of the factors, so the products $\widehat F_{l,t} \widehat \psi_l$ and $F_{l,t}^* \psi_l$ remain invariant to sign changes.
Together with Theorem \ref{thm:consistency}, a direct consequence is that
$T^{-1} \sum_{t=1}^T \| \sum_{l=1}^K ( \widehat F_{l,t} \widehat \psi_l - F_{l,t}^* \psi_l ) \| = O_P(T^{-1/2})$,
since, for any given $t=1,...,T$, the estimation error $|\widehat F_{l,t} - s_l F_{l,t}^*|$ is bounded from above by $\|\mu - \widehat \mu \| + \|Y_t - \mu \| \|\widehat \psi_l - s_l \psi_l\|$.

\subsection{Estimation of the number of factors}\label{sec:estimK}
To determine the number of factors $K$, we exploit their dynamic VAR structure, defined in Assumption \ref{as:VAR}, and construct an information criterion for its selection.
The advantage of this approach is that it not only aids in the estimation of K but also provides the opportunity to estimate the number of lags of the VAR structure, which is an essential step for empirical applications.

The temporal dynamics of the curve process $Y_t$ and its latent factors are characterized by the $K\times pK$ matrix of autoregressive coefficients denoted by $\bm{A} = [A_1, \ldots, A_p]$, as defined in Assumption \ref{as:VAR}(a).
We employ the standard conditional least squares (LS) estimator to estimate $\bm A$.
For a selected number of factors $J$ and lags $m$, the unknown $K\times 1$ vectors $F_t^*$ are replaced with the $J \times 1$ vectors of sample scores $\widehat F_t^{(J)} = (\widehat F_{1,t}, \ldots, \widehat F_{J,t})'$.
The LS estimator is given by
\begin{align} \label{eq:LS}
	\bm{\widehat A}_{(J,m)} =  \widehat\Gamma_{(J,m)} \widehat\Sigma_{(J,m)}^{-1}
\end{align}
with $\widehat\Gamma_{(J,m)}= T^{-1}\sum_{t=m+1}^T \widehat{F}_t^{(J)}(\bm{\widehat x}_{t-1}^{(J,m)})'$ and $\widehat\Sigma_{(J,m)}= T^{-1}\sum_{t=m+1}^T \bm{\widehat x}_{t-1}^{(J,m)}(\bm{\widehat x}_{t-1}^{(J,m)})'$, where the stacked vector of lagged sample scores is $\bm{\widehat x}_{t-1}^{(J,m)}= ((\widehat F_{t-1}^{(J)})', \ldots, (\widehat F_{t-m}^{(J)})')'$.\footnote{In the Supplementary Material, Technical Appendix~A.1, we show that under the conditions of Theorem~\ref{thm:Bias} the population counterpart of $\widehat\Sigma_{(J,m)}$ is positive definite for any $J\leq K_{\max}$ (including the overselection case $J>K$), implying that $\widehat\Sigma_{(J,m)}$ is invertible with probability tending to one.}
Conditional on the selected number of factors and lags, the one-step ahead curve predictor is expressed as
\begin{equation*}
	\widehat{Y}_{t|t-1}^{(J,m)}(r) =  \widehat \mu(r) +  \big(\widehat \Psi^{(J)}(r)\big)' \bm{\widehat A}_{(J,m)} \bm{\widehat x}_{t-1}^{(J,m)},  \qquad \widehat \Psi^{(J)}(r) = \big(\widehat \psi_1(r), \ldots, \widehat \psi_J(r)\big),
\end{equation*}
and the corresponding mean squared error (MSE) is given by
\begin{equation} \label{eq:MSE}
MSE_T(J,m)=\frac{1}{T-m}\sum_{t=m+1}^{T}\big\Vert Y_{t}-\widehat{Y}_{t|t-1}^{(J,m)}\big\Vert^2.
\end{equation}

Given that $MSE_T(J,m)$ depends on both the number of selected factors and the number of lags, it can be used to construct a consistent information criterion.
To achieve this, we first must examine how $MSE_T$ behaves with respect to $J$ and $m$, considering two sources of uncertainty: one from estimating the model parameter functions and the other from estimating the vector autoregression itself.
The model parameter functions that enter $MSE_T$ (i.e., $\mu$ and $\psi_l$) are invariant with respect to $J$ and $m$ and, according to Theorem \ref{thm:consistency}, are estimated with parametric rates of convergence.
Hence, it remains crucial to understand how the asymptotic properties of the conditional LS estimator and therefore $MSE_T$ are impacted by a misspecified number of factors and lags.

The population coefficient matrix, $\bm{A}$, and the conditional LS estimator matrix, $\bm{\widehat A}_{(J,m)}$, are of different dimensions.
To align the $K \times Kp$ matrix $\bm{A}$ with the $J \times Jm$ matrix $\bm{\widehat A}_{(J,m)} = [\widehat A_1^{(J)}, \ldots, \widehat A_m^{(J)}]$, we transform them into matrices of order $J^* \times J^*m^*$ with $J^*=\max\{J,K\}$ and $m^* = \max\{m,p\}$ by inserting zeros where their dimensions do not match.
Using the completion matrix
$$
\bm R_{J,K} = \begin{cases}
		\big[ \bm I_J, \bm 0_{J,K-J} \big], & \text{if} \ J < K, \\
		\bm I_J, & \text{if} \ J \geq K,
	\end{cases}
$$
where $\bm 0_{J,K-J}$ is the $J \times (K-J)$ matrix of zeros and $\bm I_J$ is the identity matrix, we define the aligned LS estimator
\begin{align*}
	\bm{\widehat A}^* =
	\begin{cases}
		\big[\bm R_{J,K}' \widehat A_1^{(J)} \bm R_{J,K}, \ldots, \bm R_{J,K}' \widehat A_m^{(J)} \bm R_{J,K}, \bm 0_{J^*,(p-m)J^*} \big], & \text{if} \ m<p, \\
		\big[\bm R_{J,K}' \widehat A_1^{(J)} \bm R_{J,K}, \ldots, \bm R_{J,K}' \widehat A_m^{(J)} \bm R_{J,K} \big], & \text{if} \ m \geq p.
	\end{cases}
\end{align*}

\noindent
Given that the loadings are only identified and correctly estimated up to a sign change, we condition our notation on the selected signs and consider the sign-adjusted matrices $\widetilde A_i = \bm S A_i \bm S$ for $i=1, \ldots, p$, with the sign transformation matrix $\bm S = \diag(s_1, \ldots, s_K)$, where $s_l = \sign(\langle\widehat \psi_l, \psi_l \rangle)$.
The VAR process can be written as $\bm S F_t^* = \sum_{i=1}^p \widetilde A_i \bm S F_{t-i}^* + \bm S \eta_t$ since $\bm S \bm S = \bm I_K$, and the aligned sign-adjusted stacked population coefficient matrix is
\begin{align*}
	\bm A^* = \begin{cases}
		\big[\bm R_{K,J}'\widetilde A_1 \bm R_{K,J}, \ldots, \bm R_{K,J}'\widetilde A_p \bm R_{K,J}, \bm 0_{J^*, (m-p) J^*} \big], & \text{if} \ m > p, \\
		\big[\bm R_{K,J}'\widetilde A_1 \bm R_{K,J}, \ldots, \bm R_{K,J}' \widetilde A_p \bm R_{K,J} \big], & \text{if} \ m \leq p.
	\end{cases}
\end{align*}

\begin{theorem}\label{thm:Bias}
Let Assumptions \ref{as:components}--\ref{as:VAR} hold true, and let $p_{max}$ and $K_{max}$ be bounded integers such that $p_{max} \geq p$, $K_{max} \geq K$.
Furthermore, for any $t$, the covariance operator of $Y_t$ has infinitely many positive eigenvalues.
Then, for any selected numbers of lags $m\leq p_{max}$ and factors $J\leq K_{max}$, as $T \to \infty$:
\begin{itemize}
	\item[(a)] If $J \geq K$ and $m \geq p$, $\Vert \bm{\widehat{A}}^* - \bm{A}^{*}  \Vert_{M}=O_p(T^{-1/2})$;
	\item[(b)] If $J < K$, $m < p$, or both, $\plim_{T\to\infty} \Vert \bm{\widehat{A}}^* - \bm{A}^{*} \Vert_{M} > 0$.
\end{itemize}
\end{theorem}

Theorem \ref{thm:Bias} shows that the consistency of the LS estimator hinges on the condition that $J\geq K$ and $m\geq p$. If either $J$ or $m$ is smaller than the actual values, the vector autoregression cannot be consistently estimated using the conditional LS estimator. This underscores the importance of the simultaneous selection of $K$ and $p$ when employing the LS estimator. For example, if the selected number of factors exceeds $K$ and the chosen lags are $m<p$, the LS estimator is biased. However, it is consistent when $m\geq p$.

The central insight from Theorem \ref{thm:Bias} is that the MSE is asymptotically minimized when $J \geq K$ and $m \geq p$. Specifically, a model estimated with $K+j$ factors and $p+i$ lags for $i,j>0$ cannot asymptotically fit worse than a model with $K$ factors and $p$ lags.
Once the threshold with the true $K$ and $p$ is met, an increase in the number of selected factors and lags does not impact the asymptotic MSE, but it may lead to parameter proliferation and a loss of efficiency.
Consequently, we propose an MSE-based information criterion for estimating $K$ and $p$ of the form
\begin{equation} \label{eq:infCrit}
\mathrm{CR}_T(J,m)= f ( MSE_T(J,m) ) + g_T(J,m),
\end{equation}
where $g_T(J,m)$ serves as a penalty term for overfitting the model, and $f(\cdot)$ is a strictly increasing and continuous function.
Then, the numbers of factors and lags are estimated as
$$
	(\widehat K, \, \widehat p ) = \argmin_{\substack{J = 1, \ldots, K_{max} \\ m = 1, \ldots, p_{max}}} \mathrm{CR}_T(J,m).
$$

\begin{theorem}\label{thm:InformCriteria}
Under the conditions of Theorem \ref{thm:Bias}, let $g_T(J,m)$ be strictly monotonically increasing in both arguments $J$ and $m$ such that $g_T(J,m)\to 0$ and $Tg_T(J,m)\to \infty$ for all $1\leq J\leq K_{max}$ and $1\leq m\leq p_{max}$, as $T\to\infty$.
Then, $\lim_{T\to\infty} \mathrm{P}(\widehat{K}=K,\widehat{p}=p)=1$.
\end{theorem}

The results of Theorem \ref{thm:InformCriteria} indicate that penalized MSE-based information criteria select both the correct number of factors and lags with probability 1.
The crucial element for the consistent estimation of $K$ and $p$ is a penalty term that vanishes at an appropriate rate to ensure that an overparameterized model is not chosen.
Commonly employed penalty terms from established information criteria in multivariate time series analysis, such as the Bayesian Information Criterion (BIC) and the Hannan-Quinn Criterion (HQC), meet the conditions outlined in Theorem \ref{thm:InformCriteria}.
Furthermore, it is also standard practice to use a logarithmic transformation to put all terms of the generic information criterion \eqref{eq:infCrit} onto the same scale.
These practical considerations naturally lead us to propose two types of information criteria for estimating $K$ and $p$.
First, a BIC-type estimator is formulated as
\begin{equation}
(\widehat K_{\text{bic}}, \widehat p_{\text{bic}})
= \argmin_{\substack{J = 1, \ldots, K_{max} \\ m = 1, \ldots, p_{max}}}
\log \big( MSE_T(J,m) \big) + Jm\frac{\log(T)}{T},
\label{eq:CR-BIC}
\end{equation}
where the term $Jm$ is a complexity penalty that increases with the selected factor dimension and lag order, and $T^{-1}\log(T)$ is the penalization rate.
Second, the HQC-type estimator employs a lower penalization rate and is formulated as
\begin{equation}
(\widehat K_{\text{hqc}}, \widehat p_{\text{hqc}})
= \argmin_{\substack{J = 1, \ldots, K_{max} \\ m = 1, \ldots, p_{max}}}
\log \big( MSE_T(J,m) \big) + 2Jm\frac{\log(\log(T))}{T}.
\label{eq:CR-HQC}
\end{equation}
Both \eqref{eq:CR-BIC} and \eqref{eq:CR-HQC} satisfy the conditions from Theorem~\ref{thm:InformCriteria} and therefore provide consistent estimators for $K$ and $p$.
In practice, the minimization problem can be solved by grid search, where $K_{max}$ and $p_{max}$ must be selected.

For reference, \citet{aue2015} develop a functional final prediction error (fFPE) criterion to jointly choose the number of functional PCA scores and the VAR order when forecasting stationary functional time series.
In our notation, their penalty has the form
$\log((T+mJ)/(T-mJ))$, whose first-order expansion is $2mJ/T$.
This AIC-type penalty aligns with \citet{aue2015}'s forecasting objective; however, as shown in Theorem~\ref{thm:InformCriteria}, it does not satisfy the requirement \(T\,g_T(J,m)\to\infty\) and therefore cannot deliver consistent joint selection of the factor dimension and lag order in our settings.

\subsection{Practical guidance} \label{sec:practical}

This section details the practical implementation of our estimation method and information criterion.
Our primary objective is to present a procedure that can be easily executed using existing software.
Building upon the theoretical foundations established in Section 3, we offer two approaches for implementing the information criterion: one based on the analytical representation of the expression in \eqref{eq:infCrit}, and the other based on a graphical representation.
Both methods require numerical integration for computing empirical eigenfunctions and eigenvalues.
Additionally, we provide an estimation and prediction algorithm that outlines  how one can execute an empirical analysis of a factor model of functional time series.
Our accompanying R package facilitates the execution of all proposed steps.

\paragraph{Analytical representation.}
Given the selected number of factors $J$ and lags $m$, the fitted factor and error components are $\widehat \chi_t^{(J)}(r) = \widehat \mu(r) + \sum_{l=1}^J \widehat F_{l,t} \widehat \psi_l(r)$ and $\hat \epsilon_t^{(J)}(r) = Y_t(r) - \widehat \chi_t^{(J)}(r)$,
where $Y_t(r) = \widehat \chi_t^{(J)}(r) + \hat \epsilon_t^{(J)}(r)$.
The one-step ahead curve predictor can be written as $\widehat{Y}_{t|t-1}^{(J,m)}(r) = \widehat{\mu}(r) + \sum_{l=1}^{J} \widehat{F}_{l,t|t-1}\widehat{\psi}_l(r)$ with $\widehat{F}_{t|t-1}^{(J)} = (\widehat{F}_{1,t|t-1}, \ldots, \widehat{F}_{J,t|t-1})' = \bm{\widehat A}_{(J,m)} \bm{\widehat x}_{t-1}^{(J,m)}$, and the functional forecast error is $Y_t(r) - \widehat{Y}_{t|t-1}^{(J,m)}(r) = \sum_{l=1}^{J} \hat{\eta}_{l,t}\widehat{\psi}_l(r) + \hat \epsilon_t^{(J)}(r)$,
where $\hat{\eta}_{l,t}=\widehat{F}_{l,t}-\widehat{F}_{l,t|t-1}$ are the VAR residuals.
From the orthonormality of the estimated loading functions, the MSE given in \eqref{eq:MSE} simplifies to
\begin{align}
	\frac{1}{T-m} \sum_{t=m+1}^{T} \bigg( \sum_{l=1}^{J}\hat{\eta}_{l,t}^2 + \big\|\hat \epsilon_t^{(J)}\big\|^2  \bigg)
	\approx \tr \big(\widehat{\Sigma}_{\eta}^{(J,m)}\big) + \int_a^b \frac{1}{T} \sum_{t=1}^T \big(\hat \epsilon_t^{(J)}(r)\big)^2 \dd r. \label{eq:MSEsimplified}
\end{align}
The advantage of the expression \eqref{eq:MSEsimplified} over the MSE in \eqref{eq:MSE} is that all components can be easily computed.
In particular, $\widehat{\Sigma}_{\eta}^{(J,m)}$ is the least squares estimator of ${\Sigma}_{\eta}$ obtained by fitting a VAR($m$) model based on the $J$-variate time series $\widehat{F}_t^{(J)}$.
The integral residual sample variance $\int_a^b \frac{1}{T} \sum_{t=1}^T (\hat \epsilon_t^{(J)}(r))^2 \dd r$ equals the sum of all eigenvalues of the sample covariance operator of the residual curves $\hat \epsilon_t^{(J)}$.

\paragraph{Graphical representation.}

\begin{figure}[t]
  \begin{center}
  \vspace*{-10ex}
    \includegraphics[width=0.48\textwidth]{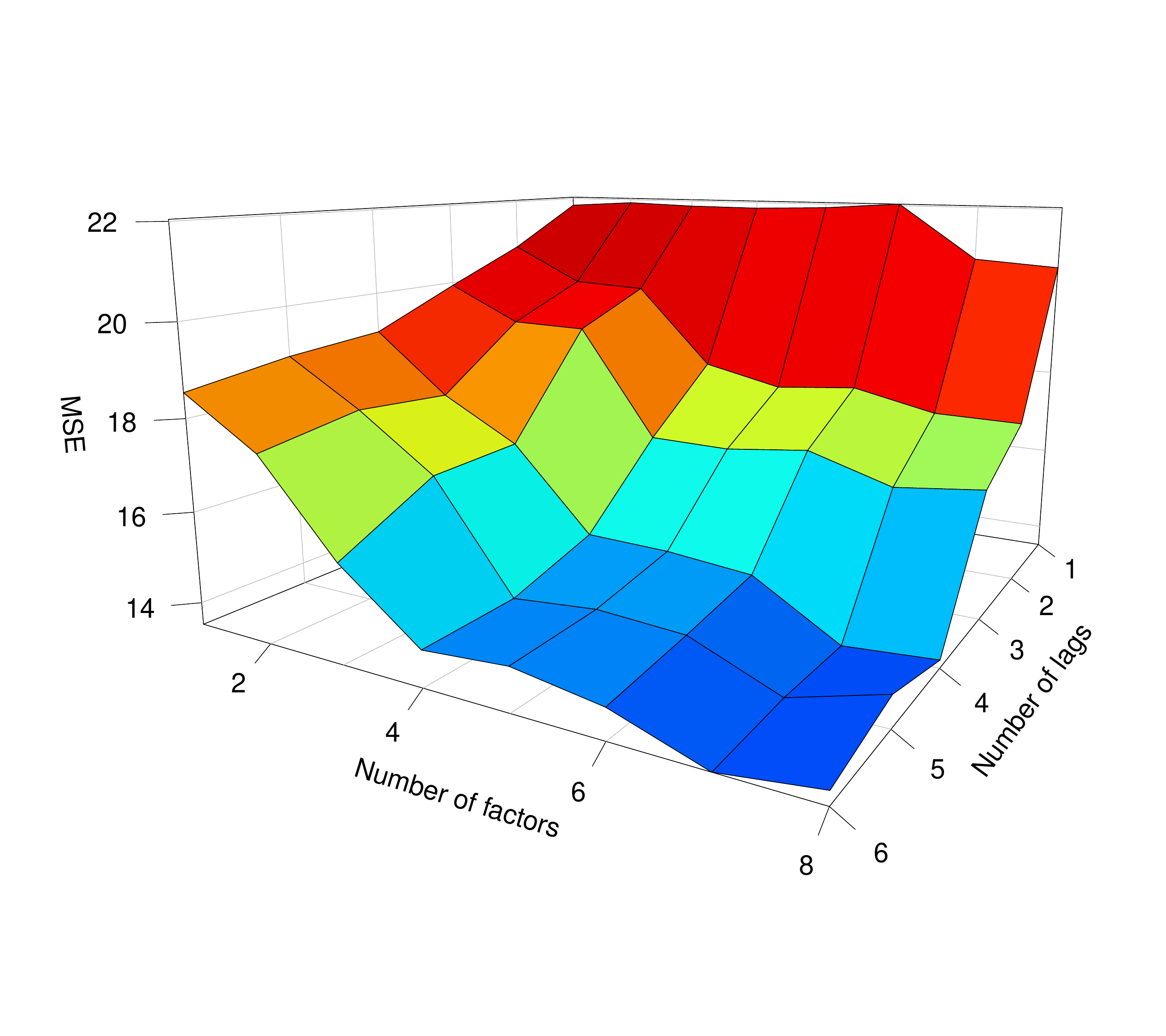}
    \vspace*{-11ex}
  \end{center}
  \caption{Graphical representation of the MSE using simulated data ($K=p=4$)}
  \label{fig:MSE}
\end{figure}

A careful inspection of the proof of Theorem \ref{thm:InformCriteria} shows that the MSE reaches its asymptotic minimum when $J \geq K$ and $m \geq p$.
This result can be used to select $(K,p)$ graphically, similar to the concept of the scree plot.
More precisely, one can plot $MSE_T(J,m)$ for various combinations of $J$ and $m$ and choose the minimum vertex of a rectangular surface with respect to $J$ and $m$ for which the MSE remains ``flat''.
For this purpose, expression \eqref{eq:MSEsimplified} can be used.
Figure \ref{fig:MSE} shows an example illustrating an MSE surface.
This figure suggests that $\widehat{K}=4$ and $\widehat{p}=4$ should be selected.

The graphical approach has an advantage over the analytical expressions presented in \eqref{eq:CR-BIC} and \eqref{eq:CR-HQC}, since it does not require the specification of the penalty term. However, it cannot be automated when it comes to multiple model selection (for instance, in Monte Carlo simulations). Furthermore, it often becomes a subjective decision of a researcher where the smallest point of the MSE rectangular ``flat'' area is since the estimated MSE will also fluctuate in this area in finite samples.

\paragraph{Estimation and prediction algorithm.} \

\noindent\textbf{Step 1:} Estimation of parameter functions. Compute the sample mean function $\widehat \mu(r)$ and the cumulative autocovariance function $\widehat d(r,s)$ from the observed curves $Y_1(r), \ldots, Y_T(r)$. Fix $K_{max}$ large enough and compute eigencomponents $\{(\widehat \lambda_l, \widehat \psi_l)\}_{l=1}^{K_{max}}$ and sample scores $\widehat{F}_{l,t} = \langle Y_t - \widehat{\mu}, \widehat{\psi}_l \rangle$, $l=1,\ldots,K_{max}$ as estimates for the factors.\medskip

\noindent\textbf{Step 2:} Estimation of $K$, $p$, and the factor dynamics. Fix $p_{max}$ large enough, compute $MSE_T(J,m)$ from \eqref{eq:MSEsimplified} for $J=1, \ldots, K_{max}$ and $m=1, \ldots, p_{max}$, and select $K$ and $p$ according to the BIC or HQC criterion in \eqref{eq:CR-BIC}--\eqref{eq:CR-HQC}. Estimate a VAR($\widehat p$) model by the LS estimator given in \eqref{eq:LS}, yielding $[\widehat A_1^{(\widehat K)}, \ldots, \widehat A_{\widehat p}^{(\widehat K)}] = \widehat{\bm{A}}_{(\widehat K, \widehat p)}$.\medskip

\noindent\textbf{Step 3:} Fitted curves and forecasting. Compute the fitted curves for $t=1, \ldots, T$ as follows: $\widehat Y_t(r) = \widehat \mu(r) + \sum_{l=1}^{\widehat K} \widehat F_{l,t} \widehat \psi_l(r)$. Compute the VAR($\widehat p$) factor forecasts as $\widehat F_{T+1|T}^{(\widehat{K})} = \sum_{i=1}^{\widehat{p}} \widehat A_i^{(\widehat{K})}  \widehat{F}_{T+1-i|T}^{(\widehat{K})}$ with $\widehat F_{T+j|T}^{(\widehat K)} = \widehat F_{T+j}^{(\widehat K)}$ for $j \leq 0$.
The $h$-step curve forecast is given by $\widehat Y_{T+h|T}^{(\widehat{K},\widehat{p})}(r) = \widehat \mu(r) + \big(\widehat \Psi^{(\widehat K)}(r)\big)' \widehat F_{T+h|T}^{(\widehat{K})}$.

\paragraph{Discretely observed curves and dense designs.}
Throughout the paper we treat $Y_1,\ldots,Y_T$ as elements of $H$.
In practice, however, curves are observed on a discrete grid.
In this paper we focus on the common case of dense and regular grids.
In this setting, our procedure does not require a separate smoothing step: all inner products and operator estimators are computed directly from the grid via standard numerical integration.
This is justified by standard functional data results: under dense regular designs, the key population objects used by our method (the mean function, (auto-)covariance operators, and their eigen-elements) can be estimated with the same first-order behavior as under fully observed curves.
For example, \citet{hall2006,li2010,zhang2016,kneip2020} provide conditions such as $N/T^{1/4}\to\infty$ and mild smoothness to characterize dense grids.
In our empirical part, for instance, curves are observed on an equidistant grid with more than $N=100$ points per curve, with $T$ ranging from 90 to 460.

For sparse or irregular designs, or very low signal-to-noise ratios, a preliminary smoothing or reconstruction step (e.g., via basis expansions; see \citealp{ramsay2005}) may be useful or necessary.
The precise boundary between sparse and dense designs depends on smoothness, measurement noise, and dimensionality (see \citealt{Guo2025b} for a non-asymptotic analysis of such phase transitions in high-dimensional functional data).
Related autocovariance-based methods accommodating measurement-error contamination are developed by \citet{chen2022}.
Extending our methodology to such settings would require accounting for the first-stage reconstruction error and is therefore not a straightforward extension of our analysis; we leave this for future research.

\paragraph{Some useful diagnostics.} Finally, in empirical applications it is useful to complement the proposed procedure with diagnostic checks on the observed curves and on the fitted finite-dimensional score dynamics. Functional stationarity tests, such as \citet{HORVATH2014}, can be used as a diagnostic for the classical stationarity of the curve process, although this is stronger than what is required for the global second-moment quantities used in our analysis. In addition, white-noise tests for functional residuals, such as \citet{zhang2016white}, and standard stability and residual serial-correlation checks for the fitted VAR score process can help assess the adequacy of the selected finite-dimensional dynamic specification.

\section{Simulations} \label{sec:simulations}

We evaluate the finite sample properties of our estimators through Monte Carlo simulations.
Functional time series are generated in the space spanned by the first 19 Fourier basis functions,
given as $v_1(r) = 1$, $v_{2j}(r) = \sqrt 2 \sin(2 j \pi r)$, and $v_{2j+1}(r) = \sqrt 2 \cos(2 j \pi r)$ for $j=1, \ldots, 9$.
We use a subset $\mathcal I$ of these Fourier functions to specify the loading functions as $(\psi_1, \ldots, \psi_K) = (v_j, \ j \in \mathcal I)$,
while the remaining functions are  used to form the error component.
This leads to the following data generating procedure:
\begin{align} \label{eq:simulationprocess}
Y_t(r) = \sum_{l=1}^K F_{l,t} \psi_l(r)+ \sum_{j \in \mathcal I^c} e_{j,t} v_j(r),
\end{align}
where $\mathcal I^c = \{1, \ldots, 19\} \setminus \mathcal I$.
Here, the first term represents the factor component \eqref{eq:factorcomponent}, while the second term represents the error component.
Their stochastic nature is generated through a random score vector $e_t = (e_{1,t}, \ldots, e_{19,t})' \sim \mathcal N(0,\text{diag}(1, 2^{-1}, \ldots, 19^{-1}))$, which is independent across $t=1, \ldots, T$.
The elements with indices $\mathcal I^c$ of this vector are used to form the error term, as given in \eqref{eq:simulationprocess}, and the factors are defined as $F_t = (F_{1,t}, \ldots, F_{K,t})' = A(L)\eta_t$ with $\eta_t' = (\eta_{1,t}, \ldots, \eta_{K,t}) = (e_{j,t}, \ j \in \mathcal I)$.

\begin{figure}[!t]
\caption{Estimation uncertainty of $\widehat \psi_1$ in models M1--M3}
\vspace{-4ex}
\begin{center}
\includegraphics[scale=0.21]{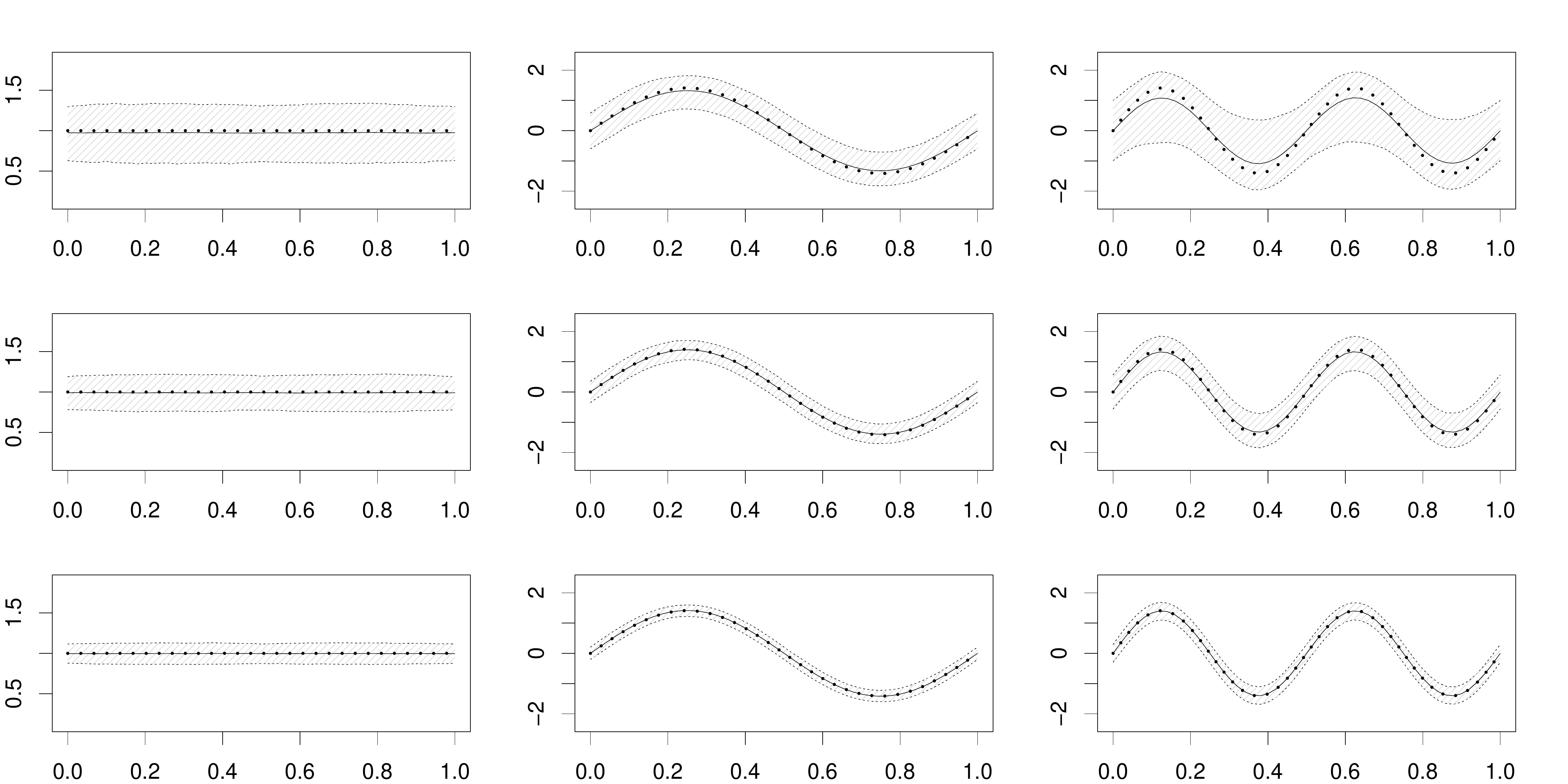}

\parbox{15cm}{ \vspace*{0.5ex} \scriptsize
Note: This panel shows the estimation uncertainty of the sign corrected estimator $s_1 \widehat{\psi}_1$ based on 10,000 Monte Carlo replications using $q_0=1$  for models M1--M3 from Table \ref{tab:modelspecs}, organized by column, and different sample sizes $T=100, 200, 500$, organized by row.
Each plot includes a dotted line indicating the true value of $\psi_1$, a solid line for the sample mean curve of all estimates, and dashed lines marking the pointwise 5\% and 95\% sample quantiles of the estimates.
}
\label{fig:psi1hat}
\end{center}
\end{figure}

We consider five scenarios, M1--M5, for the dynamics of the factors, represented by $A(L)$, and subsets $\mathcal I$ of Fourier basis functions that form the factor space and indicate the number of the factors, detailed in Table \ref{tab:modelspecs}.
It is noteworthy that the scenarios we investigate encompass empirically relevant and challenging cases where the error component can have a higher variance compared to the factors. The single-factor models M1--M3 have different loading functions and factor variances, model M4 uses a two-factor setup, and M5 is similar to the three-factor setup used in \cite{aue2015}.
Figure \ref{fig:psi1hat} and Table \ref{tab:biasRMSE} confirm our theoretical findings and illustrate the consistency of $\widehat \psi_l$ as well as the BIC-type and HQ-type information criteria from equations \eqref{eq:CR-BIC} and \eqref{eq:CR-HQC}.
The estimators $\widehat K_{bic}$, $\widehat p_{bic}$, $\widehat K_{hqc}$, and $\widehat p_{hqc}$ provide a good approximation of the true parameters for reasonable sample sizes.

\begin{table}[!t]
\caption{Model specifications for the Monte Carlo simulations}
\begin{center}
\footnotesize
\begin{tabular}{cllll}
\toprule
Model & $K$ & $H_F = \text{span}(v_l, l \in \mathcal I)$ & $p$  & Lag polynomial $A(L)$ \\
\midrule
M1 & 1 & $\mathcal I = \{1\}$ & 2  & $1 - 0.4L - 0.4 L^2$ \\
M2 & 1 & $\mathcal I= \{2\}$ & 2  & $1 - 0.4L - 0.4 L^2$ \\
M3 & 1 & $\mathcal I = \{4\}$ & 2  & $1 - 0.4L - 0.4 L^2$ \\
\vspace*{0.5ex}
M4 & 2 & $\mathcal I = \{2,3\}$ & 3  & $\bm I_2 -
\begin{psmallmatrix} 0.6 & -0.2 \\ 0.0 & \phantom{-}0.2 \end{psmallmatrix} L -
\begin{psmallmatrix} -0.25 & -0.1 \\ \phantom{-}0.00 & -0.1 \end{psmallmatrix} L^2 -
\begin{psmallmatrix} 0.6 & -0.25 \\ 0.0 & \phantom{-}0.85 \end{psmallmatrix} L^3$ \\
\vspace*{0.5ex}
M5 & 3 & $\mathcal I = \{3,4,5\}$ & 1  & $\bm I_3 -
\begin{psmallmatrix} -0.05 & -0.23 & \phantom{-}0.76 \\
\phantom{-}0.80 & -0.05 & \phantom{-}0.04 \\
\phantom{-}0.04 & \phantom{-}0.76 & \phantom{-}0.23\end{psmallmatrix} L$ \\
\bottomrule
\end{tabular}
\parbox{14.1cm}{ \vspace{0.5ex} \scriptsize
Note:
This table outlines the specifications for model \eqref{eq:simulationprocess} used in the results for Table \ref{tab:biasRMSE} and Figure \ref{fig:psi1hat}.
}
\end{center}
\label{tab:modelspecs}
\end{table}

\begin{table}[t]
\caption{Finite sample performances of models M1--M5}
\begin{center}
\scriptsize
\begin{tabular}{@{}lrrrrrrrrrrrrrrr@{}}
\toprule
 & \multicolumn{1}{|c}{}  & \multicolumn{4}{|c}{RMSE} & \multicolumn{4}{|c}{Bias}  & \multicolumn{4}{|c}{\% of false selection}  \\
Model & \multicolumn{1}{|c}{$\widehat \chi$-err } & \multicolumn{1}{|r}{$\widehat K_{bic}$}  & $\widehat K_{hqc}$ & $\widehat p_{bic}$ & $\widehat p_{hqc}$ & \multicolumn{1}{|r}{$\widehat K_{bic}$}  & $\widehat K_{hqc}$ & $\widehat p_{bic}$ & $\widehat p_{hqc}$ & \multicolumn{1}{|r}{$\widehat K_{bic}$}  & $\widehat K_{hqc}$ & $\widehat p_{bic}$ & $\widehat p_{hqc}$ \\
\midrule
M1 \\
$T=100$ & 0.10 & 0.02 & 0.06 & 0.80 & 0.62 & 0.00 & 0.00 & 0.64 & 0.38 & 0.00 & 0.00 & 0.64 & 0.39 \\
$T=200$ & 0.04 & 0.01 & 0.02 & 0.37 & 0.18 & 0.00 & 0.00 & 0.13 & 0.03 & 0.00 & 0.00 & 0.13 & 0.03 \\
$T=500$ & 0.01 & 0.00 & 0.00 & 0.00 & 0.03 & 0.00 & 0.00 & 0.00 & 0.00 & 0.00 & 0.00 & 0.00 & 0.00 \\
M2 \\
$T=100$ & 0.18 & 0.09 & 0.20 & 0.98 & 0.91 & -0.01 & -0.04 & 0.96 & 0.84 & 0.01 & 0.03 & 0.96 & 0.84 \\
$T=200$ & 0.06 & 0.03 & 0.08 & 0.85 & 0.57 & 0.00 & -0.01 & 0.72 & 0.32 & 0.00 & 0.01 & 0.72 & 0.32 \\
$T=500$ & 0.02 & 0.00 & 0.01 & 0.12 & 0.03 & 0.00 & 0.00 & 0.02 & 0.00 & 0.00 & 0.00 & 0.02 & 0.00 \\
M3 \\
$T=100$ & 0.39 & 0.17 & 0.36 & 1.00 & 0.99 & -0.03 & -0.12 & 1.00 & 0.99 & 0.03 & 0.11 & 1.00 & 0.99 \\
$T=200$ & 0.14 & 0.14 & 0.27 & 1.00 & 0.94 & -0.02 & -0.07 & 0.99 & 0.88 & 0.02 & 0.07 & 0.99 & 0.88 \\
$T=500$ & 0.03 & 0.03 & 0.08 & 0.77 & 0.28 & 0.00 & -0.01 & 0.59 & 0.08 & 0.00 & 0.01 & 0.59 & 0.08 \\
M4 \\
$T=100$ & 1.00 & 0.88 & 0.81 & 1.64 & 1.07 & 0.77 & 0.42 & 1.35 & 0.56 & 0.78 & 0.62 & 0.68 & 0.29 \\
$T=200$ & 0.51 & 0.59 & 0.81 & 0.47 & 0.07 & 0.14 & -0.32 & 0.11 & 0.00 & 0.32 & 0.37 & 0.06 & 0.00 \\
$T=500$ & 0.22 & 0.46 & 0.73 & 0.00 & 0.00 & -0.15 & -0.32 & 0.00 & 0.00 & 0.12 & 0.23 & 0.00 & 0.00 \\
M5 \\
$T=100$ & 0.30 & 0.31 & 0.46 & 0.00 & 0.00 & 0.02 & -0.16 & 0.00 & 0.00 & 0.09 & 0.17 & 0.00 & 0.00 \\
$T=200$ & 0.13 & 0.12 & 0.34 & 0.00 & 0.00 & -0.01 & -0.11 & 0.00 & 0.00 & 0.02 & 0.10 & 0.00 & 0.00 \\
$T=500$ & 0.05 & 0.08 & 0.30 & 0.00 & 0.00 & -0.01 & -0.08 & 0.00 & 0.00 & 0.01 & 0.08 & 0.00 & 0.00 \\
\bottomrule
\end{tabular}
 \parbox{14.5cm}{ \vspace{0.4ex} \scriptsize
Note: The simulation results are derived using different sample sizes $T$ across models M1--M5 (Table \ref{tab:modelspecs}), with 10,000 Monte Carlo replications. The first column shows the average estimation error $\|\widehat{\chi}_t - \chi_t\|$ over all observations and replications, where the BIC estimator for $K$ is used to compute $\widehat \chi_t$.
Subsequent columns report bias (average deviation from the true value), RMSE (square root of the mean squared error), and the frequency of false selection (share of replications with $\widehat K\neq K$ or $\widehat p\neq p$) for the BIC and HQC estimators in \eqref{eq:CR-BIC}–\eqref{eq:CR-HQC}. We use $q_0=1$ and cap the search at $K_{\max}=8$ factors and $p_{\max}=8$ lags.

}
\end{center}
\label{tab:biasRMSE}
\end{table}

\section{Empirical applications} \label{sec:applications}

\begin{figure}[t]
\caption{Log mortality rate curves for U.S. males and U.S. Treasury yield curves}
\begin{center}
\includegraphics[scale=0.6]{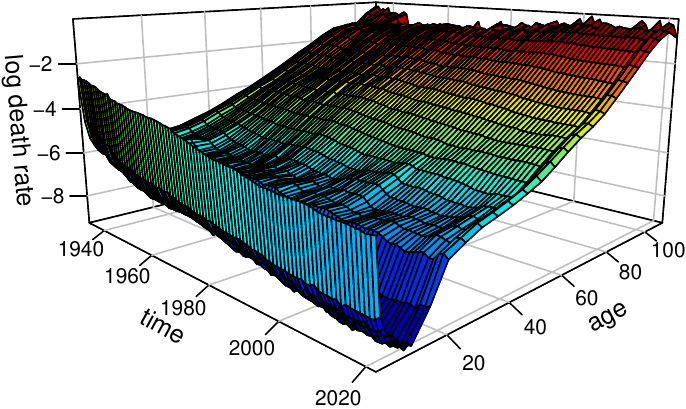}
\includegraphics[scale=0.6]{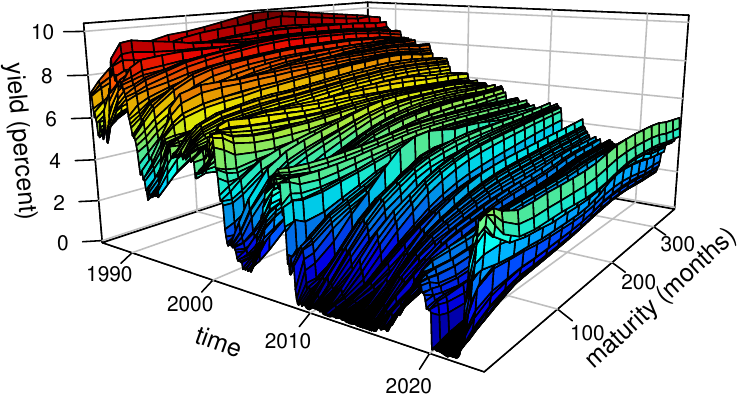}
\end{center}
\vspace{-2ex} \hspace{3ex}
\parbox{15cm}{  \scriptsize
Note: The figure depicts the two functional time series datasets used in the empirical applications.
}
\label{fig-3dplots}
\end{figure}

We apply our methodology to two datasets that naturally align with a functional time series perspective: yearly mortality rate curves and monthly bond yield curves (see Figure \ref{fig-3dplots}).
The first dataset consists U.S.\ log mortality rate curves of the male population, defined as $Y_t(r) = \log(D_t(r)/P_t(r))$,
where $D_t(r)$ represents the number of deaths in calendar year $t$ for individuals aged $r$, and $P_t(r)$ denotes the corresponding population of age $r$. This dataset is available from the Human Mortality Database\footnote{Accessible via the R package \texttt{demography}, though website registration at \url{https://www.mortality.org} is required.}, comprising $T=90$ yearly curves, each observed at 111 equidistant points spanning ages 0 to 110.

The second dataset, from \cite{liu2021}\footnote{Data source: \url{https://sites.google.com/view/jingcynthiawu/yield-data}.}, consists of reconstructed annualized continuously-compounded zero-coupon U.S.\ Treasury yield curves. Each curve is given by $Y_t(r)$, where $t$ denotes the calendar month and $r$ the time to maturity in months. The dataset spans November 1985 to December 2023 ($T=458$), with each curve containing 360 equidistant observations corresponding to maturities from 1 to 360 months.

\subsection{Mortality curve forecasting}
\begin{figure}[t]
\caption{VAR(1) forecasting results for the mortality data}
\begin{center}
\includegraphics[scale=0.66]{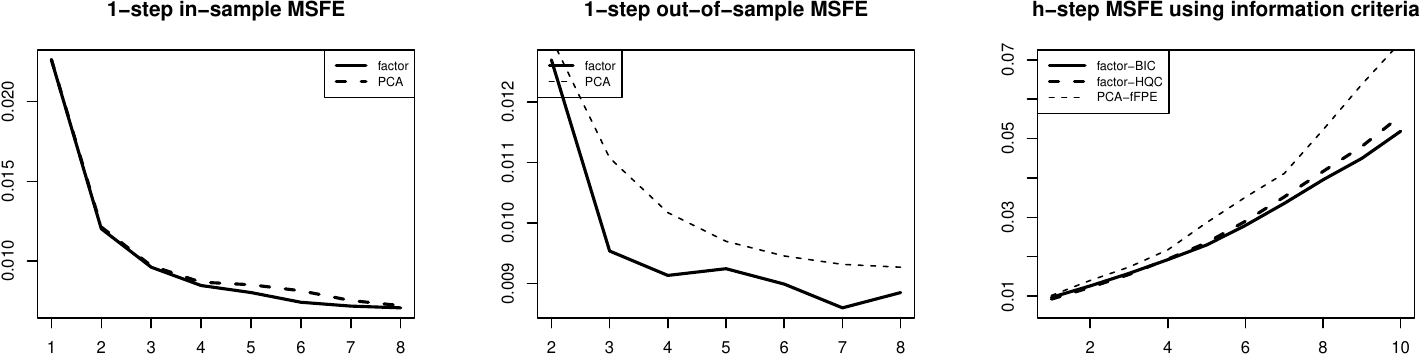}
\end{center}
\vspace{-1ex}
\parbox{15.8cm}{  \scriptsize
Note: The three panels show: (left) in-sample MSEs for factor-based $(q_0=1)$ and PCA-based VAR(1) predictions across number of selected components; (middle) out-of-sample rolling MSFEs with VAR(1) across number of selected components; (right) $h$-step rolling out-of-sample MSFEs using rolling selected components and lags across forecast horizon $h$.
}
\label{fig:mortalityforecasts}
\end{figure}

\begin{figure}[t]
\caption{Factor loadings and principal components for the mortality data}
\begin{center}
\includegraphics[scale=0.6]{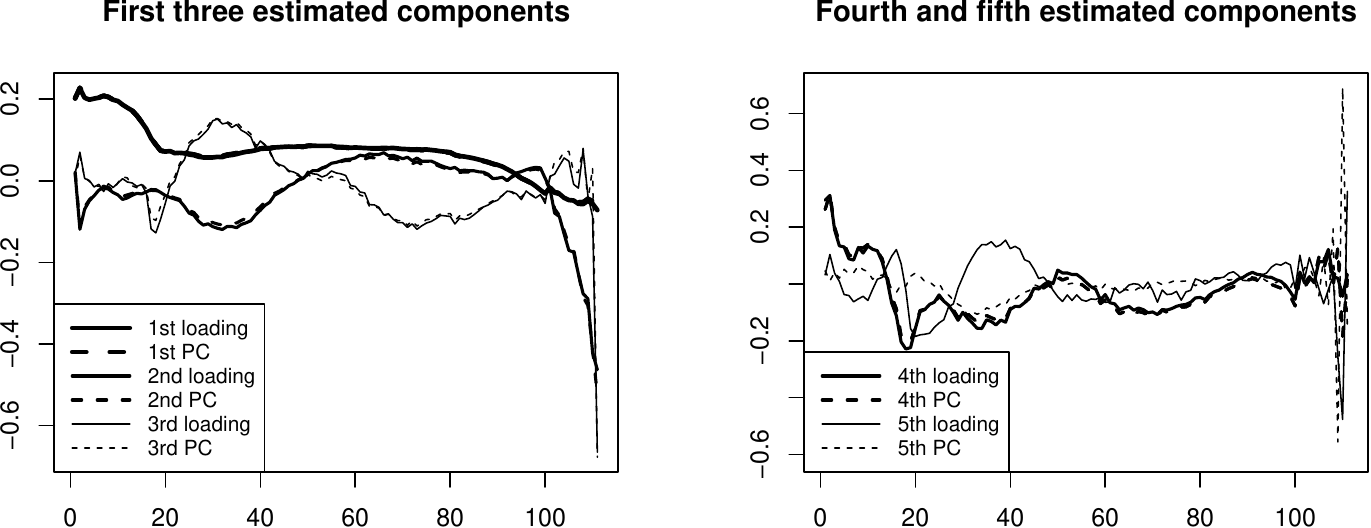}
\end{center}
\vspace{-1ex}
\parbox{15.8cm}{  \scriptsize
Note: The plots show the first five estimated factor loadings and functional principal components (PC).
}
\label{fig:mortalityloadings}
\end{figure}

Mortality rates, when viewed across age groups over time, naturally take the form of functional curves, making them well-suited for functional time series analysis.
U.S. mortality rate curves have been widely used in the literature, including in \cite{kokoszka2017}, to illustrate functional PCA-based forecasting methods developed in \cite{hyndman2007} and \cite{aue2015}.
Their approach relies on a truncated Karhunen-Loève expansion $Y_t(r) \approx \widehat \mu(r) + \sum_{l=1}^J g_{lt} \widehat \phi_l(r)$ with $g_{lt} = \langle Y_t - \widehat \mu, \widehat \phi_l \rangle$,
where $\widehat{\phi}_l$ are the sample functional principal components of $Y_1, \ldots, Y_T$.
The forecasted curve is $\widehat{\mu}(r) + \sum_{l=1}^{J} \widehat{g}_{l,T+h|T} \widehat{\phi}_l(r)$,
where $\widehat{g}_{l,T+h|T}$ are predicted scores obtained from a multivariate time series model for the vectors of sample functional principal component scores $(\widehat g_{1t}, \ldots, \widehat g_{Jt})'$.
To determine the optimal truncation parameter $J$, \cite{aue2015} propose a functional final prediction error (fFPE) information criterion.
While this approach is widely used and serves as a proven forecasting method, its key limitation is that the first $J$ functional principal components do not necessarily align with the most predictable components of the functional time series.

Using the full sample, we obtain $\widehat{K}_{\text{bic}} = 6$, $\widehat{K}_{\text{hqc}} = 7$, and $\widehat{p}_{\text{bic}} = \widehat{p}_{\text{hqc}} = 1$, while the fFPE criterion suggests using $J = 8$ functional principal component scores with one lag.
The in-sample mean squared forecast errors (MSFE) for $\widehat{K}_{\text{hqc}} = 7$ factors is lower than for $J=8$ principal components (as suggested by the fFPE criterion), which indicates that our approximate functional factor model provides a more parsimonious dynamic representation than the PCA-based approach.
Furthermore, the left panel of Figure \ref{fig:mortalityforecasts} confirms that factor-based predictions yield uniformly lower in-sample MSFE than PCA-based predictions for the same number of included components ranging from 1 to 8.
The difference becomes noticeable when higher-order components are included.
This is because the first three factor loadings closely resemble the first three functional principal components, whereas the fourth and fifth factor loadings deviate significantly from the PCA counterparts (see Figure \ref{fig:mortalityloadings}).

This difference becomes more prominent when we evaluate out-of-sample forecasts using a rolling window approach with a window size of 50 observations.
Model parameters are estimated within the rolling training sample, and $h$-step out-of-sample MSFEs are computed.
The second plot of Figure \ref{fig:mortalityforecasts} shows that factor-based predictions more clearly outperform PCA-based predictions for 1-step forecasts in the VAR(1) model when ranging the number of included factors from 1 to 8.
Furthermore, when turning to $h$-step ahead forecasts where $h=1,...,10$, we find another argument why structural representation of the dynamics is important.
The right panel of Figure \ref{fig:mortalityforecasts} demonstrates that factor-based predictions using a VAR($p$) model with $K$ factors, where $K$ and $p$ are selected via the BIC or HQC criterion, increasingly outperform forecasts from a VAR($p$) model with $J$ functional principal component scores, where $J$ and $p$ are chosen using the fFPE criterion of \cite{aue2015} with increasing $h$.
In conclusion, by focusing on factors that capture the underlying dynamics of the process, we obtain better forecasting performance compared to the traditional PCA-based approach.

\subsection{Yield curve modeling}

\begin{figure}[t]
\caption{Loading functions of the DNS model and yield curve data}
\begin{center}
\includegraphics[scale=0.43]{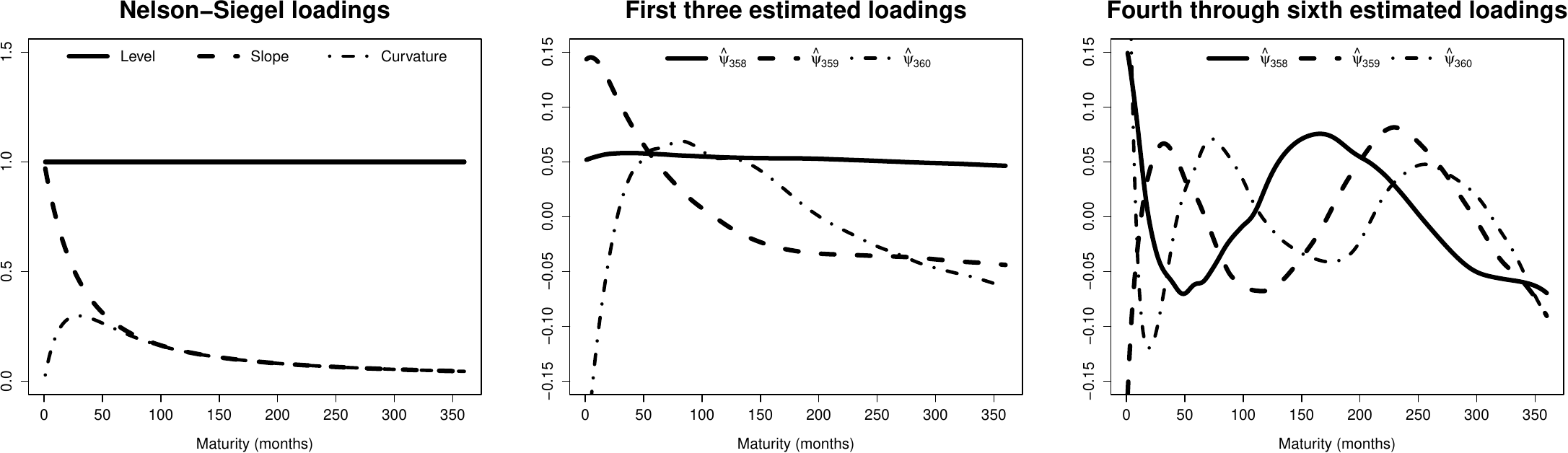}
\end{center}
\vspace{-1ex} \hspace{1ex}
\parbox{15.5cm}{  \scriptsize
Note: The figure displays the three standard loading functions from the dynamic Nelson-Siegel model from \eqref{eq:dns} (left), the first three estimated loading functions from our functional factor model (middle), and the fourth through sixth estimated loading functions (right).
}
\label{fig-yieldloadings}
\end{figure}

\begin{figure}[t]
\caption{Rolling estimation of K and p for the yield curve data}
\begin{center}
\includegraphics[scale=0.67]{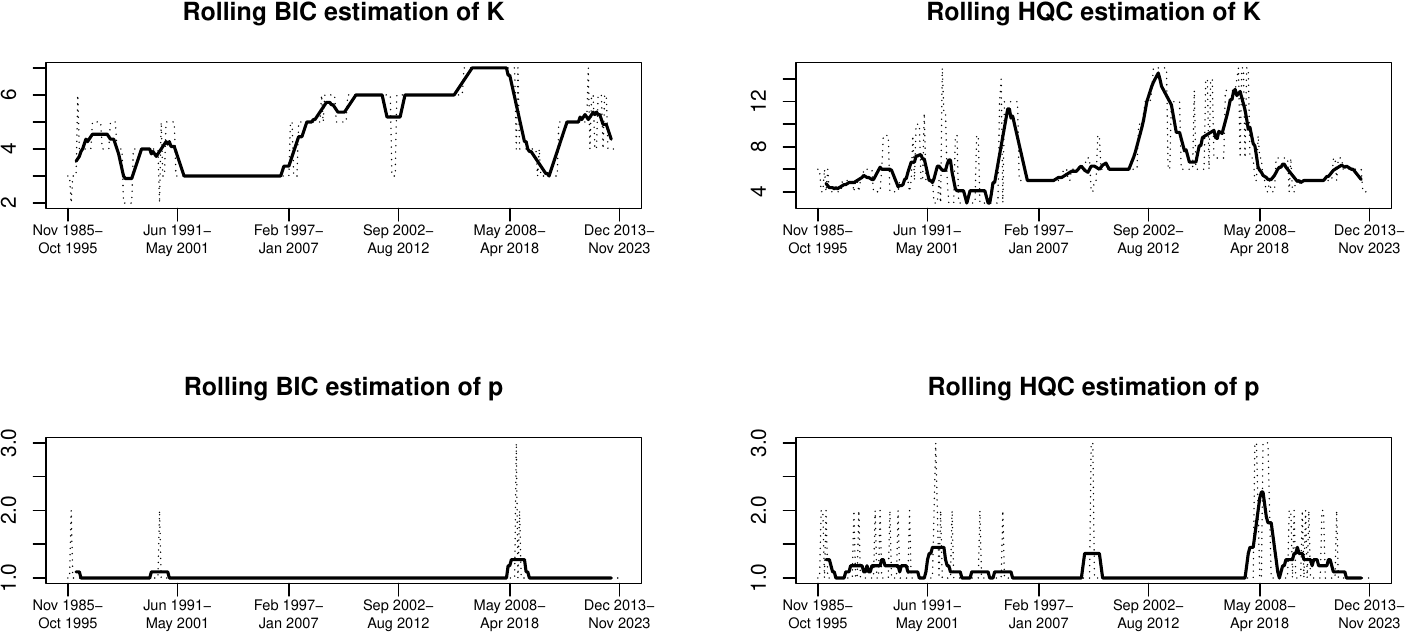}
\end{center}
\vspace{-1ex}
\parbox{15.8cm}{  \scriptsize
Note: The figure shows rolling window estimates (120 months) of the number of factors $K$ (top) and lags $p$ (bottom) over time, selected by BIC and HQC criteria. The dotted lines show the actual estimates while the solid lines indicate the simple two-sided moving average filter of order 11.
}
\label{fig-rollingK}
\end{figure}

\begin{table}[t!]
\caption{Rolling out-of-sample MSFEs for the yield curve data}
\begin{center}
\footnotesize
\begin{tabular}{@{}lrrrrrrrrr@{}}
\toprule
horizon & maturity & BIC & HQC & PCA & BIC & HQC & PCA & DNS & DNS \\
 &  & OLS & OLS & OLS & lasso & lasso & lasso & p=1 & p=2 \\
\midrule
\multicolumn{10}{c}{rolling window: 120 months} \\
\midrule
1-step & short & \textbf{0.918} & \textbf{0.903} & 1.332 & \textbf{0.983} & \textbf{0.994} & 1.103 & 1.701 & 1.710 \\
& medium & 1.213 & 1.185 & 1.748 & 1.215 & 1.208 & 1.206 & 1.758 & 1.672 \\
& long & 1.159 & 1.190 & 1.836 & 1.120 & 1.157 & 1.181 & 1.327 & 1.375 \\
3-step & short & \textbf{0.915} & \textbf{0.936} & 1.317 & \textbf{0.882} & \textbf{0.876} & \textbf{0.946} & 1.113 & 1.076 \\
& medium & 1.255 & 1.260 & 1.671 & 1.039 & 1.010 & 1.035 & 1.337 & 1.284 \\
& long & 1.362 & 1.416 & 1.872 & 1.027 & 1.031 & 1.049 & 1.265 & 1.359 \\
6-step & short & 1.120 & 1.127 & 1.569 & \textbf{0.887} & \textbf{0.899} & \textbf{0.934} & 1.067 & 1.077 \\
& medium & 1.470 & 1.485 & 2.009 & \textbf{0.997} & \textbf{0.978} & \textbf{0.988} & 1.284 & 1.295 \\
& long & 1.615 & 1.691 & 2.337 & \textbf{0.992} & \textbf{0.996} & 1.006 & 1.296 & 1.424 \\
12-step & short & 2.627 & 2.626 & 4.142 & \textbf{0.922} & \textbf{0.928} & \textbf{0.934} & 1.092 & 1.086 \\
& medium & 3.194 & 3.219 & 5.154 & \textbf{0.998} & \textbf{0.975} & \textbf{0.970} & 1.256 & 1.248 \\
& long & 3.291 & 3.406 & 5.433 & \textbf{0.999} & \textbf{0.990} & 1.001 & 1.369 & 1.478 \\
\midrule
\multicolumn{10}{c}{rolling window: 240 months} \\
\midrule
1-step & short & \textbf{0.897} & \textbf{0.858} & 1.161 & \textbf{0.986} & \textbf{0.939} & 1.081 & 1.805 & 1.803 \\
& medium & 1.039 & \textbf{0.945} & 1.280 & \textbf{0.992} & \textbf{0.934} & 1.176 & 2.033 & 1.833 \\
& long & 1.090 & 1.090 & 1.373 & 1.047 & 1.028 & 1.102 & 1.295 & 1.312 \\
3-step & short & \textbf{0.803} & \textbf{0.816} & \textbf{0.982} & \textbf{0.915} & \textbf{0.914} & \textbf{0.923} & 1.038 & \textbf{0.939} \\
& medium & 1.000 & 1.022 & 1.220 & \textbf{0.951} & \textbf{0.942} & 1.008 & 1.302 & 1.156 \\
& long & 1.178 & 1.195 & 1.393 & \textbf{0.988} & \textbf{0.989} & 1.020 & 1.158 & 1.179 \\
6-step & short & \textbf{0.797} & \textbf{0.899} & 1.019 & \textbf{0.922} & \textbf{0.924} & \textbf{0.937} & \textbf{0.936} & \textbf{0.883} \\
& medium & 1.028 & 1.172 & 1.396 & \textbf{0.957} & \textbf{0.955} & \textbf{0.993} & 1.156 & 1.091 \\
& long & 1.324 & 1.403 & 1.776 & \textbf{0.986} & \textbf{0.989} & 1.013 & 1.162 & 1.211 \\
12-step & short & \textbf{0.835} & \textbf{0.978} & 1.173 & \textbf{0.945} & \textbf{0.943} & \textbf{0.944} & \textbf{0.907} & \textbf{0.878} \\
& medium & 1.037 & 1.227 & 1.665 & \textbf{0.966} & \textbf{0.950} & \textbf{0.970} & 1.080 & 1.050 \\
& long & 1.430 & 1.564 & 2.344 & \textbf{0.991} & \textbf{0.986} & \textbf{0.998} & 1.183 & 1.233 \\
\bottomrule
\end{tabular}
\end{center}
\vspace{-1.7ex} \hspace{3ex}
\parbox{14.6cm}{  \scriptsize
Note: The table reports out-of-sample MSFEs relative to the random walk forecast across forecast horizons and maturity segments (short: $\leq$12 months, medium: 12-24 months, long: $>$24 months) using 120- and 240-month rolling windows. Bold entries indicate better performance than the random walk benchmark. Forecast methods include functional factor models using BIC/HQC criteria ($q_0=1$) with OLS and lasso estimation, PCA-based forecasts with fFPE criterion, and DNS models with VAR(1) and VAR(2) dynamics.
}
\label{tbl-rolling_msfe}
\end{table}

The dynamic Nelson-Siegel (DNS) framework introduced first by \cite{nelson1987} and further developed by \cite{diebold2006} has emerged as a workhorse model in the financial econometrics literature and has been the basis for many modifications and extensions (see \citealt{svensson1995}, \citealt{christensen2009}, \citealt{lengwiler2010}, and \citealt{diebold2013}).
Functional factor models also find their role in the analysis of the term structure of bond yields.
For instance, functional data models for yield curves have been explored (see \citealt{hays2012}, \citealt{bardsley2017}, \citealt{sen2019}, and \citealt{horvath2022}).
More recently, \citet{lihao2025} applies the cumulative autocovariance-operator approach of \citet{bathia2010} to serially dependent yield curves and develops a forecasting-oriented procedure: the extracted scores are forecast using a VAR model, with the dimension and lag order chosen by forecast-performance criteria (MSPE and cross-validation).
In contrast, our purpose in this application is to illustrate the structural identification, estimation framework and the information-criterion selection developed in Sections~\ref{sec:model}--\ref{sec:estimation}.
If the primary goal is a forecasting-focused analysis (including tuning and evaluation designs tailored to forecast performance), we refer interested readers to \citet{lihao2025}.

Central to the models of the Nelson-Siegel class is the assumption that the bond yield $Y_t(r)$ with time to maturity $r \in [a, b]$ at point in time $t$ follows a strict factor model framework that incorporates an additive discrete white noise component.
The curve process is represented as $Y_t(r_i) = \widetilde{\chi}_t(r_i) + \widetilde{\epsilon}_{i,t}$, where $\widetilde{\chi}_t$ denotes a finite-dimensional factor component, $\widetilde{\epsilon}_{i,t}$ is white noise across $i$ and $t$, and $r_1, \ldots, r_N$ form the discrete grid of available maturities.
The shape of the loading functions and the number of factors are treated as pre-specified parameters.
Specifically, the DNS model assumes the three-factor structure
\begin{equation} \label{eq:dns}
\widetilde \chi_t(r) = F_{1,t} + F_{2,t} \frac{1 - e^{-\xi r}}{\xi r} + F_{3,t} \Big( \frac{1 - e^{-\xi r}}{\xi r} - e^{- \xi r} \Big)
\end{equation}
with \cite{diebold2006} suggesting the decay parameter value $\xi = 0.0609$.
The factors, $F_{1,t}, F_{2,t}, F_{3,t}$, are estimated through ordinary least squares at each time $t$ using the available maturities.

\cite{lengwiler2010} and \cite{nielsen2024} have already highlighted that the three-factor DNS model fails to fully capture the dynamics of yield curve data, suggesting the need for additional or more complex loading functions. Our information criteria support these findings, estimating the number of factors as $\widehat{K}_{\text{bic}} = 5$ and $\widehat{K}_{\text{hqc}} = 10$, with the number of lags estimated as $\widehat{p}_{\text{bic}} = \widehat{p}_{\text{hqc}} = 1$.
In Figure \ref{fig-yieldloadings}, we compare the first six estimated loading functions with the pre-specified loadings from the DNS model.
The first three estimated loadings show similarities to the DNS loadings in terms of magnitude and curvature and have similar economic interpretations: the first factor represents long-term effects, the second short-term, and the third medium-term. The fourth, fifth and sixth factors mediate between short, long, and medium-term effects.

However, a key finding of our analysis is that the estimated number of factors and lags varies significantly depending on the time period. A rolling window analysis with a 120-month (10-year) window reveals that during periods of economic stability, 3-4 factors suffice, whereas during economic crises, substantially more factors are required.
Figure \ref{fig-rollingK} shows that the BIC criterion suggests around four factors until the mid-1990s, a period that includes the 1990–1991 recession and the 1994 bond market crisis. During the relatively stable mid-2000s, the estimated number of factors drops to three. Following the 2007 housing bubble collapse, it increases to seven before fluctuating between three and five in the late 2010s and early 2020s.
These findings highlight the sensitivity of factor specifications to prevailing economic conditions. Rather than fixing the number of factors a priori, we strongly recommend that practitioners adopt a data-driven approach to determine the appropriate specification.

To assess the impact of additional factors on forecasting accuracy, we employ a rolling window approach similar to \cite{diebold2006}, performing sequential monthly out-of-sample yield curve forecast comparisons. At each step, we use a fixed rolling window of $w=120$ and $w=240$ for the training period. Specifically, the $h$-step-ahead forecast for time $t$ is based on data from periods $t-h-w$ to $t-h$.

In addition to comparing our forecasts with those from the dynamic DNS model, we include as a benchmark the naive random walk forecast, which is simply the observed curve of $h$ periods before. \cite{diebold2006} and \cite{caldeira2025} previously noted that DNS yield curve forecasts offer little to no improvement over the naive benchmark.
Their results suggest that any potential improvements from DNS over the random walk are limited to short- and medium-term interest rates, with no significant gains observed for long-term rates.
Therefore, when evaluating the MSFEs, we consider short-term interest rates (up to 12 months to maturity), medium-term interest rates (13 months to 24 months to maturity), and long-term interest rates (25 months to 360 months to maturity) separately.

Table \ref{tbl-rolling_msfe} confirms that the DNS model rarely outperforms the random walk forecast. In contrast, functional factor-based forecasts, where $K$ and $p$ are selected using the BIC or HQC criterion, consistently outperform the random walk for short-term interest rate prediction. However, as illustrated in Figure \ref{fig-rollingK}, the selected values of $K$ and $p$ can become excessively large in certain periods, leading to overparameterization. Consequently, this can result in suboptimal forecasts, particularly for PCA-based predictions using the fFPE criterion, which often tends to select a large number of components and lags.

To address this issue, we also include forecasts from an L1 shrinkage-estimated VAR model, where $K$ and $p$ are selected based on the HQC criterion. The shrinkage parameter is tuned within each rolling training sample using 10-fold cross-validation, as implemented by default in the glmnet R package. Our results indicate that lasso estimators for the factor-based VAR model outperform all other models, including the random walk, particularly for longer forecasting horizons.

\section{Conclusion}
\label{sec:conc}
This paper provides an in-depth study of the factor model for functional time series, including its identification, estimation, and prediction. From a practical point of view, the approximate functional factor model is an attractive modeling framework for infinite-dimensional temporal data, as it allows analyses and predictions via a low-dimensional factor component of the data. Our results are useful for a broad range of applications in which the number of factors is unknown, and the error component potentially has strong cross-correlation and is weakly correlated with the common component. We have developed a simple-to-use novel method, yielding consistent estimates of the number of factors and their dynamics. A Monte Carlo study and two empirical illustrations show that our method provides an attractive modeling and predictive framework.

Several methodological problems await further analysis. The first is to develop the distributional and inferential theory for the estimators beyond the consistency results obtained in this paper. For instance, in the empirical illustration of yield curves, it might be interesting to provide confidence bands or test some restrictions on the loading functions. The second is to go beyond the weakly stationary assumption on the factors, for instance, by allowing some factors to have short memory while others are permitted to have long memory (persistence).
Finally, the third is to develop a predictive methodology for the factors using semiparametric or nonparametric models.

\section*{Acknowledgments}
We thank Jörg Breitung, Juan Carlos Escanciano, Joachim Freyberger, Tobias Hartl, Justus Henseler, Alois Kneip, Malte Knüppel, Dominik Liebl, Alexander Mayer, Daan Opschoor, and Luis Winter for their valuable comments and suggestions, and Justin Franken for his assistance with software implementations.
This work was supported by the Deutsche Forschungsgemeinschaft (DFG) under project number 511905296. Moreover, the first author received financial support from the University of Bonn's Argelander Grants, while the second author was funded by Juan de la Cierva Incorporación, grant number IJC2019-041742-I.
Additionally, we acknowledge the use of the CHEOPS HPC cluster for parallel computing.

\section*{Disclosure Statement}
The authors report there are no competing interests to declare.

\section*{Data Availability Statement}

The U.S.\ log mortality rate data is publicly available from the Human Mortality Database (\url{https://www.mortality.org}) with free registration. The yield curve data is available from \cite{liu2021} at \url{https://sites.google.com/view/jingcynthiawu/yield-data}.
All code and analysis scripts are available in the accompanying R package
at \url{https://github.com/ottosven/dffm}.

\newpage

\begin{center}
{\large\bf SUPPLEMENTARY MATERIAL TO} \\ \vspace{2ex}
{\Large \textbf{Approximate Factor Models for Functional Time Series}} \\
{\large by Sven Otto and Nazarii Salish}
\end{center}

\renewcommand{\thesection}{A}
\setcounter{equation}{0} \renewcommand{\theequation}{A.\arabic{equation}}
\setcounter{lemma}{0} \renewcommand{\thelemma}{A.\arabic{lemma}}
\appendix

\onehalfspacing

\section{Technical Appendix}

\subsection{Notations and Definitions} \label{sec:notations}

In this section, we provide a detailed description of the notations used in this appendix.

\subsubsection*{Norms}
For a function $g \in H = L^2([a,b])$, the squared $L^2$ norm is $\|g\|^2 = \int_a^b g(r)^2 \dd r$. The squared Euclidean norm for a vector $\bm{a} \in \mathbb{R}^n$ is $\|\bm a \|^2_2 = \bm a' \bm a$, and the squared Frobenius norm for a matrix $\bm{A} \in \mathbb{R}^{n \times k}$ with entries $A_{ij}$ is $\|A\|^2_M = \sum_{i=1}^n \sum_{j=1}^k A_{ij}^2$. For an integral operator $\mathcal{T}: H \to H$ with kernel function $\tau(r,s)$, the squared Hilbert-Schmidt norm is $\|\mathcal T\|^2_\mathcal S = \int_a^b \int_a^b \tau(r,s)^2 \dd s \dd r$.

\subsubsection*{Orthogonalized factor model representation}
The transformed factors introduced in Section \ref{sec:model} are defined by the projection coefficients $F_{l,t}^* = \langle Y_t - \mu, \psi_l \rangle$ for $l = 1, \ldots, K$ and satisfy the relation $F_{l,t}^* = F_{l,t} + \langle \epsilon_t, \psi_l \rangle$.
Assumption \ref{as:components}(c) also holds for the $K$-variate process $F_t^* = (F_{1,t}^*, \ldots, F_{K,t}^*)'$ since the error function is a martingale difference sequence by Assumption \ref{as:components}(a).
The factor model has the orthogonalized representation
\begin{align}
	Y_t(r) = (\Psi(r))'F_t^* + \epsilon_t^*(r), \label{eq:transformedmodel}
\end{align}
where the orthogonalized error term is defined as
$$
	\epsilon_t^*(r) := \epsilon_t(r) - \sum_{l=1}^K \langle \epsilon_t, \psi_l \rangle \psi_l(r).
$$
Note that $\epsilon_t^*$ satisfies $\int_a^b \Psi(r) \epsilon_t^*(r) \dd r = 0$ for all $t$, which implies that $\epsilon_t^*$ takes values in $H_F^\perp = \Span(\psi_1, \ldots, \psi_K)^\perp$.
Note that $\|\epsilon_t^*\| \leq (K+1) \|\epsilon_t\|$, and Assumption \ref{as:components}(a) implies that $\epsilon_t^*$ is a martingale difference sequence with respect to $\{ \epsilon_{t-1}^*, F_{t-1}^*, \epsilon_{t-2}^*, F_{t-2}^*, \ldots \}$.
Assumption \ref{as:VAR} implies that $\sup_{r \in [a,b]} E[(\epsilon_t^*(r))^4] < \infty$, $\sup_{r \in [a,b]} E[(Y_t(r))^4] < \infty$ and $E[(F_{l,t}^*)^4] < \infty$ for any $t$ and $l=1, \ldots, K$.

\subsubsection*{Sign-adjusted factor model representation}
The signs of the loading functions are not identified.
Therefore, we condition our notation on the sign transformation matrix $\bm S = \diag(s_1, \ldots, s_K)$, where $s_l = \sign(\langle \widehat \psi_l, \psi_l \rangle)$ denotes the selected sign of the $l$-th sample eigenfunction.
The sign-adjusted vectors of loadings and factors are defined as $\widetilde \Psi(r) =  (\Psi(r))' \bm S$ and $\widetilde F_t = \bm S F_t^*$, and their $l$-th components are $\widetilde \psi_l = s_l \psi_l$ and $\widetilde F_{l,t} = s_l F_{l,t}^*$.
Assumptions \ref{as:components}--\ref{as:VAR} are not affected by the sign-adjustment and also hold for $\widetilde \Psi(r)$ and $\widetilde F_t$.
Given that $\bm{S} \bm{S} = \bm{I}_K$ and $\widetilde A_i = \bm{S} A_i \bm{S}$, the VAR($p$) model of Assumption \ref{as:VAR}(a) can be written as
$$
	\widetilde F_t = \sum_{i=1}^p \widetilde A_i \widetilde F_{t-i} + \bm S \eta_t.
$$

To streamline the notation, we define the sign-adjusted stacked coefficient matrix $\bm{\widetilde A} := [\widetilde A_1, \ldots, \widetilde A_p]$ and the lagged factors vector $\bm{\widetilde x_{t-1}} := (\widetilde F_{t-1}', \ldots, \widetilde F_{t-p}')'$. The VAR($p$) equation then becomes
$$
    \widetilde F_t = \bm{\widetilde A} \bm{\widetilde x}_{t-1} + \bm S \eta_t.
$$

Given that $\widetilde F_t$ follows a linear process with innovations $\bm S \eta_t$ forming a martingale difference sequence, we obtain the normal equation $\widetilde \Gamma = \bm{\widetilde A} \widetilde \Sigma$ and the population moment representation
$$
    \bm{\widetilde A} = \widetilde \Gamma \widetilde \Sigma^{-1} =
    \begin{bmatrix}
    E[\widetilde F_t \widetilde F_{t-1}'] & \ldots & E[\widetilde F_t \widetilde F_{t-p}']
    \end{bmatrix}
    \begin{bmatrix} E[\widetilde F_{t-1} \widetilde F_{t-1}'] & \ldots & E[\widetilde F_{t-1} \widetilde F_{t-p}'] \\ \vdots & & \vdots \\ E[\widetilde F_{t-p} \widetilde F_{t-1}'] & \ldots & E[\widetilde F_{t-p} \widetilde F_{t-p}'] \end{bmatrix}^{-1}.
$$

Combining the factor model equation with the newly introduced orthogonalized and sign-adjusted notations, the model has the dynamic representation
\begin{equation} \label{eq:dynamicrepresentation}
    Y_t(r) = \mu(r) + (\widetilde \Psi(r))' \widetilde F_t + \epsilon_t^*(r) = \mu(r) + (\widetilde \Psi(r))' \bm{\widetilde A} \bm{\widetilde x}_{t-1} + (\Psi(r))' \eta_t + \epsilon_t^*(r).
\end{equation}

\subsubsection*{VAR coefficient matrix estimator}

For the selected numbers of factors $J$ and lags $m$, we define the $J \times 1$ vector of sample scores $\widehat F_t^{(J)} = (\widehat F_{1,t}, \ldots, \widehat F_{J,t})'$, where the $l$-th component is $\widehat F_{l,t} = \langle Y_t - \widehat \mu, \widehat \psi_l \rangle$.
The LS estimator can be represented using the stacked vector of lagged sample scores $\bm{\widehat x}_{t-1}^{(J,m)}= ((\widehat F_{t-1}^{(J)})', \ldots, (\widehat F_{t-m}^{(J)})')'$ and the matrices
$$
\widehat\Gamma_{(J,m)}= \frac{1}{T} \sum_{t=m+1}^T \widehat{F}_t^{(J)}(\bm{\widehat x}_{t-1}^{(J,m)})', \quad \widehat\Sigma_{(J,m)}= \frac{1}{T} \sum_{t=m+1}^T \bm{\widehat x}_{t-1}^{(J,m)}(\bm{\widehat x}_{t-1}^{(J,m)})'.
$$
In the scenario of overselection ($J \geq K$ and $m \geq p$), the estimator is
$$
\bm{\widehat A}^* = \bm{\widehat A}_{(J,m)} = [\widehat A_1^{(J)}, \ldots, \widehat A_m^{(J)}] = \widehat \Gamma_{(J,m)} \widehat \Sigma_{(J,m)}^{-1}.
$$
In cases of lag underselection and factor underselection, we have
 $$
 \bm{\widehat A}^* = \begin{cases}
 	\big[\bm R_{J,K}' \widehat A_1^{(J)} \bm R_{J,K}, \ldots, \bm R_{J,K}' \widehat A_m^{(J)} \bm R_{J,K}, \bm 0_{J^*, (p-m)J^*} \big] & m < p, \\
 	\big[\bm R_{J,K}' \widehat A_1^{(J)} \bm R_{J,K}, \ldots, \bm R_{J,K}' \widehat A_m^{(J)} \bm R_{J,K} \big] & J < K \ \text{and} \ m \geq p,
 \end{cases}
 $$
 where
$J^* = \max\{J,K\}$ and
$$
\bm R_{J,K} = \begin{cases}
		\big[ \bm I_J, \bm 0_{J,K-J} \big], & \text{if} \ J < K, \\
		\bm I_J, & \text{if} \ J \geq K.
	\end{cases}
$$

\subsubsection*{Aligned VAR population coefficients}

For the underselection scenarios, the aligned population coefficient matrix is defined as
$$
 \bm{A}^* = \begin{cases}
 	\big[\bm R_{K,J}'\widetilde A_1 \bm R_{K,J}, \ldots, \bm R_{K,J}' \widetilde A_p \bm R_{K,J} \big] & m < p, \\
 	\big[\widetilde A_1 , \ldots, \widetilde A_p, \bm 0_{J, (m-p)J} \big] & J < K \ \text{and} \ m \geq p.
 \end{cases}
$$
The overselection scenario ($J \geq K$ and $m \geq p$) requires a more complex notation.
Our proof of Theorem \ref{thm:Bias} hinges on the identification of appropriate population counterparts for $\widehat \Gamma_{(J,m)}$ and $\widehat \Sigma_{(J,m)}$ that satisfy the equation $\bm{A}^* = \widetilde \Gamma^* (\widetilde \Sigma^*)^{-1}$.
In the special case $J = K$, the aligned matrix is represented as
$$
	\bm{A}^* =
	\begin{bmatrix}
		E[\widetilde F_t \widetilde F_{t-1}'] & \ldots & E[\widetilde F_t \widetilde F_{t-m}']
	\end{bmatrix}
	\begin{bmatrix}
		E[\widetilde F_{t-1} \widetilde F_{t-1}'] & \ldots & E[\widetilde F_{t-1} \widetilde F_{t-m}'] \\
		\vdots & & \vdots \\
E[\widetilde F_{t-m} \widetilde F_{t-1}']	 & \ldots & E[\widetilde F_{t-m} \widetilde F_{t-m}']	
	\end{bmatrix}^{-1},
$$
where the last $K(m-p)$ columns consist solely of zeros, which follows from the fact that the partial autocorrelation function for a VAR($p$) process is zero for lags exceeding $p$.
For the case $J > K$, we introduce the auxiliary block matrices
$$
	G_{ij} :=
	\begin{cases} \begin{bmatrix}
	E[\widetilde F_{t-i} \widetilde F_{t-j}'] & \bm 0_{K,J-K} \\
	\bm 0_{J-K, K} & \bm 0_{J-K, J-K}
	\end{bmatrix} & \text{for} \ i \neq j, \\ \vspace*{-2ex} \\
	\begin{bmatrix} E[\widetilde F_{t-i} \widetilde F_{t-i}'] & \bm 0_{K, J-K} \\
	\bm 0_{J-K, K} & \widehat V_i
	\end{bmatrix} & \text{for} \ i = j,
	\end{cases}
$$
where
$$
\widehat V_i := \frac{1}{T} \sum_{t=m+1}^T
\begin{pmatrix}
	\widehat F_{K+1,t-i} \\ \vdots \\ \widehat F_{J,t-i}
	\end{pmatrix}	
	\begin{pmatrix}
	\widehat F_{K+1,t-i} \\ \vdots \\ \widehat F_{J,t-i}
	\end{pmatrix}'.
$$
Since the covariance operator of $Y_t$ has infinitely many positive eigenvalues for any $t$, the entries of $\widehat V_i$ are uniformly bounded away from zero for all $i = 1, \ldots, m$.
Together with Assumptions \ref{as:components}(b) and \ref{as:VAR}, it follows that $G_{ii}$ is symmetric and uniformly positive definite.
Then, a population counterpart of $\widehat \Sigma_{(J,m)}$ can be defined as
$$
\widetilde \Sigma^* := \begin{bmatrix}
		G_{11} & \ldots & G_{1m} \\
		\vdots & & \vdots \\
		G_{m1} & \ldots & G_{mm}
	\end{bmatrix},
$$
which is symmetric and uniformly positive definite as well.

Notice that, for the column and row indices $(j_1,j_2) \in \mathcal I_i = \{(i-1)J+K+1, \ldots, iJ\}$, the entries of $\widehat \Sigma_{(J,m)}$ and $\widetilde \Sigma^*$ coincide for all $i = 1, \ldots, m$ because $\widehat V_i$ are also the corresponding block entries of $\widehat \Sigma_{(J,m)}$.
This is a convenient notation because the corresponding block diagonal entries of the difference $\widehat \Sigma_{(J,m)} - \widetilde \Sigma^*$ are zero by definition.
Moreover, focusing on the column and row indices from the set $\mathcal{I}_i$ for $i=1, \ldots, m$, the inverse $(\widetilde \Sigma^*)^{-1}$ has the same block structure as $\widetilde \Sigma^*$ with
$$
(\widetilde \Sigma^*)^{-1} = \begin{bmatrix}
		* & \bm 0 & * \\ \bm 0 & \widehat V_{i} & \bm 0 \\ * & \bm 0 & *
	\end{bmatrix}^{-1} = \begin{bmatrix}
		* & \bm 0 & * \\
		\bm 0 & \widehat V_{i}^{-1} & \bm 0 \\
		* & \bm 0 & *
	\end{bmatrix},
$$
where the entries of the diagonal blocks $\widehat V_{i}$ do not affect any other entry of the inverse of that matrix  (see, e.g., \cite{luetkepohl1996} Section 3.5.3).
We define the population equivalent of $\widehat \Gamma_{(J,m)}$ as
$$\widetilde \Gamma^* := \begin{bmatrix}
G_{01} & \ldots & G_{0m}
\end{bmatrix},$$
which is a $J \times Jm$ matrix consisting of zeros at the columns with indices from the index set $\mathcal I = \cup_{i=1}^m \mathcal I_i$.
Consequently, the entries of the product $\widetilde \Gamma^* (\widetilde \Sigma^*)^{-1}$ do not depend on $\widehat V_i$ and are zero for the column and row indices $(j_1,j_2) \in \mathcal I_i$ for all $i=1, \ldots, m$.
Therefore,
$$
		\bm{A}^* =
\widetilde \Gamma^* (\widetilde \Sigma^*)^{-1}		
		 =
		 \begin{bmatrix}
G_{01} & \ldots & G_{0m}
\end{bmatrix}
	\begin{bmatrix}
		G_{11} & \ldots & G_{1m} \\
		\vdots & & \vdots \\
G_{m1}	 & \ldots & G_{mm}	
	\end{bmatrix}^{-1}.
$$

\subsubsection*{Aligned one-step ahead curve predictor}
The estimated one-step ahead curve predictor can be written as
\begin{align} \label{eq:estimatedpredictorcurve}
	\widehat{Y}_{t|t-1}^{(J,m)}(r) =  \widehat \mu(r) +  \big(\widehat \Psi^{(J)}(r)\big)' \bm{\widehat A}_{(J,m)} \bm{\widehat x}_{t-1}^{(J,m)}
	= \widehat \mu(r) +  \big(\widehat \Psi^{(J^*)}(r)\big)' \bm{\widehat A}^* \bm{\widehat x}_{t-1}^{(J^*,m^*)},
\end{align}
where $J^*=\max\{J,K\}$ and $m^* = \max\{m,p\}$, and its population counterpart is
\begin{align} \label{eq:predictorcurve}
	\widetilde Y_{t|t-1}(r) = \mu(r) + (\widetilde \Psi(r))' \bm{\widetilde A} \bm{\widetilde x}_{t-1} = \mu(r) + (\widetilde \Psi^{(J^*)}(r))' \bm{A}^* \bm{\widetilde x}^{(J^*,m^*)}_{t-1},
\end{align}
where we set $\widetilde \Psi^{(J^*)}(r) = \widetilde \Psi(r)$ for $J \leq K$, and $\widetilde \Psi^{(J^*)}(r) = ((\widetilde \Psi(r))', \bm 0_{J-K}')'$ for $J > K$.

\subsection{Proof of Proposition \ref{prop:optimality}}

Since $(\lambda_l, \psi_l)$ are the eigencomponents of $D$, we have
\begin{align*}
	&\lim_{T \to \infty} \sum_{\tau=1}^{q_0} \int_a^b \bigg( \frac{1}{T}  \sum_{t=\tau+1}^T Cov\big[\langle Y_t - \mu, \psi_l \rangle, Y_{t-\tau}(q) \big] \bigg)^2 \dd q \\
	&= \int_a^b \int_a^b \sum_{\tau=1}^{q_0} \int_a^b c_\tau(r,q) c_\tau(s,q) \psi_l(s) \psi_l(r) \dd s \dd r \dd q \\
	&= \int_a^b \int_a^b d(r,s) \psi_l(s) \psi_l(r) \dd s \dd r
	= \lambda_l.
\end{align*}
Furthermore, we have $\sum_{j=1}^{K} \langle \psi_j, f\rangle^2 \leq \|f\|^2 = 1$ and $\sum_{j=1}^{l-1} \langle \psi_j, f\rangle^2 = 0$ because $f \in \Span(\psi_1, \ldots, \psi_{l-1})^\perp$.
Recall that $d(r,s) = \sum_{l=1}^K \lambda_l \psi_l(r) \psi_l(s)$.
Then,
\begin{align*}
	&\lim_{T \to \infty} \sum_{\tau=1}^{q_0} \int_a^b \bigg( \frac{1}{T}  \sum_{t=\tau+1}^T Cov\big[\langle Y_t - \mu, f \rangle, Y_{t-\tau}(q)\big] \bigg)^2 \dd q \\
	&= \int_a^b \int_a^b \int_a^b c_\tau(r,q) c_\tau(s,q) f(s) f(r) \dd s \dd r \dd q
	= \int_a^b \int_a^b d(r,s) f(s) f(r) \dd s \dd r \\
	&= \sum_{j=1}^K \lambda_j \langle \psi_j, f\rangle^2
	\leq \lambda_l \sum_{j=l}^K \langle \psi_j, f\rangle^2
	\leq \lambda_l,
\end{align*}
which implies the first statement.
The second statement follows from the fact that $\int_a^b d(r,s) g(s) \dd s = 0$ because $g \in H_F^\perp$ and $\langle g, \psi_l\rangle = 0$ for all $l=1, \ldots, K$.

\subsection{Auxiliary Lemmas}

For the proofs of Theorems \ref{thm:consistency}--\ref{thm:InformCriteria} we require some additional lemmas.
Lemma \ref{lem:L1} is needed for Theorems \ref{thm:consistency}--\ref{thm:InformCriteria},  Lemma \ref{lem:L2} is needed for Theorems \ref{thm:Bias}--\ref{thm:InformCriteria}, Lemma \ref{lem:L3} is needed for Theorem \ref{thm:Bias}, Lemma \ref{lem:L4} is needed for Lemma \ref{lem:L5}, and Lemma \ref{lem:L5} is needed for Theorem \ref{thm:InformCriteria}.
The proofs of the lemmas can be found in Sections \ref{sec:proofL1}--\ref{sec:proofL4}.

\begin{lemma}\label{lem:L1}
Under Assumptions \ref{as:components}--\ref{as:VAR}, for any $0 \leq h < \infty$ and $i_1, i_2 = 0, \ldots, h$, as $T \to \infty$, we have
	\begin{align*}
		E\Bigg[\bigg\| \frac{1}{T} \sum_{t=h+1}^T \Big( \widetilde F_{t-i_1} \widetilde F_{t-i_2}' - E\big[\widetilde F_{t-i_1} \widetilde F_{t-i_2}'\big] \Big) \bigg\|_M^2  \Bigg] = O(T^{-1}).
	\end{align*}
\end{lemma}

\begin{lemma} \label{lem:L2}
Under the conditions of Theorem \ref{thm:Bias}, for any $0 \leq h < \infty$ and $i_1, i_2 = 0, \ldots, h$, as $T \to \infty$,
$$\bigg\| \frac{1}{T} \sum_{t=h+1}^T \Big( \widehat F_{t-i_1}^{(K)} \big(\widehat F_{t-i_2}^{(K)}\big)' - \widetilde F_{t-i_1} \widetilde F_{t-i_2}' \Big) \bigg\|_M = O_P(T^{-1/2}).$$
\end{lemma}

\begin{lemma} \label{lem:L3}
Under the conditions of Theorem \ref{thm:Bias}, for any $0 \leq h < \infty$, $i, j = 0, \ldots, h$, and $J \geq K$, as $T \to \infty$,
$$\bigg\| \frac{1}{T} \sum_{t=h+1}^T \Big( \widehat F_{t-i}^{(J)} \big(\widehat F_{t-j}^{(J)}\big)' - \widehat G_{t,i,j} \Big) \bigg\|_M = O_P(T^{-1/2}),$$
where
$$
	\widehat G_{t,i,j} :=
	\begin{cases} \begin{bmatrix}
	\widehat F_{t-i}^{(K)} (\widehat F_{t-j}^{(K)})' & \bm 0_{K,J-K} \\
	\bm 0_{J-K, K} & \bm 0_{J-K, J-K}
	\end{bmatrix} & \text{for} \ i \neq j, \\ \vspace*{-2ex} \\
	\begin{bmatrix} \widehat F_{t-i}^{(K)} (\widehat F_{t-i}^{(K)})' & \bm 0_{K, J-K} \\
	\bm 0_{J-K, K} & \widehat V_i
	\end{bmatrix} & \text{for} \ i = j,
	\end{cases}
$$
and $\widehat V_i$ is defined as in Section \ref{sec:notations}.
\end{lemma}

\begin{lemma} \label{lem:L4}
Under the conditions of Theorem \ref{thm:Bias}, for any $K \leq J \leq K_{max}$ and $p \leq m \leq p_{max}$, as $T \to \infty$,
\begin{itemize}
	\item[(a)] $\| T^{-1} \sum_{t=m+1}^T \widehat{\bm x}_{t-1}^{(J,m)}  \eta_{t}'  \|_M = O_P(T^{-1/2})$
	\item[(b)] $\sum_{l=1}^{J} \| T^{-1} \sum_{t=m+1}^T   \widehat{\bm x}_{t-1}^{(J,m)} \langle \widehat \psi_l, \epsilon_t^* \rangle \|_M = O_P(T^{-1/2})$
	\item[(c)] $\| T^{-1} \sum_{t=m+1}^T  \widehat{\bm x}_{t-1}^{(J,m)} ( \bm{\widetilde x}_{t-1} - \widehat{\bm x}_{t-1}^{(K,p)})' \|_M = O_P(T^{-1/2})$
\end{itemize}
\end{lemma}

\begin{lemma} \label{lem:L5}
Let $\Theta_P(\cdot)$ denote the exact order Landau symbol, that is, $a_T = \Theta_P(1)$ if and only if $a_T = O_P(1)$ and $a_T^{-1} = O_P(1)$.
Under the conditions of Theorem \ref{thm:Bias}, for any $J \leq K_{max}$ and $m \leq p_{max}$, as $T \to \infty$,
\begin{itemize}
	\item[(a)] $\displaystyle \frac{1}{T} \sum_{t=m^*+1}^T \big\| \widehat Y_{t|t-1}^{(K,p)} - \widehat Y_{t|t-1}^{(J,m)} \big\|^2 = \begin{cases} O_P(T^{-1}) & \text{if} \ J\geq K \ \text{and} \ m \geq p, \\ \Theta_P (1) & \text{otherwise,} \end{cases}$
	\item[(b)]
	$ \displaystyle \frac{1}{T} \sum_{t=m^*+1}^T \big\langle Y_t - \widetilde Y_{t|t-1}, \widehat Y_{t|t-1}^{(K,p)} - \widehat Y_{t|t-1}^{(J,m)} \big\rangle
	= \begin{cases} O_P(T^{-1}) & \text{if} \ J\geq K \ \text{and} \ m \geq p, \\ O_P(T^{-1/2}) & \text{otherwise,} \end{cases}$
	\item[(c)] $ \displaystyle	\frac{1}{T} \sum_{t=m^*+1}^T \big\langle \widetilde Y_{t|t-1} - \widehat Y_{t|t-1}^{(K,p)}, \widehat Y_{t|t-1}^{(K,p)} - \widehat Y_{t|t-1}^{(J,m)} \big\rangle
	= \begin{cases} O_P(T^{-1}) & \text{if} \ J\geq K \ \text{and} \ m \geq p, \\ O_P(T^{-1/2}) & \text{otherwise.} \end{cases}$
\end{itemize}
\end{lemma}

\subsection{Proof of Theorem \ref{thm:consistency}}

\subsubsection*{Proof of Theorem \ref{thm:consistency}(a)}

Using sign-adjusted version of equation \eqref{eq:transformedmodel},
\begin{equation}
	Y_t(r) = \mu(r) + (\widetilde{\Psi}(r))'\widetilde F_t + \epsilon_t^*(r), \label{eq:transformedmodelSignAdj}
\end{equation}
we have
\begin{align*}
		\big(\widehat \mu(r) - \mu(r)\big)^2
		&= \bigg((\widetilde{\Psi}(r))' \bigg( \frac{1}{T} \sum_{t=1}^T \widetilde F_t \bigg) + \frac{1}{T} \sum_{t=1}^T \epsilon_t^*(r)\bigg)^2 \\
		&= \frac{1}{T^2} \bigg(\sum_{t=1}^T \widetilde F_t  \bigg)' (\widetilde{\Psi}(r))(\widetilde{\Psi}(r))' \bigg(\sum_{t=1}^T \widetilde F_t  \bigg)  + \frac{1}{T^2} \bigg( \sum_{t=1}^T \epsilon_t^*(r) \bigg)^2 \\
		&\quad
		 + \frac{2}{T^2} \bigg(\sum_{t=1}^T \widetilde F_t  \bigg)' \bigg(\sum_{t=1}^T \widetilde{\Psi}(r) \epsilon_t^*(r) \bigg).
\end{align*}	
Since $\int_a^b \widetilde{\Psi}(r) \epsilon_t^*(r) \dd r = 0$ and $\int_a^b \widetilde{\Psi}(r) (\widetilde{\Psi}(r))' \dd r = \bm I_{K}$, it follows that
\begin{align} \label{eq:thm1a.keyexpression}
	E\big[\| \widehat \mu - \mu \|^2 \big]
	= \frac{1}{T^2} \sum_{t,h=1}^T E\big[\widetilde F_t' \widetilde F_h\big] + \frac{1}{T^2} \sum_{t,h=1}^T E\big[ \langle \epsilon_t^*, \epsilon_h^* \rangle \big].
\end{align}
The first term of \eqref{eq:thm1a.keyexpression} satisfies
$$
\frac{1}{T^2} \sum_{t,h=1}^T E[\widetilde F_t' \widetilde F_h]  = \frac{1}{T^2} \sum_{t,h=1}^T \tr\big(E[\widetilde F_t \widetilde F_h']\big)
= \frac{1}{T^2} \sum_{t=1}^T \sum_{i,j=0}^\infty \tr\big(B_i E[\eta_{t-i} \eta_{t-i}'] B_j'\big) = O(T^{-1})
$$
since
$$
	\frac{1}{T} \sum_{t=1}^T \sum_{i,j=0}^\infty \tr\big(B_i E[\eta_{t-i} \eta_{t-i}'] B_j'\big) \overset{T \to \infty}{\longrightarrow} \sum_{i,j=0}^\infty \tr\big(B_i \Sigma_\eta B_j' \big) \leq \big\|\Sigma_\eta \big\|_M \sum_{i,j=0}^\infty \big\|B_i \big\|_M  \big\|B_j \big\|_M < \infty
$$
by Assumption \ref{as:VAR}(a) and the Cauchy-Schwarz inequality for the trace.
For the second term of \eqref{eq:thm1a.keyexpression}, note that $\epsilon_t^*$ is a m.d.s with respect to $\{\epsilon_{t-1}^*,\epsilon_{t-2}^*,...\}$ (see the discussion in Section \ref{sec:notations}). Hence,
$$
\frac{1}{T^2} \sum_{t,h=1}^T E\big[ \langle \epsilon_t^*, \epsilon_h^* \rangle \big]
= \frac{1}{T^2} \sum_{t=1}^T E \big[\|\epsilon_t^*\|^2\big]=O(T^{-1})
$$
by Assumption \ref{as:VAR}(b).
The assertion follows by Markov's inequality.

\subsubsection*{Proof of Theorem \ref{thm:consistency}(b)}

Consider the demeaned curves $Y_t^\mu(r) := Y_t(r) - \mu(r)$ and $\widehat Y_t^{\mu}(r) := Y_t(r) - \widehat \mu(r)$
yielding
$$
	c_\tau(r,s) = \lim_{T \to \infty} \frac{1}{T} \sum_{t=\tau+1}^T E[Y_t^\mu(r) Y_{t-\tau}^\mu(s)], \quad
	\widehat c_\tau(r,s) = \frac{1}{T} \sum_{t=\tau+1}^T \widehat Y_t^\mu(r) \widehat Y_{t-\tau}^\mu(s).
$$
Let $\widetilde C_\tau$ be the integral operator with kernel
\begin{align*}
	\widetilde c_\tau(r,s) = \frac{1}{T} \sum_{t=\tau+1}^{T} Y_t^\mu(r) Y_{t-\tau}^\mu(s)
\end{align*}
such that, by the triangle inequality,
\begin{align}
	\| \widehat C_\tau - C_\tau \|_{\mathcal S} \leq \| \widehat C_\tau - \widetilde C_\tau \|_{\mathcal S} + \| \widetilde C_\tau - C_\tau \|_{\mathcal S}. \label{eq:C-consistency-aux1}
\end{align}
It remains to show that the right-hand side of \eqref{eq:C-consistency-aux1} is $O_P(T^{-1/2})$.
For the first term of \eqref{eq:C-consistency-aux1}, we decompose
\begin{align*}
	&\widehat Y_t^\mu(r) \widehat Y_{t-\tau}^\mu(s) - Y_t^\mu(r)Y_{t-\tau}^\mu(s) \\
	&= \widehat Y_t^\mu(r)(\widehat Y_{t-\tau}^\mu(s) - Y_{t-\tau}^\mu(s)) + Y_{t-\tau}^\mu(s) (\widehat Y_t^\mu(r) - Y_t^\mu(r)) \\
	&= \widehat Y_t^\mu(r)(\mu(s) - \widehat \mu(s)) + Y_{t-\tau}^\mu(s) (\mu(r) - \widehat \mu(r)) \\
	&= (Y_t^\mu(r) + \mu(r) - \widehat \mu(r) )(\mu(s) - \widehat \mu(s)) + Y_{t-\tau}^\mu(s) (\mu(r) - \widehat \mu(r)).
\end{align*}
Then,
$$	\| \widehat C_\tau - \widetilde C_\tau \|_{\mathcal S}^2 \leq  \|A_1\|_S^2 + \|A_2\|_S^2+2\|A_1\|_S\|A_2\|_S,
$$
where $A_1$, $A_2$ are the integral operators with kernels defined as:
\begin{align*}
	a_1(r,s) &= \frac{1}{T} \sum_{t=\tau+1}^{T} (Y_t^\mu(r) + \mu(r) - \widehat \mu(r) )(\mu(s) - \widehat \mu(s)),  \\
	a_2(r,s) &= \frac{1}{T} \sum_{t=\tau+1}^{T} Y_{t-\tau}^\mu(s) (\mu(r) - \widehat \mu(r)).
\end{align*}
Hence, to derive the rate of convergence of $\| \widehat C_\tau - \widetilde C_\tau \|_{\mathcal S}$ it suffices to obtain rates of convergence for $\|A_1\|_S$ and $\|A_2\|_S$.
By Theorem \ref{thm:consistency}(a) and Minkowski's inequality, we have
\begin{align*}
	\|A_1\|_S = \bigg\| \frac{1}{T} \sum_{t=\tau+1}^{T} (Y_t^\mu + \mu - \widehat \mu) \bigg\| \ \|\mu - \widehat \mu\|  \leq \frac{1}{T} \sum_{t=\tau+1}^{T} \left\|Y_t^\mu + \mu - \widehat \mu\right\|\|\mu - \widehat \mu\| = O_P(T^{-1/2})
\end{align*}
because $T^{-1} \sum_{t=\tau+1}^{T} E\|(Y_t^\mu + \mu - \widehat \mu)\|\leq  T^{-1} \sum_{t=\tau+1}^{T} E\left\|Y_t^\mu\right\|+ E \|\mu - \widehat \mu\|=O(1)$.
By the identical arguments, it follows that $\|A_2\|_S=O_P(T^{-1/2})$, and $\| \widehat C_\tau - \widetilde C_\tau \|_{\mathcal S}=O_P(T^{-1/2})$.
For the second term of \eqref{eq:C-consistency-aux1}, we use the notation of equation \eqref{eq:transformedmodelSignAdj} yielding
\begin{align*}
	Y_t^\mu(r) Y_{t-\tau}^\mu(s) &= \big( (\widetilde{\Psi}(r))'\widetilde F_t + \epsilon_t^*(r) \big) \big( \widetilde F_{t-\tau}'\widetilde{\Psi}(s) + \epsilon_{t-\tau}^*(s) \big), \\
	E[Y_t^\mu(r) Y_{t-\tau}^\mu(s)] &= (\widetilde{\Psi}(r))' E[\widetilde F_t\widetilde F_{t-\tau}'] \Psi(s) +  (\widetilde{\Psi}(r))'E[\widetilde F_{t} \epsilon_{t-\tau}^*(s) ],
\end{align*}
where the last equality follows from the m.d.s.\ property of $\epsilon_t^*$.
By the definition of the operators,
\begin{align*}
	\big( \widetilde c_\tau(r,s) - c_\tau(r,s) \big)^2 = \lim_{T\to \infty} \Big( \sum_{i=1}^4 b_i(r,s) \Big)^2,
\end{align*}
where
\begin{align*}
	b_1(r,s) &= \frac{1}{T} \sum_{t=\tau+1}^{T} (\widetilde{\Psi}(r))' \Big( \widetilde F_t \widetilde F_{t-\tau}' - E[\widetilde F_t \widetilde F_{t-\tau}'] \Big) \widetilde{\Psi}(s), \\
	b_2(r,s) &= \frac{1}{T} \sum_{t=\tau+1}^{T}  (\widetilde \Psi(r))'\Big( \widetilde F_{t} \epsilon_{t-\tau}^*(s)  - E[\widetilde F_{t} \epsilon_{t-\tau}^*(s) ] \Big), \\
	b_3(r,s) &= \frac{1}{T} \sum_{t=\tau+1}^{T}   \epsilon_{t}^*(r) \widetilde F_{t-\tau}'\widetilde{\Psi}(s), 	 \quad
	b_4(r,s) = \frac{1}{T} \sum_{t=\tau+1}^{T}  \epsilon_{t}^*(r) \epsilon_{t-\tau}^*(s).
\end{align*}
Let $B_i$ be the integral operator with kernel function $b_i(r,s)$.
Then, by the triangle inequality,
$$
	\big\| \widetilde C_\tau - C_\tau \big\|_{\mathcal S}^2
	= \int_a^b \int_a^b \big( \widetilde c_\tau(r,s) - c_\tau(r,s) \big)^2 \dd s \dd r
	\leq \lim_{T \to \infty} \sum_{i,j=1}^4 \big\|B_i\|_{\mathcal S} \big\|B_j\|_{\mathcal S}.
$$
Hence, it remains to show that $\|B_i\|_{\mathcal S} = O_P(T^{-1/2})$ for $i=1, \ldots, 4$.
For the first operator, define
$$
	G_{1} := \frac{1}{T} \sum_{t=\tau + 1}^{T} \big( \widetilde F_t \widetilde F_{t-\tau}' - E[\widetilde F_t \widetilde F_{t-\tau}'] \big).
$$
By the Cauchy-Schwarz inequality,
\begin{align*}
	(b_1(r,s))^2 = \big( (\widetilde{\Psi}(r))' G_{1} \widetilde{\Psi}(s) \big)^2 = (\widetilde{\Psi}(r))' G_{1} \widetilde{\Psi}(s) (\widetilde{\Psi}(s))'  G_{1}' \widetilde{\Psi}(r),
\end{align*}
and, since the loading functions are orthonormal,
\begin{align*}
	\|B_1\|_{\mathcal S}^2
	&= \int_a^b (\widetilde{\Psi}(r))' G_{1} G_{1}' \widetilde{\Psi}(r) \dd r
	= \int_a^b \tr\Big( G_{1} G_{1}' \widetilde{\Psi}(r)(\widetilde{\Psi}(r))' \Big) \dd r \\
	&= \tr\big( G_{1} G_{1}'\big) = \|G_{1}\|_M^2 = O_P(T^{-1}),
\end{align*}
where the last step follows from Lemma \ref{lem:L1}.
For the second operator,
$$
	\|B_2\|_{\mathcal S} = \sqrt{ \int_a^b \Big\| \frac{1}{T} \sum_{t=\tau+1}^T \widetilde F_t \epsilon_{t-\tau}^*(s) \Big\|_2^2 \dd s } = O_P(T^{-1/2})
$$
since $E[\|B_2\|_{\mathcal S}] = O(T^{-1/2})$ by Assumption \ref{as:VAR}(b) and Jensen's inequality.
For the third operator,
$$
	\|B_3\|_{\mathcal S}^2 = \int_a^b \sum_{l=1}^K \bigg( \frac{1}{T} \sum_{t=\tau+1}^T \widetilde F_{l,t-\tau} \epsilon_t^*(s) \bigg)^2 \dd s
$$
and $E[(T^{-1} \sum_{t=\tau+1}^T \widetilde F_{l,t-\tau} \epsilon_t^*(s))^2] = T^{-2} \sum_{t=\tau+1}^T E[\widetilde F_{l,t}^2 (\epsilon_t^*(s))^2] = O(T^{-1})$ by Assumption \ref{as:components}(a) and the fact that the fourth moments are bounded.
Hence, $\|B_3\|_{\mathcal S} = O_P(T^{-1/2})$.
Finally, for the fourth operator,
\begin{align} \label{eq:proofCaux2}
	\|B_4\|_{\mathcal S}^2 = \frac{1}{T^2} \sum_{t=\tau+1}^T  \|\epsilon_t^*\|^2 \|\epsilon_{t-\tau}^*\|^2 + \frac{2}{T^2} \sum_{t=\tau+2}^{T} \sum_{h=\tau+1}^{t-1} \langle \epsilon_t^*, \epsilon_{t-h}^*\rangle \langle  \epsilon_{t-\tau}^*, \epsilon_{t-\tau-h}^* \rangle,
\end{align}
where the first term of \eqref{eq:proofCaux2} is $O_P(T^{-1})$.
As discussed in the proof of Theorem \ref{thm:consistency}(a), from Assumption \ref{as:components}(a) it follows that $\epsilon_t^*$ is m.d.s with respect to $\{\epsilon_{t-1}^*,\epsilon_{t-2}^*,...\}$. Hence, for the second term, $E[\langle \epsilon_t^*, \epsilon_{t-h}^*\rangle \langle  \epsilon_{t-\tau}^*, \epsilon_{t-\tau-h}^* \rangle] = 0$ for $\tau>0$ and $h=\tau+1,...,t-1$ by m.d.s property and the law of iterated expectation, and the second term is $O_P(T^{-1})$. Consequently, $\|B_4\|_{\mathcal S} = O_P(T^{-1/2})$, and the right-hand side of \eqref{eq:C-consistency-aux1} is $O_P(T^{-1/2})$ as well.

\subsubsection*{Proof of Theorem \ref{thm:consistency}(c)}
We have the decomposition
\begin{align*}
	&\widehat c_\tau(r,q) \widehat c_\tau(s,q) - c_\tau(r,q)c_\tau(s,q) \\
	&= ( \widehat c_\tau(r,q) - c_\tau(r,q)) \widehat c_\tau(s,q) + (\widehat c_\tau(s,q) - c_\tau(s,q)) c_\tau(r,q) \\
	&= ( \widehat c_\tau(r,q) - c_\tau(r,q)) ( \widehat c_\tau(s,q) - c_\tau(s,q)) \\
	&\quad + ( \widehat c_\tau(r,q) - c_\tau(r,q)) c_\tau(s,q) + (\widehat c_\tau(s,q) - c_\tau(s,q)) c_\tau(r,q),
\end{align*}
which implies that $\widehat D - D = \sum_{\tau=1}^{q_0} ( A_{\tau,1} + A_{\tau,2} + A_{\tau,3} )$,
where $A_{\tau,1}$ is the integral operator with kernel function
$$
	a_{\tau,1}(r,s) = \int_a^b ( \widehat c_\tau(r,q) - c_\tau(r,q)) ( \widehat c_\tau(s,q) - c_\tau(s,q)) \dd q,
$$
$A_{\tau,2}$ is the integral operator with kernel function
$$
	a_{\tau,2}(r,s) = \int_a^b ( \widehat c_\tau(r,q) - c_\tau(r,q)) c_\tau(s,q) \dd q,
$$
and $A_{\tau,3}$ is the integral operator with kernel function $a_{\tau,3}(r,s) = a_{\tau,2}(s,r)$.
The Cauchy-Schwarz inequality implies
\begin{align*}
	(a_{\tau,1}(r,s))^2 &\leq \bigg( \int_a^b (\widehat c_\tau(r,q) - c_\tau(r,q))^2 \dd q \bigg) \bigg( \int_a^b (\widehat c_\tau(s,q) - c_\tau(s,q))^2 \dd q \bigg) \\
	(a_{\tau,2}(r,s))^2 &\leq  \bigg( \int_a^b ( \widehat c_\tau(r,q) - c_\tau(r,q))^2 \dd q  \bigg) \bigg( \int_a^b (c_\tau(s,q))^2 \dd q \bigg).
\end{align*}
Assumption \ref{as:components}(b) implies that $\int_a^b (c_\tau(r,q))^2 \dd q = (\Psi(r))' M (\Psi(r))$, which is continuous in $r$ by Assumption \ref{as:components}(c).
Therefore, $\sup_{r \in [a,b]}\int_a^b (c_\tau(r,q))^2 \dd q < \infty$, and it follows that
$$
\|A_{\tau,1}\|_{\mathcal S} + \|A_{\tau,2}\|_{\mathcal S} + \|A_{\tau,3}\|_{\mathcal S} = O_P(\| \widehat C_\tau - C_\tau \|_{\mathcal S}),
$$
and
$$
	\| \widehat D - D \|_{\mathcal S} \leq \sum_{\tau = 1}^{q_0} \big( \|A_{\tau,1}\|_{\mathcal S} + \|A_{\tau,2}\|_{\mathcal S} + \|A_{\tau,3}\|_{\mathcal S} \big) = O_P(T^{-1/2}).
	$$

\subsubsection*{Proof of Theorem \ref{thm:consistency}(d)}
Lemma 2.2 in \cite{horvath2012} implies $|\widehat \lambda_l - \lambda_l| \leq \Vert \widehat D - D \Vert_\mathcal{S}$
for all $l=1, \ldots, K$ and $|\widehat \lambda_l| \leq \Vert \widehat D - D \Vert_\mathcal{S}$ for all $l > K$. Then, the result follows from (c).

\subsubsection*{Proof of Theorem \ref{thm:consistency}(e)}
Lemma 2.3 in \cite{horvath2012} implies
\begin{align*}
	\max_{1 \leq l \leq K} \Vert s_l \widehat \psi_l - \psi_l \Vert \leq \frac{2 \sqrt 2}{\alpha} \Vert \widehat D - D \Vert_\mathcal{S},
\end{align*}
where $	\alpha = \min \{ \lambda_1 - \lambda_2, \lambda_2 - \lambda_3, \ldots, \lambda_{K-1} - \lambda_K, \lambda_K \}$, and the result follows from (c).

\subsection{Proof of Theorem \ref{thm:Bias}}

\subsubsection*{Proof of Theorem \ref{thm:Bias}(a)}
With the notations introduced in Section \ref{sec:notations},
we have
$$
	\big\|\widehat \Gamma_{(J,m)} - \widetilde \Gamma^* \big\|_M^2 = \sum_{i=1}^m \Big\| \frac{1}{T} \sum_{t=m+1}^T \widehat F_t^{(J)} (\widehat F_{t-i}^{(J)})'  - \bm R_{K,J}'E[\widetilde F_t \widetilde F_{t-i}']\bm R_{K,J} \Big\|_M^2,
$$
and the triangle inequality implies
$$
	\Big\| \frac{1}{T} \sum_{t=m+1}^T \widehat F_t^{(J)} (\widehat F_{t-i}^{(J)})'  - \bm R_{K,J}'E[\widetilde F_t \widetilde F_{t-i}']\bm R_{K,J} \Big\|_M \leq M_1 + M_2 + M_3
$$
with
\begin{align*}
	M_1 &= \Big\| \frac{1}{T} \sum_{t=m+1}^T \widehat F_t^{(J)} (\widehat F_{t-i}^{(J)})'  - \bm R_{K,J}'\widehat F_t^{(K)} (\widehat F_{t-i}^{(K)})' \bm R_{K,J} \Big\|_M, \\
	M_2 &= \Big\| \frac{1}{T} \sum_{t=m+1}^T \bm R_{K,J}'\widehat F_t^{(K)} (\widehat F_{t-i}^{(K)})' \bm R_{K,J} - \bm R_{K,J}'\widetilde F_t \widetilde F_{t-i}' \bm R_{K,J} \Big\|_M, \\
	M_3 &= \Big\| \frac{1}{T} \sum_{t=m+1}^T \bm R_{K,J}'\widetilde F_t \widetilde F_{t-i}' \bm R_{K,J}  - \bm R_{K,J}'E[\widetilde F_t \widetilde F_{t-i}']\bm R_{K,J} \Big\|_M.
\end{align*}
Using the notation introduced in Lemma \ref{lem:L3}, we have $\bm R_{K,J}'\widehat F_t^{(K)} (\widehat F_{t-i}^{(K)})' \bm R_{K,J} = \widehat G_{t,0,i}$ with $i \geq 1$. Then, Lemma \ref{lem:L3} implies $M_1 = O_P(T^{-1/2})$. Due to the block structure of zeros generated by $\bm R_{K,J}$, we have $M_2 = \| T^{-1} \sum_{t=m+1}^T \widehat F_t^{(K)} (\widehat F_{t-i}^{K)})' - \widetilde F_t (\widetilde F_{t-i})' \|_M = O_P(T^{-1/2})$ by Lemma \ref{lem:L2}, and $M_3 = \| T^{-1} \sum_{t=m+1}^T \widetilde F_t (\widetilde F_{t-i})' - E[\widetilde F_t (\widetilde F_{t-i})'] \|_M = O_P(T^{-1/2})$ by Lemma \ref{lem:L1}, which implies
$$
\big\|\widehat \Gamma_{(J,m)} - \widetilde \Gamma^* \big\|_M = O_P(T^{-1/2}).
$$
Analogously,
$$
\big\|\widehat \Sigma_{(J,m)} - \widetilde \Sigma^* \big\|_M^2 = \sum_{i_1,i_2=1}^m \Big\| \frac{1}{T} \sum_{t=m+1}^T \widehat F_{t-i_1}^{(J)} (\widehat F_{t-i_2}^{(J)})' - G_{i_1,i_2} \Big\|_M^2,
$$
where
$$
	\Big\| \frac{1}{T} \sum_{t=m+1}^T \widehat F_{t-i_1}^{(J)} (\widehat F_{t-i_2}^{(J)})' - G_{i_1,i_2} \Big\|_M \leq M_4 + M_5 + M_6,
$$
and, by Lemmas \ref{lem:L1}--\ref{lem:L3}, the terms on the right hand side satisfy
\begin{align*}
	M_4 &= \Big\| \frac{1}{T} \sum_{t=m+1}^T \widehat F_{t-i_1}^{(J)} (\widehat F_{t-i_2}^{(J)})' - \widehat G_{t,i_1,i_2} \Big\|_M = O_P(T^{-1/2}), \\
	M_5 &= \Big\| \frac{1}{T} \sum_{t=m+1}^T \widehat F_{t-i_1}^{(K)} (\widehat F_{t-i_2}^{(K)})' - \widetilde F_{t-i_1} \widetilde F_{t-i_2}' \Big\|_M = O_P(T^{-1/2}), \\
	M_6 &= \Big\| \frac{1}{T} \sum_{t=m+1}^T \widetilde F_{t-i_1} \widetilde F_{t-i_2}' - E[\widetilde F_{t-i_1} \widetilde F_{t-i_2}'] \Big\|_M = O_P(T^{-1/2}).
\end{align*}
Consequently, $\|\widehat \Sigma_{(J,m)} - \widetilde \Sigma^* \|_M = O_P(T^{-1/2})$.
Note that $\|\widehat \Gamma_{(J,m)}\|_M = O_P(1)$, and, since $\widetilde \Sigma^*$ is uniformly positive definite, $\|(\widetilde \Sigma^*)^{-1}\|_M = O_P(1)$.
Following the proof of Lemma 3 in \cite{berk1974} we define $q = \widehat \Sigma^{-1}_{(J,m)}  - (\widetilde \Sigma^*)^{-1}$. Then,
\begin{align*}
	q = (\widetilde \Sigma^{-1}_{(J,m)} + q) ((\widetilde \Sigma^*)^{-1} - \widehat \Sigma_{(J,m)}) (\widetilde \Sigma^*)^{-1},
\end{align*}
which implies that
\begin{align} \label{eq:berktrick}
	\|q\|_M^2 \leq \frac{\|(\widetilde \Sigma^*)^{-1}\|_M^2 \|(\widetilde \Sigma^*)^{-1} - \widehat \Sigma_{(J,m)}\|_M}{1 - \|(\widetilde \Sigma^*)^{-1}\|_M \|(\widetilde \Sigma^*)^{-1} - \widehat \Sigma_{(J,m)}\|_M}.
\end{align}
The numerator of \eqref{eq:berktrick} is $O_P(T^{-1/2})$, and the denominator is bounded away from zero, which implies that
$$
	\big\|\widehat \Sigma^{-1}_{(J,m)}  - (\widetilde \Sigma^*)^{-1}\big\|_M = O_P(T^{-1/2}).
$$
Consider the decomposition
$$
	\bm{\widehat{A}}^* - \bm{A}^* = \widehat \Gamma_{(J,m)} (\widehat \Sigma^{-1}_{(J,m)} - (\widetilde \Sigma^*)^{-1}) + (\widehat \Gamma_{(J,m)} - \widetilde \Gamma^* ) (\widetilde \Sigma^*)^{-1}.
$$
Then, by the triangle inequality and by putting together all rates, we obtain
\begin{align*}
	\big\| \bm{\widehat{A}}^* - \bm{A}^{\ast} \big\|_M
	&\leq  \big\|\widehat \Gamma_{(J,m)}\big\|_M
	\big\| \widehat \Sigma^{-1}_{(J,m)} - (\widetilde \Sigma^*)^{-1} \big\|_M
	+ \big\| \widehat \Gamma_{(J,m)} - \widetilde \Gamma^* \big\|_M  \big\|(\widetilde \Sigma^*)^{-1} \big\|_M \\
	&= O_P(T^{-1/2}).
\end{align*}

\subsubsection*{Proof of Theorem \ref{thm:Bias}(b)}

In the first scenario, $m<p$, we have
\begin{align*}
	\bm{A}^* &= \big[\bm R_{K,J}'\widetilde A_1 \bm R_{K,J}, \ldots, \bm R_{K,J}' \widetilde A_p \bm R_{K,J} \big], \\
	\bm{\widehat A}^* &= \big[\bm R_{J,K}' \widehat A_1^{(J)} \bm R_{J,K}, \ldots, \bm R_{J,K}' \widehat A_m^{(J)} \bm R_{J,K}, \bm 0_{J^*, (p-m)J^*} \big].
\end{align*}
Then, for any $T$,
\begin{align*}
	\big\Vert \bm{\widehat{A}}^{\ast} - \bm{A}^{\ast} \big\Vert_{M}^2
   &= \sum_{i=1}^{m}\big\Vert \bm R_{J,K}' \widehat A_i^{(J)} \bm R_{J,K} - \bm R_{K,J}' \widetilde A_i \bm R_{K,J} \big\Vert_{M}^2 + \sum_{i=m+1}^{p} \big\Vert \bm R_{K,J}' \widetilde A_i \bm R_{K,J} \big\Vert_{M}^2 \\
   &\geq \sum_{i=m+1}^{p} \big\Vert \bm R_{K,J}' \widetilde A_i \bm R_{K,J} \big\Vert_{M}^2 = \sum_{i=m+1}^{p}\big\Vert{A}_i \big\Vert_{M}^2 >0,
\end{align*}
where the last inequality follows from the fact that $A_p \neq 0$ by Assumption \ref{as:VAR}(a).

\medskip

\noindent
In the scenario $J<K$ and $m \geq p$ we have
\begin{align*}
	\bm{A}^* = \big[\widetilde A_1 , \ldots, \widetilde A_p, \bm 0_{J, (m-p)J} \big], \quad
	\bm{\widehat A}^* = \big[\bm R_{J,K}' \widehat A_1^{(J)} \bm R_{J,K}, \ldots, \bm R_{J,K}' \widehat A_m^{(J)} \bm R_{J,K} \big].
\end{align*}
The first $Kp$ elements of the last row of $\bm A^*$ coincide with those of $\bm{\widetilde A}$ and are given by $\bm a = E[\bm{\widetilde x}_{t-1} \bm{\widetilde x}_{t-1}']^{-1} E[\bm{\widetilde x}_{t-1} \widetilde F_{K,t}]$,
which are the population coefficients of the regression of $\widetilde F_{K,t}$ on $\widetilde F_{t-1}, \ldots, \widetilde F_{t-p}$. Assumption \ref{as:components}(b) implies that $\|\bm a\|_2 > 0$.
The first $Kp$ elements of the last row of $\bm{\widehat A}^*$ are zero since $J<K$. Therefore,
$$
\big\Vert \bm{\widehat{A}}^{\ast} - \bm{A}^{\ast} \big\Vert_{M}^2
   \geq \|\bm a\|_2 > 0.
$$

\subsection{Proof of Theorem \ref{thm:InformCriteria}}

Note that $\widehat K$ and $\widehat p$ are discrete random variables.
Therefore, to prove Theorem \ref{thm:InformCriteria}, it is sufficient to show that
\begin{align*}
	\lim\limits_{T\to\infty} \mathrm{P}\Big(\mathrm{CR}_{T}(J,m)<\mathrm{CR}_T(K,p)\Big)=0
\end{align*}
for all $J\leq K_{max}$ and $m\leq p_{max}$.
From the definition of the information criterion, we have
\begin{eqnarray*}
  \mathrm{CR}_{T}(J,m)-\mathrm{CR}_T(K,p)= MSE_T(J,m)-MSE_T(K,p)+ g_T(J,m)-g_T(K,p).
\end{eqnarray*}
Without loss of generality, we prove the result for the case when $f(x)=x$ as the proof for any other strictly increasing transformation $f(\cdot)$ is identical.
Hence, it remains to show that
\begin{eqnarray*}
\lim_{T \to \infty} \mathrm{P}\Big(MSE_T(K,p)-MSE_T(J,m)>g_T(J,m)-g_T(K,p)\Big) = 0.
\end{eqnarray*}
We split the proof into case I, the case of overselection ($J\geq K$ and $m \geq p$), and case II, the case of underselection ($J < K$ or $m < p$ or both).
In any of the two cases, we have
$$
MSE_T(J,m) = \frac{1}{T-m} \sum_{t=m+1}^T \big\| Y_t - \widehat Y_{t|t-1}^{(J,m)} \big\|^2 = O_P(1),
$$
which follows from representations \eqref{eq:dynamicrepresentation} and \eqref{eq:estimatedpredictorcurve} together with Theorem \ref{thm:consistency}(a), Theorem \ref{thm:Bias}, Lemma \ref{lem:L2}, and the fact that enough moments are bounded.
Moreover, we have $MSE_T(J,m) - T^{-1}(T-m) MSE_T(J,m) = O_P(T^{-1})$, and
\begin{align*}
&\frac{T-m}{T} MSE_T(J,m) - \frac{T-p}{T} MSE_T(K,p) - \frac{1}{T} \sum_{t=m^*+1}^T \Big( \big\| Y_t - \widehat Y_{t|t-1}^{(J,m)} \big\|^2  -  \big\| Y_t - \widehat Y_{t|t-1}^{(K,p)} \big\|^2 \Big) \\
&= \frac{1}{T} \sum_{t=m+1}^{m^*} \big\| Y_t - \widehat Y_{t|t-1}^{(J,m)} \big\|^2 - \frac{1}{T} \sum_{t=p+1}^{m^*} \big\| Y_t - \widehat Y_{t|t-1}^{(K,p)} \big\|^2  = O_P(T^{-1}),
\end{align*}
where $m^* = \max\{m,p\}$, which implies that, for any of the two cases I and II,
$$
MSE_T(J,m) - MSE_T(K,p) = \frac{1}{T} \sum_{t=m^*+1}^T \Big( \big\| Y_t - \widehat Y_{t|t-1}^{(J,m)} \big\|^2  -  \big\| Y_t - \widehat Y_{t|t-1}^{(K,p)} \big\|^2 \Big) + O_P(T^{-1}).
$$
Hence, it remains to study $T^{-1} \sum_{t=m^*+1}^T ( \| Y_t - \widehat Y_{t|t-1}^{(J,m)} \|^2  -  \| Y_t - \widehat Y_{t|t-1}^{(K,p)} \|^2)$.
A useful decomposition is obtained by adding and subtracting $\widehat Y_{t|t-1}^{(K,p)}$, i.e.,
\begin{align*}
	&\big\| Y_t - \widehat Y_{t|t-1}^{(J,m)} \big\|^2  -  \big\| Y_t - \widehat Y_{t|t-1}^{(K,p)} \big\|^2 \\
	&= \big\| Y_t - \widehat Y_{t|t-1}^{(K,p)} + \widehat Y_{t|t-1}^{(K,p)} - \widehat Y_{t|t-1}^{(J,m)} \big\|^2 - \big\| Y_t - \widehat Y_{t|t-1}^{(K,p)} \big\|^2 \\
	&= \big\| \widehat Y_{t|t-1}^{(K,p)} - \widehat Y_{t|t-1}^{(J,m)} \big\|^2 + 2 \big\langle Y_t - \widehat Y_{t|t-1}^{(K,p)}, \widehat Y_{t|t-1}^{(K,p)} - \widehat Y_{t|t-1}^{(J,m)} \big\rangle,
\end{align*}
and Lemma \ref{lem:L5} implies
\begin{align*}
	\frac{1}{T} \sum_{t=m^*+1}^T \Big( \big\|Y_t - \widehat Y_{t|t-1}^{(J,m)}\big\|^2 - \big\| \widehat Y_{t|t-1}^{(K,p)} - \widehat Y_{t|t-1}^{(J,m)} \big\|^2 \Big) = \begin{cases} O_P(T^{-1}) & \text{for case I}, \\ \Theta_P (1) & \text{for case II}, \end{cases}
\end{align*}
so that
\begin{align}
	MSE_T(J,m) - MSE_T(K,p) = \begin{cases} O_P(T^{-1}) & \text{for case I}, \\ \Theta_P (1) & \text{for case II}. \end{cases} \label{eq:thm4aux1}
\end{align}
For case I, if $J \geq K$ and $m>p$ or $J>K$ and $m \geq p$, we have $g_T(J,m)-g_T(K,p)>0$, which converges to zero at a slower rate than $T^{-1}$.
This follows from the condition that $T g_T(J,m) \to \infty$ as $T \to \infty$ for all $J$ and $m$.
Thus, $\mathrm{P}(\mathrm{CR}_{T}(J,m)<\mathrm{CR}_T(K,p))\to 0$ as $T\to\infty$.
This result is trivially satisfied if $(J,m) = (K,p)$.
For case II, \eqref{eq:thm4aux1} implies $\plim_{T\to\infty} (MSE_T(K,p)-MSE_T(J,m)) > 0$.
Since $\lim_{T\to\infty} (g_T(J,m)-g_T(K,p))=0$, which is implied by the condition that $g_T(J,m) \to 0$ for all $J$ and $m$, it follows that
$\mathrm{P}(\mathrm{CR}_{T}(J,m)<\mathrm{CR}_T(K,p))\to 0$ as $T\to\infty$, which concludes the proof of the theorem.

\subsection{Proof of Lemma \ref{lem:L1}} \label{sec:proofL1}

The expression of interest can be rewritten as
\begin{align}
	 &E\Bigg[\bigg\| \frac{1}{T} \sum_{t=h+1}^T \widetilde F_{t-i_1} \widetilde F_{t-i_2}' - E\Big[\widetilde F_{t-i_1} \widetilde F_{t-i_2}'\Big] \bigg\|_M^2\Bigg] \nonumber \\
	 &=\frac{1}{T^2} \sum_{t,s=h+1}^T \sum_{m,l=1}^K  Cov\big[\widetilde F_{m,t-i_1} \widetilde F_{l,t-i_2}, \widetilde F_{m,s-i_1}\widetilde F_{l,s-i_2}\big]. \label{eq:problem}
\end{align}
Since the VAR($p$) process $\widetilde F_t$ is causal by Assumption \ref{as:VAR}(a), it has the vector moving average representation $\widetilde F_{t} = \sum_{j=0}^\infty B_j \eta_{t-j}$,
where $\sum_{j=0}^\infty \left\Vert B_{j} \right\Vert_M < \infty$, or, equivalently
\begin{align*}
	\widetilde F_{l,t} = \sum_{j=0}^\infty \sum_{k=1}^{K}b_j^{(l,k)} \eta_{k,t-j},
\end{align*}
where $b_j^{(l,k)}$ is the $(l,k)$ element of the matrix $B_j$ and $\eta_{k,t-j}$ is the $k$-th element of the vector $\eta_{t-j}$.
Then, $Cov[\widetilde F_{m,t-i_1} \widetilde F_{l,t-i_2}, \widetilde F_{m,s-i_1}\widetilde F_{l,s-i_2}]$ is equal to
$$
\sum_{j_1,j_2,j_3,j_4=0}^\infty \sum_{k_1,k_2,k_3,k_4=1}^{K}  b_{j_1}^{(m,k_1)}b_{j_2}^{(l,k_2)}b_{j_3}^{(m,k_3)}b_{j_4}^{(l,k_4)} Cov\big[\eta_{k_1,t-i_1-j_1}\eta_{k_2,t-i_2-j_2},\eta_{k_3,s-i_1-j_3}\eta_{k_4,s-i_2-j_4}\big].
$$
By Assumption \ref{as:VAR}(a), there exists a constant $\kappa < \infty$ such that
\begin{align*}
\lim_{T \to \infty}  \sup_{j_1, j_2, j_3, j_4 \in \mathbb N} \frac{1}{T}  \bigg| \sum_{t,s=h+1}^T Cov\big[\eta_{k_1,t-i_1-j_1}\eta_{k_2,t-i_2-j_2},\eta_{k_3,s-i_1-j_3}\eta_{k_4,s-i_2-j_4}\big] \bigg| \leq \kappa.
\end{align*}
Equation \eqref{eq:problem} and the triangle inequality imply
$$
	\lim_{T \to \infty} E\Bigg[\bigg\| \frac{1}{\sqrt T} \sum_{t=h+1}^T \widetilde F_{t-i_1} \widetilde F_{t-i_2}' - E\Big[\widetilde F_{t-i_1} \widetilde F_{t-i_2}'\Big] \bigg\|_M^2\Bigg] \leq K^6 \kappa \bigg( \sum_{j=0}^\infty \|B_j\|_{\infty} \bigg)^4 < \infty,
$$
where $ \Vert A \Vert_{\infty}=\max_{i,j}\{|a_{i,j}|\}$ is the maximum norm satisfying the matrix inequality $\Vert A \Vert_{\infty}\leq\Vert A \Vert_{M}$ (see, e.g., \citealt{luetkepohl1996} Section 8.5.2).
Consequently, \eqref{eq:problem} is $O(T^{-1})$.

\subsection{Proof of Lemma \ref{lem:L2}} \label{sec:proofL2}

We split our proof in two parts. In the first part we focus on the estimation error for the mean function and show that
\begin{align} \label{eq:lem2part1}
	\bigg\| \frac{1}{T} \sum_{t=h+1}^T \widehat F_{t-i_1}^{(K)} \big(\widehat F_{t-i_2}^{(K)}\big)' - \breve F_{t-i_1} \breve F_{t-i_2}'   \bigg\|_M = O_P(T^{-1/2}),
\end{align}
where $\breve F_t = (\breve F_{1,t}, \ldots, \breve F_{K,t})'$ with $\breve F_{l,t} = \langle Y_t - \mu, \widehat \psi_l \rangle$, and in the second part, we show that
\begin{align} \label{eq:lem2part2}
	\bigg\| \frac{1}{T} \sum_{t=h+1}^T\breve F_{t-i_1} \breve F_{t-i_2}'   - \widetilde F_{t-i_1} \widetilde F_{t-i_2}'  \bigg\|_M = O_P(T^{-1/2}).
\end{align}
The final result then follows by the triangle inequality.
To show equation \eqref{eq:lem2part1}, we define $\breve R_t = \widehat F_t^{(K)} - \breve F_t$ and decompose the expression of interest as
\begin{align*}
	\| \widehat F_{t-i_1}^{(K)} \big(\widehat F_{t-i_2}^{(K)}\big)' - \breve F_{t-i_1} \breve F_{t-i_2}' \|_M^2
	&= \| \breve R_{t-i_1} (\widehat F_{t-i_2}^{(K)})' + \breve F_{t-i_1} \breve R_{t-i_2}' \|_M^2 \\
	&= \| \breve R_{t-i_1} \breve R_{t-i_2}' + \breve R_{t-i_1} \breve F_{t-i_2}' + \breve F_{t-i_1} \breve R_{t-i_2}' \|_M^2 \\
	&= \|\breve R_{t-i_1}\|_2^2 \| \breve R_{t-i_2}\|_2^2
	 + \|\breve R_{t-i_1}\|_2^2 \|\breve F_{t-i_2} \|_2^2
	 + \|\breve F_{t-i_1}\|_2^2 \|\breve R_{t-i_2}\|_2^2,
\end{align*}
where the last step follows by the definitions of the Frobenius and the Euclidean norm.
The Cauchy-Schwarz inequality and the fact that the estimated loadings have unit norm imply
$$ \|\breve R_{t}\|_2^2 = \sum_{l=1}^K |\widehat F_{l,t} - \breve F_{l,t}|^2 = \sum_{l=1}^K |\langle \mu - \widehat \mu, \widehat \psi_l \rangle|^2 \leq K \|\mu - \widehat \mu \|^2$$
for any $t$.
Consequently, by the triangle inequality,
\begin{align*}
&\bigg\| \frac{1}{T} \sum_{t=h+1}^T \widehat F_{t-i_1}^{(K)} \big(\widehat F_{t-i_2}^{(K)}\big)' - \breve F_{t-i_1} \breve F_{t-i_2}'   \bigg\|_M \\
	&\leq \frac{1}{T} \sum_{t=h+1}^T \big\| \widehat F_{t-i_1}^{(K)} \big(\widehat F_{t-i_2}^{(K)}\big)' - \breve F_{t-i_1} \breve F_{t-i_2}' \big\|_M \\
	&\leq \|\mu - \widehat \mu \| \frac{\sqrt{K}}{T} \sum_{t=h+1}^T \sqrt{K \|\mu - \widehat \mu \|^2 + \|\breve F_{t-i_1}\|_2^2 + \|\breve F_{t-i_2} \|_2^2},
\end{align*}
and \eqref{eq:lem2part1} follows by Theorem \ref{thm:consistency}(a) and the fact that enough moments of the factors are bounded.
Analogously, to show \eqref{eq:lem2part2}, we define $R_t := \breve F_t - \widetilde F_t$ and decompose
$$
 \breve F_{t-i_1} \breve F_{t-i_2}' - \widetilde F_{t-i_1} \widetilde F_{t-i_2}' = R_{t-i_1} R_{t-i_2}' + R_{t-i_1} \widetilde F_{t-i_2}' + \widetilde F_{t-i_1} R_{t-i_2}'.
$$
By the triangle inequality, it remains to show that
\begin{align} \label{eq:lem2keyresult}
	\bigg\| \frac{1}{T} \sum_{t=h+1}^T R_{t-i_1} R_{t-i_2}'  \bigg\|_M = O_P(T^{-1/2}), \quad \bigg\| \frac{1}{T} \sum_{t=h+1}^T \widetilde F_{t-i_1} R_{t-i_2}'  \bigg\|_M = O_P(T^{-1/2}).
\end{align}
Note that the $l$-th entry of $R_{t}$ is $R_{l,t} = \langle Y_t - \mu, \widehat \psi_l - \widetilde \psi_l \rangle$.
Then, the Cauchy-Schwarz inequality implies
\begin{align*}
	&\bigg\| \frac{1}{T} \sum_{t=h+1}^T R_{t-i_1} R_{t-i_2}'  \bigg\|_M^2
	= \sum_{k,l=1}^K \bigg( \frac{1}{T} \sum_{t=h+1}^T \langle Y_{t-i_1} - \mu, \widehat \psi_k - \widetilde \psi_k \rangle  \langle Y_{t-i_2} - \mu, \widehat \psi_l - \widetilde \psi_l \rangle \bigg)^2 \\
	&\leq \sum_{k,l=1}^K \bigg( \frac{1}{T} \sum_{t=h+1}^T \langle Y_{t-i_1} - \mu, \widehat \psi_k - \widetilde \psi_k \rangle^2 \bigg) \bigg( \frac{1}{T} \sum_{t=h+1}^T \langle Y_{t-i_2} - \mu, \widehat \psi_l - \widetilde \psi_l \rangle^2 \bigg) \\
	&\leq \bigg( \sum_{k,l=1}^K \|\widehat \psi_k - \widetilde \psi_k\|^2 \|\widehat \psi_l - \widetilde \psi_l\|^2 \bigg) \bigg( \frac{1}{T} \sum_{t=h+1}^T \| Y_{t-i_1} - \mu \|^2 \bigg) \bigg( \frac{1}{T} \sum_{t=h+1}^T \| Y_{t-i_2} - \mu \|^2 \bigg).
\end{align*}
Theorem \ref{thm:consistency}(e), Slutsky's theorem, and the fact that $T^{-1} \sum_{t=h+1}^T\| Y_{t-i_1} - \mu \|^2 = O_P(1)$ since enough moments are bounded imply that the term above is $O_P(T^{-2})$, which implies the first statement in \eqref{eq:lem2keyresult}.
Similarly,
\begin{align*}
	&\bigg\| \frac{1}{T} \sum_{t=h+1}^T \widetilde F_{t-i_1} R_{t-i_2}'  \bigg\|_M^2 = \sum_{k,l=1}^K \Big\langle \frac{1}{T} \sum_{t=h+1}^T \widetilde F_{k,t-i_1}  (Y_{t-i_2} - \mu), \widehat \psi_l - \widetilde \psi_l \Big\rangle^2 \\
	 &\leq \bigg( \sum_{l=1}^K \|\widehat \psi_l - \widetilde \psi_l\|^2 \bigg)  \bigg( \sum_{k=1}^K \Big\| \frac{1}{T} \sum_{t=h+1}^T \widetilde F_{k,t-i_1}  (Y_{t-i_2} - \mu) \Big\|^2 \bigg) = O_P(T^{-1}),
\end{align*}
where the last step follows from Theorem \ref{thm:consistency}(e) and the fact that enough moments are bounded.
Consequently, \eqref{eq:lem2keyresult} and \eqref{eq:lem2part2} hold, and the assertion follows.

\subsection{Proof of Lemma \ref{lem:L3}} \label{sec:proofL3}

The expression $T^{-1} \sum_{t=h+1}^T (\widehat F_{t-i}^{(J)} \big(\widehat F_{t-j}^{(J)}\big)' - \widehat G_{t,i,j})$ can be partitioned into four blocks: the top left block is $\bm 0_{K,K}$, the top right block is $T^{-1} \sum_{t=h+1}^T \widehat F_{t-i}^{(K)} (\widehat F_{K+1,t-j}, \ldots, \widehat F_{J,t-j})$, the bottom left block is $T^{-1} \sum_{t=h+1}^T (\widehat F_{K+1,t-i}, \ldots, \widehat F_{J,t-i})'\widehat F_{t-j}^{(K)}$, and, for the case $i \neq j$, the bottom right block is $T^{-1} \sum_{t=h+1}^T (\widehat F_{K+1,t-i}, \ldots, \widehat F_{J,t-i})'(\widehat F_{K+1,t-j}, \ldots, \widehat F_{J,t-j})$.
If $i = j$, the bottom right block is $\bm 0_{J-K, J-K}$.
Hence, the assertion is equivalent to
$$
	\frac{1}{T} \sum_{t=h+1}^T \widehat F_{k,t-i} \widehat F_{l,t-j} = O_P(T^{-1/2})
$$
for all index combinations $k=K+1, \ldots, J$, $l = 1, \ldots, J$, and $i,j$ that satisfy either $i \neq j$ or $l \leq K$.
Using the notation $\breve F_{l,t} = \langle Y_t - \mu, \widehat \psi_l \rangle$ with $\widehat F_{l,t} = \breve F_{l,t} + \langle \mu - \widehat \mu, \widehat \psi_k \rangle$, Theorem \ref{thm:consistency}(a) implies that it remains to show
\begin{align}	\label{eq:lem3keyresult2}
	\frac{1}{T} \sum_{t=h+1}^T \breve F_{k,t-i} \breve F_{l,t-j} = O_P(T^{-1/2}).
\end{align}
Inserting the sign-adjusted model representation $Y_{t} = \mu + \sum_{m=1}^K \widetilde F_{m,t} \widetilde \psi_m + \epsilon_t^*$, we get, for any $l = 1, \ldots, J$,
$$
	\breve F_{l,t-j} = \langle Y_{t-j} - \mu, \widehat \psi_l \rangle
	= \sum_{m=1}^K \widetilde F_{m,t-j} \langle \widetilde \psi_m, \widehat \psi_l \rangle + \langle \epsilon_{t-j}^*, \widehat \psi_l \rangle.
$$
Since $k > K$, we have $\langle \widehat \psi_m, \widehat \psi_k \rangle = 0$, which yields
$$
\breve F_{k,t-i} = \sum_{m=1}^K \widetilde F_{m,t-i} \langle \widetilde \psi_m, \widehat \psi_k \rangle + \langle \epsilon_{t-i}^*, \widehat \psi_k \rangle
	= \sum_{m=1}^K \widetilde F_{m,t-i} \langle \widetilde \psi_m - \widehat \psi_m, \widehat \psi_k \rangle + \langle \epsilon_{t-i}^*, \widehat \psi_k \rangle.
$$
Hence, the left-hand-side of \eqref{eq:lem3keyresult2} is $T^{-1} \sum_{t=h+1}^T \breve F_{k,t-i} \breve F_{l,t-j} = H_1 + H_2 + H_3 + H_4$ with
\begin{align*}
	H_1 &= \frac{1}{T} \sum_{t=h+1}^T \sum_{m_1, m_2 = 1}^K \widetilde F_{m_1, t-i} \widetilde F_{m_2, t-j} \langle \widetilde \psi_{m_1}-\widehat \psi_{m_1}, \widehat \psi_k \rangle \langle \widetilde \psi_{m_2}, \widehat \psi_l \rangle, \\
	H_2 &= \frac{1}{T} \sum_{t=h+1}^T \sum_{m= 1}^K \widetilde F_{m,t-i} \langle \widetilde \psi_m -  \widehat \psi_m, \widehat \psi_k \rangle \langle \epsilon_{t-j}^*, \widehat \psi_l \rangle, \\
	H_3 &= \frac{1}{T} \sum_{t=h+1}^T \sum_{m=1}^K \widetilde F_{m,t-j} \langle \widetilde \psi_m, \widehat \psi_l \rangle \langle \epsilon_{t-i}^*, \widehat \psi_k \rangle, \\
	H_4 &= \frac{1}{T} \sum_{t=h+1}^T \langle \epsilon_{t-i}^*, \widehat \psi_k \rangle \langle \epsilon_{t-j}^*, \widehat \psi_l \rangle.
\end{align*}
In what follows, we will use the facts that the loadings and sample eigenfunctions are of unit norm and that all random variables have bounded fourth moments. Moreover, we will apply the Cauchy-Schwarz and the triangle inequality multiple times.
For the first two summands, Theorem \ref{thm:consistency}(e) implies
\begin{align*}
	|H_1| &\leq \sum_{m_1, m_2=1}^K \Big\|\widetilde \psi_{m_1} - \widehat \psi_{m_1}  \Big\|\cdot \Big| \frac{1}{T} \sum_{t=h+1}^T \widetilde F_{m_1, t-i} \widetilde F_{m_2, t-j} \Big| = O_P(T^{-1/2}), \\
	|H_2| &\leq \sum_{m= 1}^K \Big\|\widetilde \psi_{m} - \widehat \psi_{m}  \Big\|\cdot \Big| \frac{1}{T} \sum_{t=h+1}^T \widetilde F_{m,t-i} \langle \epsilon_{t-j}^*, \widehat \psi_l \rangle \Big| = O_P(T^{-1/2}).
\end{align*}
For the third summand, we have
$$
	|H_3| \leq \sum_{m=1}^K \Big| \frac{1}{T} \sum_{t=h+1}^T \widetilde F_{m,t-j} \langle \epsilon_{t-i}^*, \widehat \psi_k \rangle \Big| \leq \sum_{m=1}^K \Big\| \frac{1}{T} \sum_{t=h+1}^T \widetilde F_{m,t-j} \epsilon_{t-i}^*(s) \Big\|.
$$
For $j \leq i$, Assumption \ref{as:VAR}(b) implies
$$
	E[|H_3|] \leq K \cdot \sup_{s \in [a,b]} E\bigg[ \Big\| \frac{1}{T} \sum_{t=h+1}^T \widetilde F_{t-j} \epsilon_{t-i}^*(s) \Big\|_2 \bigg] = O(T^{-1/2}),
$$
and Markov's inequality implies $H_3 = O_P(T^{-1/2})$.
For $j > i$, note that, by Assumption \ref{as:components}(a), $\epsilon_t^*$ as a martingale difference sequence with respect to $\{\epsilon_{t-1}^*, \widetilde F_{t-1}, \epsilon_{t-2}^*, \widetilde F_{t-2}, \ldots \}$, which implies that $E[\widetilde F_{m,t_1-j} \widetilde F_{m, t_2 - j} \epsilon_{t_1-i}^*(r) \epsilon_{t_2-i}^*(r)] = 0$ for any $t_2 < t_1$. Then, since the fourth moments are bounded,
\begin{align*}
	&H_3^2 \leq \int_a^b  \bigg( \frac{1}{T} \sum_{t=h+1}^T \widetilde F_{m,t-j} \epsilon_{t-i}^*(r) \bigg)^2  \dd r \\
	&= \int_a^b \frac{1}{T^2} \sum_{t=h+1}^T \big( \widetilde F_{m,t-j} \epsilon_{t-i}^*(r)\big)^2 \dd r  +  \int_a^b \frac{2}{T^2} \sum_{t_1=h+2}^T \sum_{t_2=h+1}^{t_1-1} \widetilde F_{m,t_1-j} \widetilde F_{m, t_2 - j} \epsilon_{t_1-i}^*(r) \epsilon_{t_2-i}^*(r) \dd r \\
	&= O_P(T^{-1}).
\end{align*}
Finally, for the fourth summand, let us first consider the case $i \neq j$. Then,
\begin{align}
	|H_4|^2
	&\leq \Big\| \frac{1}{T} \sum_{t=h+1}^T \epsilon_{t-i}^* \langle \epsilon_{t-j}^*, \widehat \psi_l \rangle \Big\|^2
	\leq \int_a^b \int_a^b \Big( \frac{1}{T} \sum_{t=h+1}^T \epsilon_{t-i}^*(r) \epsilon_{t-j}^*(s) \Big)^2 \dd s \dd r \nonumber \\
	&= \int_a^b \int_a^b \frac{1}{T^2} \sum_{t=h+1}^T (\epsilon^*_{t-i}(r) \epsilon^*_{t-j}(s))^2 \dd s \dd r \nonumber \\
	&\quad
	+ \int_a^b \int_a^b \frac{2}{T^2} \sum_{t_1=h+2}^T \sum_{t_2=h+1}^{t_1-1} \epsilon^*_{t_1-i}(r) \epsilon^*_{t_2-i}(r) \epsilon^*_{t_1-j}(s) \epsilon^*_{t_2-j}(s) \dd s \dd r \nonumber \\
	&= O_P(T^{-1}), \label{eq:lem3_H4result}
\end{align}
where the last step follows from the facts that $\epsilon^*_t$ has bounded fourth moments and satisfies $E[\epsilon^*_{t_1-i}(r) \epsilon^*_{t_2-i}(r) \epsilon^*_{t_1-j}(s) \epsilon^*_{t_2-j}(s)] = 0$ because of the martingale difference sequence property discussed above.
Now let us consider $i = j$, where it must be the case that $j \leq K$ and $k > K$ as discussed at the beginning of this proof.
Then, since $\langle \epsilon^*_{t-i}, \widetilde \psi_l \rangle = 0$,
$$
	|H_4| = \Big| \frac{1}{T} \sum_{t=h+1}^T \langle \epsilon^*_{t-i}, \widehat \psi_k \rangle \langle \epsilon^*_{t-i}, \widehat \psi_l - \widetilde \psi_l \rangle \Big|
	\leq \Big\| \widehat \psi_l - \widetilde \psi_l \Big\| \cdot \Big\| \frac{1}{T} \sum_{t=h+1}^T \epsilon^*_{t-i} \langle \epsilon^*_{t-i}, \widehat \psi_k \rangle \Big\|,
$$
which is $O_P(T^{-1/2})$ by Theorem \ref{thm:consistency}(e).
Finally, the triangle inequality implies \eqref{eq:lem3keyresult2}, and the assertion follows.

\subsection{Proof of Lemma \ref{lem:L4}} \label{sec:proofL4}

A useful result for this proof is that, for any $j=1, \ldots, K$,
\begin{align} \label{eq:lem5_aux1}
		\big|\widehat F_{l,t} - \widetilde F_{l,t}\big| = \big| \langle Y_{t} - \mu, \widehat \psi_l - \widetilde \psi_l \rangle + \langle \widehat \mu - \mu, \widehat \psi_l \rangle \big| \leq \big\|Y_{t} - \mu \big\| \big\| \widehat \psi_l - \widetilde \psi_l \big\| + \big\| \widehat \mu - \mu \big\|,
\end{align}
and, for any $j > K$,
\begin{align}
		&\big|\widehat F_{l,t} - \langle \epsilon_t^*, \widehat \psi_l \rangle \big|
		= \big| \langle Y_t - \widehat \mu - \epsilon_t^*, \widehat \psi_l \rangle \big|
		= \bigg| \langle \mu - \widehat \mu, \widehat \psi_l \rangle + \sum_{k=1}^K \widetilde F_{k,t} \langle \widetilde \psi_k, \widehat \psi_l \rangle \bigg| \nonumber \\
		&= \bigg| \langle \mu - \widehat \mu, \widehat \psi_l \rangle + \sum_{k=1}^K \widetilde F_{k,t} \langle \widetilde \psi_k - \widehat\psi_k, \widehat \psi_l \rangle \bigg|
		\leq  \big\| \mu - \widehat \mu \big\| + \sum_{k=1}^K \big| \widetilde F_{k,t} \big| \big\|  \widetilde \psi_k - \widehat\psi_k \big\|. \label{eq:lem5_aux2}
\end{align}
\textit{Proof of item (a):}
By the triangle inequality,
\begin{align}
	&\bigg\| \frac{1}{T} \sum_{t=m+1}^T \widehat{\bm x}_{t-1}^{(J,m)}  \eta_{t}' \bigg\|_M
	 \leq \sum_{i=1}^m \bigg\| \frac{1}{T} \sum_{t=m+1}^T \widehat F_{t-i}^{(J)}  \eta_{t}' \bigg\|_M \nonumber \\
	 &\leq \sum_{i=1}^m \bigg\| \frac{1}{T} \sum_{t=m+1}^T \widehat F_{t-i}^{(K)}  \eta_{t}' \bigg\|_M + \sum_{i=1}^m \sum_{l=K+1}^J \sum_{h=1}^K \bigg|  \frac{1}{T} \sum_{t=m+1}^T   \widehat F_{l,t-i} \eta_{h,t} \bigg|. \label{eq:lem5a_eq1}
\end{align}
For the first summand of \eqref{eq:lem5a_eq1}, we have, for any $i = 1, \ldots, m$,
\begin{align} \label{eq:lem5a_eq2}
	\bigg\| \frac{1}{T} \sum_{t=m+1}^T \widehat F_{t-i}^{(K)}  \eta_{t}' \bigg\|_M \leq \sum_{h,l=1}^K \bigg| \frac{1}{T} \sum_{t=m+1}^T \big(\widehat F_{l,t-i} - \widetilde F_{l,t-i}\big) \eta_{h,t} \bigg| + \bigg\| \frac{1}{T} \sum_{t=m+1}^T \widetilde F_{t-i} \eta_{t}' \bigg\|_M.
\end{align}
The first summand of \eqref{eq:lem5a_eq2} is $O_P(T^{-1/2})$ by \eqref{eq:lem5_aux1} and Theorem \ref{thm:consistency}.
For the second summand of \eqref{eq:lem5a_eq2}, note that $\widetilde F_t = \sum_{j=0}^\infty B_j \eta_{t-j}$ for some $B_j$ with $\sum_{j=0}^\infty \|B_j \|_M < \infty$ by Assumption \ref{as:VAR}(a).
Then, by the triangle inequality,
$$
\bigg\| \frac{1}{T} \sum_{t=m+1}^T \widetilde F_{t-i} \eta_{t}' \bigg\|_M \leq \sum_{j=0}^\infty \big\|B_j\big\|_M \sum_{h,k=1}^K \bigg| \frac{1}{T} \sum_{t=m+1}^T \eta_{h,t-i-j} \eta_{l,t} \bigg|,
$$
and, by Assumption \ref{as:VAR}(a),
$$
E \bigg[ \Big( \frac{1}{T} \sum_{t=m+1}^T \eta_{h,t-i-j} \eta_{l,t} \Big)^2 \bigg] = \frac{1}{T^2} \sum_{t=m+1}^T E[\eta_{h,t-i-j}^2 \eta_{l,t}^2] = O(T^{-1})
$$
since $E[\eta_{h,t_1-i-j} \eta_{l,t_1}\eta_{h,t_2-i-j} \eta_{l,t_2}] = 0$ for $t_1 \neq t_2$ due to the martingale difference sequence property of $\eta_t$.
Therefore, the second summand of \eqref{eq:lem5a_eq2} is $O_P(T^{-1/2})$, and, consequently, the first summand of \eqref{eq:lem5a_eq1} is $O_P(T^{-1/2})$.
For the second summand of \eqref{eq:lem5a_eq1},
\begin{align} \label{eq:lem5a_eq3}
	\bigg| \frac{1}{T} \sum_{t=m+1}^T \widehat F_{l,t-i} \eta_{h,t} \bigg| \leq \bigg| \frac{1}{T} \sum_{t=m+1}^T \big(\widehat F_{l,t-i} - \langle \epsilon_{t-i}^*, \widehat \psi_l \rangle \big) \eta_{h,t} \bigg| +
	\bigg| \frac{1}{T} \sum_{t=m+1}^T \langle \epsilon_{t-i}^*, \widehat \psi_l \rangle \eta_{h,t} \bigg|.
\end{align}
The first summand of the right-hand side of \eqref{eq:lem5a_eq3} is $O_P(T^{-1/2})$ by \eqref{eq:lem5_aux2} and Theorem \ref{thm:consistency}.
For the second summand of \eqref{eq:lem5a_eq3}, the Cauchy-Schwarz inequality yields
$$
	\bigg| \frac{1}{T} \sum_{t=m+1}^T \langle \epsilon_{t-i}^*, \widehat \psi_l \rangle \eta_{h,t} \bigg| \leq \bigg\| \frac{1}{T} \sum_{t=m+1}^T \epsilon_{t-i}^* \eta_{h,t} \bigg\| = O_P(T^{-1/2}),
$$
where the last step follows by Assumption \ref{as:VAR}(b) and the fact that $\eta_t = F_t^* - \sum_{j=1}^p A_j F_{t-j}^*$ by Assumption \ref{as:VAR}(a).
Consequently, \eqref{eq:lem5a_eq1} is $O_P(T^{-1/2})$.
\\
\noindent
\textit{Proof of item (b):}
By the triangle inequality, we have
$$
	\sum_{l=1}^{J} \bigg\| \frac{1}{T} \sum_{t=m+1}^T   \widehat{\bm x}_{t-1}^{(J,m)} \langle \widehat \psi_l, \epsilon_t^* \rangle \bigg\|_M
\leq \sum_{h,l=1}^J \sum_{i=1}^m \bigg| \frac{1}{T} \sum_{t=m+1}^T \widehat F_{h,t-i} \langle \widehat \psi_l, \epsilon_t^* \rangle  \bigg|.
$$
and it remains to show that $T^{-1} \sum_{t=m+1}^T \widehat F_{h,t-i} \langle \widehat \psi_l, \epsilon_t^* \rangle  = O_P(T^{-1/2})$ for any $i=1, \ldots, m$ and $h,l = 1, \ldots, J$.
We follow the same steps as in the proof of item (b) and first consider the case $h \leq K$, where
$$
	\bigg| \frac{1}{T} \sum_{t=m+1}^T \widehat F_{h,t-i} \langle \widehat \psi_l, \epsilon_t^* \rangle  \bigg| \leq \bigg| \frac{1}{T} \sum_{t=m+1}^T \big(\widehat F_{h,t-i} - \widetilde F_{h,t-i}\big) \langle \widehat \psi_l, \epsilon_t^* \rangle \bigg| + \bigg| \frac{1}{T} \sum_{t=m+1}^T \widetilde F_{h,t-i} \langle \widehat \psi_l, \epsilon_t^* \rangle \bigg|.
$$
The first summand is $O_P(T^{-1/2})$ by \eqref{eq:lem5_aux1} and Theorem \ref{thm:consistency}, and the second summand is $O_P(T^{-1/2})$ since $\epsilon_t^*$ is a m.d.s.\ with respect to $\{\epsilon_{t-1}^*, \widetilde F_{t-1}, \epsilon_{t-2}^*, \widetilde F_{t-2}, \ldots\}$.
For the case $h > K$, we have
$$
	\bigg| \frac{1}{T} \sum_{t=m+1}^T \widehat F_{h,t-i} \langle \widehat \psi_l, \epsilon_t^* \rangle  \bigg|
	\leq \bigg| \frac{1}{T} \sum_{t=m+1}^T \big( \widehat F_{h,t-i} - \langle \epsilon_{t-i}^*, \widehat \psi_h \rangle \big) \langle \widehat \psi_l, \epsilon_t^* \rangle  \bigg| + \bigg| \frac{1}{T} \sum_{t=m+1}^T \langle \epsilon_{t-i}^*, \widehat \psi_h \rangle \langle \widehat \psi_l, \epsilon_t^* \rangle  \bigg|,
$$
where the first summand is $O_P(T^{-1/2})$ by \eqref{eq:lem5_aux2} and Theorem \ref{thm:consistency}. The second summand is $O_P(T^{-1/2})$ analogously to \eqref{eq:lem3_H4result} in the proof of Lemma \ref{lem:L3} because $\epsilon_t^*$ is a m.d.s.\ with respect to $\{\epsilon_{t-1}^*, \epsilon_{t-2}^*, \ldots \}$.
\\
\textit{Proof of item (c):}
By the Cauchy-Schwarz and the triangle inequality,
\begin{align*}
	&\bigg\| \frac{1}{T} \sum_{t=m+1}^T \widehat{\bm x}_{t-1}^{(J,m)} ( \bm{\widetilde x}_{t-1} - \widehat{\bm x}_{t-1}^{(K,p)})' \bigg\|_M
	\leq \frac{1}{T} \sum_{t=m+1}^T \big\|\widehat{\bm x}_{t-1}^{(J,m)} \big\|_2 \big\| \bm{\widetilde x}_{t-1} - \widehat{\bm x}_{t-1}^{(K,p)} \big\|_2  \\
	&\leq \frac{1}{T} \sum_{t=m+1}^T \sum_{i=1}^p \sum_{l=1}^K \big\|\widehat{\bm x}_{t-1}^{(J,m)} \big\|_2 \big|\widetilde F_{l,t-i} - \widehat F_{l,t-i}\big| = O_P(T^{-1/2}),
\end{align*}
where the last step follows from \eqref{eq:lem5_aux1} and Theorem \ref{thm:consistency}.

\subsection{Proof of Lemma \ref{lem:L5}} \label{sec:proofL5}

We denote the overselection case ($J \geq K$ and $m \geq p$) as case I and the underselection case ($J < K$ or $m < p$ or both) as case II.\\
\textit{Proof of statement (a):} From equation \eqref{eq:estimatedpredictorcurve} the predictor curves have the representations
$$
\widehat Y_{t|t-1}^{(K,p)}(r) - \widehat \mu(r)
	= \big(\widehat \Psi^{(J^*)}(r)\big)' \bm{\widehat A}^* \bm{\widehat x}_{t-1}^{(J^*,m^*)}
$$
and
\begin{align*}
	\widehat Y_{t|t-1}^{(K,p)}(r) - \widehat \mu(r)
	&= \big(\widehat \Psi^{(K)}(r)\big)' \big( \widehat{\bm A}_{(K,p)} - \bm{\widetilde A}\big) \widehat{\bm x}_{t-1}^{(K,p)} + \big(\widehat \Psi^{(K)}(r)\big)' \bm{\widetilde A} \widehat{\bm x}_{t-1}^{(K,p)} \\
	&= \big(\widehat \Psi^{(K)}(r)\big)' \big( \widehat{\bm A}_{(K,p)} - \bm{\widetilde A}\big) \widehat{\bm x}_{t-1}^{(K,p)} + \big(\widehat \Psi^{(J^*)}(r)\big)' \bm A^* \widehat{\bm x}_{t-1}^{(J^*,m^*)}.
\end{align*}
Then, $\widehat Y_{t|t-1}^{(K,p)}(r) - \widehat Y_{t|t-1}^{(J,m)}(r) = Z_{(1)}(r) + Z_{(2)}(r)$, where
\begin{align} \label{eq:l4.z1z2}
	Z_{(1)}(r) = \big(\widehat \Psi^{(J^*)}(r)\big)' \big(\bm A^* - \widehat{\bm A}^* \big) \widehat{\bm x}_{t-1}^{(J^*,m^*)}, \quad
	Z_{(2)}(r) =\big(\widehat \Psi^{(K)}(r)\big)' \big( \widehat{\bm A}_{(K,p)} - \bm{\widetilde A} \big) \widehat{\bm x}_{t-1}^{(K,p)}.
\end{align}
To simplify the exposition we ignore the additional indices $\{t,T,J,m,K,p\}$ on which $Z_{(1)}$ and $Z_{(2)}$ depend.
To disentangle the loading vectors and matrix products, let $e_l^{(J)}$ be the $l$-th unit vector of length $J$, where the $l$-th entry of $e_l^{(J)}$ is 1, and all other entries are zeros.
For the first term, we have
\begin{align*}
	\big\|Z_{(1)} \big\|^2
	&= \int_a^b \Big( \sum_{l=1}^{J^*} \widehat \psi_l(r) \big(e_l^{(J^*)}\big)' \big(\bm A^* - \widehat{\bm A}^* \big) \widehat{\bm x}_{t-1}^{(J^*,m^*)} \Big)^2 \dd r \\
	&= \sum_{l=1}^{J^*} \Big( \big(e_l^{(J^*)}\big)' \big(\bm A^* - \widehat{\bm A}^* \big) \widehat{\bm x}_{t-1}^{(J^*,m^*)} \Big)^2 \\
	&= \big\| \big(\bm A^* - \widehat{\bm A}^* \big) \widehat{\bm x}_{t-1}^{(J^*,m^*)} \big\|_M^2 \\
	&= \tr\Big( \big(\widehat{\bm x}_{t-1}^{(J^*,m^*)}\big)' \big(\bm A^* - \widehat{\bm A}^* \big)' \big(\bm A^* - \widehat{\bm A}^* \big) \widehat{\bm x}_{t-1}^{(J^*,m^*)} \Big) \\
	&= \tr\Big(  \big(\bm A^* - \widehat{\bm A}^* \big)' \big(\bm A^* - \widehat{\bm A}^* \big) \widehat{\bm x}_{t-1}^{(J^*,m^*)} \big(\widehat{\bm x}_{t-1}^{(J^*,m^*)}\big)' \Big),
\end{align*}
and
\begin{align*}
	\frac{1}{T} \sum_{t=m^*+1}^T \big\|Z_{(1)} \big\|^2 = \tr\Big(  \big(\bm A^* - \widehat{\bm A}^* \big)' \big(\bm A^* - \widehat{\bm A}^* \big) \widehat \Sigma_{(J^*, m^*)} \Big).
\end{align*}
From the proof of Theorem \ref{thm:Bias}, $\| \widehat{\Sigma}_{(J^*, m^*)} - \widetilde \Sigma^* \|_M = o_P(1)$ and $\| \widehat{\Sigma}_{(J^*, m^*)}^{-1} - (\widetilde \Sigma^*)^{-1} \|_M = o_P(1)$.
Consider the Cholesky decompositions $\widehat \Sigma_{(J^*, m^*)} = \widehat \Omega \widehat \Omega'$ and $\widetilde \Sigma^* = \Omega \Omega'$, where $\|\Omega\|_M < \infty$ and $\|\Omega^{-1}\|_M < \infty$.
Then,
\begin{align*}
	\tr\Big(  \big(\bm A^* - \widehat{\bm A}^* \big)' \big(\bm A^* - \widehat{\bm A}^* \big) \widehat \Sigma_{(J^*, m^*)} \Big) = \big\|\big(\bm A^* - \widehat{\bm A}^* \big) \widehat \Omega\big\|_M^2,
\end{align*}
and
\begin{align*}
	\frac{\|(\bm A^* - \widehat{\bm A}^* ) \widehat \Omega\|_M^2}{\|\bm A^* - \widehat{\bm A}^* \|_M^2}
	&\leq \|\widehat \Omega\|_M^2 = O_P(1), \\
	\frac{\|\bm A^* - \widehat{\bm A}^*  \|_M^2}{\|(\bm A^* - \widehat{\bm A}^*) \widehat \Omega \|_M^2}
	&= \frac{\|(\bm A^* - \widehat{\bm A}^* ) \widehat \Omega \widehat \Omega^{-1} \|_M^2}{\|(\bm A^* - \widehat{\bm A}^* ) \widehat \Omega \|_M^2}
	\leq \|\widehat \Omega^{-1}\|_M^2 = O_P(1),
\end{align*}
which implies that $T^{-1} \sum_{t=m^*+1}^T \|Z_{(1)} \|^2$ is of exactly the same order as $\| \bm A^* - \widehat{\bm A}^* \|_M^2$.
By Theorem \ref{thm:Bias}, we have $\| \bm A^* - \widehat{\bm A}^* \|_M^2 = O_P(T^{-1})$ for case I and $\| \bm A^* - \widehat{\bm A}^* \|_M^2 = \Theta_P(1)$ for case II, which implies that
\begin{align*}
	\frac{1}{T} \sum_{t=m^*+1}^T \big\|Z_{(1)} \big\|^2 = \begin{cases} O_P(T^{-1}) & \text{for case I,} \\ \Theta_P (1) & \text{for case II.} \end{cases}
\end{align*}
For the second term, by the orthonormality of the loadings,
$$
\big\|Z_{(2)}\big\|^2 = \big\| \big( \widehat{\bm A}_{(K,p)} - \bm{\widetilde A} \big) \widehat{\bm x}_{t-1}^{(K,p)} \big\|_M^2
\leq  \big\|  \widehat{\bm A}_{(K,p)} - \bm{\widetilde A} \big\|_M^2 \big\| \widehat{\bm x}_{t-1}^{(K,p)} \big\|_M^2,
$$
and, for both cases,
$$
	\frac{1}{T} \sum_{t=m^*+1}^T \big\|Z_{(2)} \big\|^2 \leq \frac{1}{T} \sum_{t=m^*+1}^T \big\| \widehat{\bm x}_{t-1}^{(K,p)} \big\|_M^2 \big\| \widehat{\bm A}_{(K,p)} - \bm{\widetilde A} \big\|_M^2
	= O_P(T^{-1})
$$
by Theorem \ref{thm:Bias} and the fact that enough moments are bounded.
Finally, for the cross term,
\begin{align*}
	\frac{1}{T} \sum_{t=m^*+1}^T \langle Z_{(1)} , Z_{(2)}  \rangle &\leq \frac{1}{T} \sum_{t=m^*+1}^T \big\| \widehat{\bm x}_{t-1}^{(K,p)} \big\|_2 \big\| \widehat{\bm x}_{t-1}^{(J,m)}\|_M \big\| \widehat{\bm A}_{(K,p)} - \bm{\widetilde A} \big\|_M \big\|\bm A^* - \widehat{\bm A}^* \big\|_2,
\end{align*}
which is $O_P(T^{-1})$ for case I and $O_P(T^{-1/2})$ for case II by Theorem \ref{thm:Bias}.
Since
\begin{align*}
	\frac{1}{T} \sum_{t=m^*+1}^T \big\| \widehat Y_{t|t-1}^{(K,p)} - \widehat Y_{t|t-1}^{(J,m)} \big\|^2
	&= \frac{1}{T} \sum_{t=m^*+1}^T \big( \|Z_{(1)}\|^2 + \|Z_{(2)}\|^2 + 2 \langle Z_{(1)} , Z_{(2)}  \rangle \big),
\end{align*}
statement (a) follows. \\
\textit{Proof of statement (b):}
From equations \eqref{eq:dynamicrepresentation} and \eqref{eq:predictorcurve}, it follows that
$$
		Y_t(r) - \widetilde Y_{t|t-1}(r) = Z_{(3)}(r) + Z_{(4)}(r), \quad Z_{(3)}(r) = (\widetilde \Psi(r))' \eta_t, \quad  Z_{(4)}(r) = \epsilon_t^*(r).
$$
Using the definitions of $Z_{(1)}$ and $Z_{(2)}$ in \eqref{eq:l4.z1z2}, by Theorem \ref{thm:Bias} it remains to show that
$$
	\frac{1}{T} \sum_{t=m^*+1}^T \langle Z_{(1)} + Z_{(2)} , Z_{(3)} + Z_{(4)} \rangle = O_P(T^{-1/2} \|\bm A^* - \widehat{\bm A}^* \|_M).
$$
We consider the four terms $\langle Z_{(i)} , Z_{(j)} \rangle$ for $i = 1,2$ and $j = 3,4$ separately.
For the first term, by the properties of the trace,
\begin{align*}
	\langle Z_{(1)}, Z_{(3)}\rangle
	&= \int_a^b \eta_t' (\widetilde \Psi(r))(\widehat \Psi^{(J^*)}(r))' (\bm A^* - \widehat{\bm A}^*) \widehat{\bm x}_{t-1}^{(J^*,m^*)} \dd r \\
	&= \tr\bigg( \Big( \int_a^b  (\widetilde \Psi(r))(\widehat \Psi^{(J^*)}(r))' \dd r \Big) (\bm A^* - \widehat{\bm A}^*) \widehat{\bm x}_{t-1}^{(J^*,m^*)} \eta_t' \bigg),
\end{align*}
and, by the Cauchy-Schwarz inequality for the trace,
$$
	\frac{1}{T} \sum_{t=m^*+1}^T \langle Z_{(1)}, Z_{(3)}\rangle \leq \big\|\bm A^* - \widehat{\bm A}^* \big\|_M \bigg\| \frac{1}{T} \sum_{t=m^*+1}^T \widehat{\bm x}_{t-1}^{(J^*,m^*)} \eta_t' \bigg\|_M = O_P\big(T^{-1/2} \|\bm A^* - \widehat{\bm A}^* \|_M \big),
$$
where the last step follows from Lemma \ref{lem:L4}(a).
Analogously, for the second term,
\begin{align*}
	\langle Z_{(2)}, Z_{(3)}\rangle
	&= \int_a^b \eta_t' (\widetilde \Psi(r))(\widehat \Psi^{(K)}(r))' ( \widehat{\bm A}_{(K,p)} - \bm{\widetilde A}) \widehat{\bm x}_{t-1}^{(K,p)} \dd r \\
	&= \tr\bigg( \Big( \int_a^b  (\widetilde \Psi(r))(\widehat \Psi^{(K)}(r))' \dd r \Big) (\widehat{\bm A}_{(K,p)} - \bm{\widetilde A}) \widehat{\bm x}_{t-1}^{(K,p)} \eta_t' \bigg),
\end{align*}
and
$$
	\frac{1}{T} \sum_{t=m^*+1}^T \langle Z_{(2)}, Z_{(3)}\rangle \leq \big\|\widehat{\bm A}_{(K,p)} - \bm{\widetilde A} \big\|_M  \bigg\| \frac{1}{T} \sum_{t=m^*+1}^T \widehat{\bm x}_{t-1}^{(K,p)} \eta_t' \bigg\|_M = O_P\big(T^{-1/2} \|\bm A^* - \widehat{\bm A}^* \|_M \big).
$$
For the third term, we have
\begin{align*}
	\langle Z_{(1)}, Z_{(4)}\rangle
	&= \int_a^b \epsilon_t^*(r) (\widehat \Psi^{(J^*)}(r))' (\bm A^* - \widehat{\bm A}^*) \widehat{\bm x}_{t-1}^{(J^*,m^*)} \dd r \\
	&= \tr\bigg( (\bm A^* - \widehat{\bm A}^*)   \sum_{l=1}^{J^*} \widehat{\bm x}_{t-1}^{(J^*,m^*)}  \langle \epsilon_t^*, \widehat \psi_l \rangle  \bigg),
\end{align*}
and
$$
	\frac{1}{T} \sum_{t=m^*+1}^T \langle Z_{(1)}, Z_{(4)}\rangle \leq \big\|\bm A^* - \widehat{\bm A}^* \big\|_M \bigg\| \frac{1}{T} \sum_{t=m^*+1}^T \sum_{l=1}^{J^*} \widehat{\bm x}_{t-1}^{(J^*,m^*)}  \langle \epsilon_t^*, \widehat \psi_l \rangle  \bigg\|_M,
$$
which is $O_P(T^{-1/2} \|\bm A^* - \widehat{\bm A}^* \|_M )$ by Lemma \ref{lem:L4}(b).
Finally, for the fourth term,
\begin{align*}
	\langle Z_{(2)}, Z_{(4)}\rangle
	&= \int_a^b \epsilon_t^*(r) (\widehat \Psi^{(K)}(r))' ( \widehat{\bm A}_{(K,p)} - \bm{\widetilde A}) \widehat{\bm x}_{t-1}^{(K,p)} \dd r \\
	&= \tr\bigg( (\widehat{\bm A}_{(K,p)} - \bm{\widetilde A})  \sum_{l=1}^{K} \widehat{\bm x}_{t-1}^{(K,p)}  \langle \epsilon_t^*, \widehat \psi_l \rangle  \bigg),
\end{align*}
and
$$
	\frac{1}{T} \sum_{t=m^*+1}^T \langle Z_{(2)}, Z_{(4)}\rangle \leq \big\|\widehat{\bm A}_{(K,p)} - \bm{\widetilde A} \big\|_M  \bigg\| \frac{1}{T} \sum_{t=m^*+1}^T \sum_{l=1}^{K} \widehat{\bm x}_{t-1}^{(K,p)}  \langle \epsilon_t^*, \widehat \psi_l \rangle \bigg\|_M,
$$
which is $O_P(T^{-1/2} \|\bm A^* - \widehat{\bm A}^* \|_M )$ by Lemma \ref{lem:L4}(b) as well. \\
\textit{Proof of statement (c):}
We decompose
\begin{align*}
	\widetilde Y_{t|t-1}(r) - \widehat Y_{t|t-1}^{(K,p)}(r) &= \mu(r) + \big( \widetilde \Psi(r) \big)' \bm{ \widetilde A} \bm{\widetilde x}_{t-1}
	- \widehat \mu(r) - \big( \widehat \Psi^{(K)}(r) \big)' \widehat{\bm A}_{(K,p)} \widehat{\bm x}_{t-1}^{(K,p)} \\
	 &= Z_{(5)}(r) + Z_{(6)}(r) + Z_{(7)}(r) + Z_{(8)}(r),
\end{align*}
where
\begin{align*}
	Z_{(5)}(r) &= \mu(r) - \widehat \mu(r), \quad
	Z_{(6)}(r) = \big( \widetilde \Psi(r) - \widehat \Psi^{(K)}(r) \big)' \bm{ \widetilde A} \bm{\widetilde x}_{t-1}, \\
	Z_{(7)}(r) &=\big(\widehat \Psi^{(K)}(r) \big)' (\bm{ \widetilde A} - \widehat{\bm A}_{(K,p)}) \bm{\widetilde x}_{t-1}, \quad
	Z_{(8)}(r) = \big(\widehat \Psi^{(K)}(r)\big)' \widehat{\bm A}_{(K,p)} (\bm{\widetilde x}_{t-1} - \widehat{\bm x}_{t-1}^{(K,p)}).
\end{align*}
It remains to show that
\begin{align*}
	\frac{1}{T} \sum_{t=m^*+1}^T \langle Z_{(1)} + Z_{(2)} , Z_{(5)} + Z_{(6)} + Z_{(7)} + Z_{(8)} \rangle = O_P\big(T^{-1/2} \|\bm A^* - \widehat{\bm A}^* \|_M\big).
\end{align*}
We consider the four terms $\langle Z_{(1)} + Z_{(2)}, Z_{(j)} \rangle$ for $j = 5,6,7,8$ separately.
First, from the proof of statement (a),
\begin{align*}
	\frac{1}{T} \sum_{t=m^*+1}^T \big( \| Z_{(1)} \| + \| Z_{(2)} \| \big) = O_P\big(\|\bm A^* - \widehat{\bm A}^* \|_M\big),
\end{align*}
which, together with Theorem \ref{thm:consistency}(a), implies that
\begin{align*}
\bigg| \frac{1}{T} \sum_{t=m^*+1}^T \langle Z_{(1)} + Z_{(2)}, Z_{(5)}\rangle \bigg|
 &\leq \frac{1}{T} \sum_{t=m^*+1}^T \big( \| Z_{(1)} \| + \| Z_{(2)} \|  \big) \| \mu - \widehat \mu \| \\
 &=  O_P\big(T^{-1/2} \|\bm A^* - \widehat{\bm A}^* \|_M\big).
\end{align*}
For the second term, by the Cauchy-Schwarz inequality and the orthonormality of the loadings, we have
\begin{align*}
	\langle Z_{(1)}, Z_{(6)} \rangle
	&= \tr\bigg(\int_a^b \big(\widehat{\bm x}_{t-1}^{(J^*,m^*)}\big)' \big(\bm A^* - \widehat{\bm A}^* \big)' \big(\widehat \Psi^{(J^*)}(r)\big) \big( \widetilde \Psi(r) - \widehat \Psi^{(K)}(r) \big)' \bm{ \widetilde A} \bm{\widetilde x}_{t-1} \dd r \bigg) \\
	&\leq \big\| \bm{\widetilde A}  \bm{\widetilde x}_{t-1} \big(\widehat{\bm x}_{t-1}^{(J^*,m^*)}\big)' \big\|_M \big\| \bm A^* - \widehat{\bm A}^* \big\|_M \sum_{l=1}^K  \big\| \widetilde \psi_l - \widehat \psi_l \big\| \\
	\langle Z_{(2)}, Z_{(6)} \rangle
	&= \tr\bigg( \int_a^b \big(\widehat{\bm x}_{t-1}^{(K,p)}\big)' \big( \widehat{\bm A}_{(K,p)} - \bm{\widetilde A} \big)' \big(\widehat \Psi^{(K)}(r)\big) \big( \widetilde \Psi(r) - \widehat \Psi^{(K)}(r) \big)' \bm{ \widetilde A} \bm{\widetilde x}_{t-1} \dd r \bigg) \\
	&\leq \big\| \bm{\widetilde A}  \bm{\widetilde x}_{t-1} \big(\widehat{\bm x}_{t-1}^{(K,p)}\big)' \big\|_M \big\| \widehat{\bm A}_{(K,p)} - \bm{\widetilde A}\big\|_M \sum_{l=1}^K  \big\| \widetilde \psi_l - \widehat \psi_l \big\|,
\end{align*}
and Theorem \ref{thm:consistency}(e) implies
$$
	\bigg| \frac{1}{T} \sum_{t=m^*+1}^T  \langle Z_{(1)} + Z_{(2)}, Z_{(6)} \rangle \bigg|  = O_P\big(T^{-1/2} \|\bm A^* - \widehat{\bm A}^* \|_M\big).
$$
For the third term,
\begin{align*}
	\langle Z_{(1)}, Z_{(7)} \rangle
	&= \tr\bigg( \int_a^b \big(\widehat{\bm x}_{t-1}^{(J^*,m^*)}\big)' \big(\bm A^* - \widehat{\bm A}^* \big)' \big(\widehat \Psi^{(J^*)}(r)\big) \big(\widehat \Psi^{(K)}(r) \big)' (\bm{ \widetilde A} - \widehat{\bm A}_{(K,p)}) \bm{\widetilde x}_{t-1} \dd r \bigg) \\
	&\leq K \big\| \bm A^* - \widehat{\bm A}^* \big\|_M \big\| \bm{ \widetilde A} - \widehat{\bm A}_{(K,p)} \big\|_M \big\| \bm{\widetilde x}_{t-1} \big(\widehat{\bm x}_{t-1}^{(J^*,m^*)}\big)' \big\|_M,
	\\
	\langle Z_{(2)}, Z_{(7)} \rangle &= \tr\bigg( \int_a^b \big(\widehat{\bm x}_{t-1}^{(K,p)}\big)' \big( \widehat{\bm A}_{(K,p)} - \bm{\widetilde A} \big)' \big(\widehat \Psi^{(K)}(r)\big) \big(\widehat \Psi^{(K)}(r) \big)' (\bm{ \widetilde A} - \widehat{\bm A}_{(K,p)}) \bm{\widetilde x}_{t-1} \dd r \bigg) \\
	&\leq K \big\| \bm{ \widetilde A} - \widehat{\bm A}_{(K,p)} \big\|_M^2 \big\| \bm{\widetilde x}_{t-1} \big(\widehat{\bm x}_{t-1}^{(K,p)}\big)' \big\|_M,
\end{align*}
and Theorem \ref{thm:Bias} implies
$$
	\frac{1}{T} \sum_{t=m^*+1}^T  \langle Z_{(1)} + Z_{(2)}, Z_{(7)} \rangle  = O_P\big(T^{-1/2} \|\bm A^* - \widehat{\bm A}^* \|_M\big).
$$
Finally, for the fourth term,
\begin{align*}
	\langle Z_{(1)}, Z_{(8)} \rangle
	&= \tr\bigg( \int_a^b \big(\widehat{\bm x}_{t-1}^{(J^*,m^*)}\big)' \big(\bm A^* - \widehat{\bm A}^* \big)' \big(\widehat \Psi^{(J^*)}(r)\big) \big(\widehat \Psi^{(K)}(r)\big)' \widehat{\bm A}_{(K,p)} (\bm{\widetilde x}_{t-1} - \widehat{\bm x}_{t-1}^{(K,p)}) \dd r \bigg) \\
	\langle Z_{(2)}, Z_{(8)} \rangle &= \tr\bigg( \int_a^b \big(\widehat{\bm x}_{t-1}^{(K,p)}\big)' \big( \widehat{\bm A}_{(K,p)} - \bm{\widetilde A} \big)' \big(\widehat \Psi^{(K)}(r)\big) \big(\widehat \Psi^{(K)}(r)\big)' \widehat{\bm A}_{(K,p)} (\bm{\widetilde x}_{t-1} - \widehat{\bm x}_{t-1}^{(K,p)}) \dd r \bigg)
\end{align*}
and
\begin{align*}
	&\frac{1}{T} \sum_{t=m^*+1}^T  \langle Z_{(1)} + Z_{(2)}, Z_{(8)} \rangle \\
	&\leq 2 K \bigg\| \frac{1}{T} \sum_{t=m^*+1}^T (\bm{\widetilde x}_{t-1} - \widehat{\bm x}_{t-1}^{(K,p)}) \big(\widehat{\bm x}_{t-1}^{(J^*,m^*)}\big)' \bigg\|_M \big\|\bm A^* - \widehat{\bm A}^*\big\|_M \big\|\widehat{\bm A}_{(K,p)}\big\|_M,
\end{align*}
which is $O_P(T^{-1/2} \|\bm A^* - \widehat{\bm A}^* \|_M)$ by Lemma \ref{lem:L4}(c).

\newpage

\bibliographystyle{apalike}
\bibliography{references}

\end{document}